\documentclass[12pt]{article}
\usepackage{amsmath}
\usepackage{amsfonts}
\usepackage{amssymb}
\usepackage{amsthm}
\usepackage{mathtools}
\usepackage{graphicx,psfrag,epsf}
\usepackage{enumerate}
\usepackage{natbib}
\usepackage{multirow}
\usepackage{array, booktabs}
\usepackage{lscape}
\usepackage[ruled,linesnumbered]{algorithm2e}
\usepackage{hyperref}
\usepackage{bm}
\usepackage{bbm}

\newcommand{\blind}{0}

\makeatletter
\newcommand{\removelatexerror}{\let\@latex@error\@gobble}
\makeatother

\title{FastMUOD Theory Draft Paper}
\author{seguntaiwoojo}
\date{\today}

\newtheorem{definition}{Definition}
\newtheorem{proposition}{Proposition}
\newtheorem{corollary}{Corollary}
\begin{document}
\def\spacingset#1{\renewcommand{\baselinestretch}%
{#1}\small\normalsize} \spacingset{1}

\newcolumntype{P}[1]{>{\raggedright\arraybackslash}p{#1}}
\newcommand{\ra}[1]{\renewcommand{\arraystretch}{#1}}

%%%%%%%%%%%%%%%%%%%%%%%%%%%%%%%%%%%%%%%%%%%%%%%%%%%%%%%%%%%%%%%%%%%%%%%%%%%%%%
% FastMUOD Theory - Analysis of the FastMUOD FDA outlier detection method and extension to multivariate data
\if0\blind
{
  \title{\bf Multivariate Functional Outlier Detection using the FastMUOD Indices}
  \author{Oluwasegun Taiwo Ojo
  %\thanks{The authors gratefully acknowledge}\hspace{.2cm}
    \\
    \small{IMDEA Networks Institute, Leganés, Spain}\\
    \small{Universidad Carlos III de Madrid, Spain} \\ 
    Antonio  Fern\'andez Anta \\
    \small{IMDEA Networks Institute, Leganés, Spain} \\
    Marc G. Genton  \\
    \small{Statistics Program, King Abdullah University of Science and Technology, Thuwal 23955-6900, Saudi Arabia}\\
    Rosa E. Lillo  \\
    \small{uc3m-Santander Big Data Institute, Getafe, Spain}\\
\small{Department of Statistics, Universidad Carlos III de Madrid, Spain}
    }
  \maketitle
} \fi

\if1\blind
{
  \bigskip
  \bigskip
  \bigskip
  \begin{center}
    {\LARGE\bf FastMUOD Theory Draft Paper}
\end{center}
  \medskip
} \fi
\medskip
\begin{abstract}
We present definitions and properties of the fast massive unsupervised outlier detection (FastMUOD) indices, used for outlier detection (OD) in functional data. FastMUOD detects outliers by computing, for each curve, an amplitude, magnitude and shape index meant to target the corresponding types of outliers. Some methods adapting FastMUOD to outlier detection in multivariate functional data are then proposed. These include applying FastMUOD on the components of the multivariate data and using random projections. Moreover, these techniques are tested on various simulated and real multivariate functional datasets. Compared with the state of the art in multivariate functional OD, the use of random projections showed the most effective results with similar, and in some cases improved, OD performance.
\end{abstract}
\noindent%
{\it Keywords: FastMUOD, functional data, functional outlier detection, multivariate functional data, outlier classification, video data}
\spacingset{1.45}
\section{Introduction}
We consider the problem of detecting outliers in a collection of multivariate functional observations. In particular, we consider observations of the form: $\{\bm{Y}_i(t), t\in \mathcal{I}\}_{i = 1}^n$, wherein a vector $\bm{Y}_i(t)\in \mathbb{R}^d$, $d\in \mathbb{N}$, is observed at a domain point $t$ in the interval $\mathcal{I}$. Such vector-valued functional observations are increasingly observed in real-life studies and various physical and environmental applications. Thus, exploratory methods for multivariate functional data have been recently garnering considerable interest. 

Outlier detection (OD), a part of the exploratory data analysis process, involves identifying observations that differ from the bulk of the data, either because they come from a different distribution compared with the bulk or because they lie at the extremes of the distribution of the data. However, identifying outliers is more complicated when observations are functions observed on a domain, i.e., functional data. Functional observations demonstrate different outlying behaviours, e.g., a vertical shift, compared to the bulk of the data (magnitude outliers) or a horizontal shift, in which case the outlying function is not well aligned with the bulk of the data. Functional outliers can also have different shapes or follow different paths compared to the bulk of the data. \cite{Hubert2015} proposed a taxonomy for different types of functional outliers based on the different outlying behaviours they exhibit, and whether such behaviours can be observed in a small part of the domain or throughout the domain; \cite[see also][]{dai2020}.

To identify outliers among multivariate (non-functional) observations (i.e., vector observations $\bm{X}\in \mathbb{R}^d$), it is typical to order the observations, from the center outward using a notion of \emph{statistical depth}. Then, the observations having the lowest depth values can be closely examined for outlying behaviours. This procedure is convenient because most depth notions are non-parametric and they do not require any assumption concerning the underlying data distribution. 

The approach mentioned above has also caught on in the analysis of functional observations, where several OD methods are based on notions of functional depths. For example, the {\em functional boxplot} \citep{sun2011functional} uses the modified band depth \citep{lopez2009concept} to order functional observations and define a 50\% central region. Then, the outliers are functions that lie outside of the central region inflated by 1.5, similar to the classical boxplot. Other proposals around this theme include \cite{sguera2016functional} and \cite{febrero2008}, where functional depth measures were used for OD.

On the other hand, several functional OD methods are based on ``custom-built" outlyingness indices, metrics, or pseudo-depths directly targeted toward OD, instead of ordering (as with functional depth notions). Examples along this line include the {\em magnitude-shape plot} (MS-plot) \citep{Dai2018}, based on the directional outlyingness proposed by \cite{dai2019directional}; the {\em functional outlier map} (FOM), based on another (functional) directional outlyingness proposed by \cite{Rousseeuw2018}; the {\em modified shape similarity} index (MSS) proposed in \cite{huang2019decomposition}; the (robustified) {\em functional tangential angle} (rFUNTA) proposed in \cite{kunt2016}, and an earlier proposal of \cite{Hubert2015} in which the {\em bag distance} and {\em skewness adjusted projection depth} were proposed for functional OD. %While MS-plot, FOM, (rFUNTA), bag distance and the skewness adjusted projection depth all work on multivariate functional data, MSS was proposed for outlier detection in univariate functional data.   

Finally, certain functional OD procedures are based on either a combination of depth notions and outlyingness indices, or the use of more primitive methods (such as dimension reduction or transformation). Some of these include the \emph{outliergram}, based on the modified epigraph index \citep{lopez2011half} and the modified band depth; the {\em functional bagplots}, and the {\em functional highest density regions} \citep{hyndman2010rainbow}, both using the first two robust principal components of the functional data to construct plots used for detecting functional outliers. Likewise, \cite{dai2020} proposed detecting functional outliers using a sequence of (functional) data transformations, each followed by a functional boxplot to detect different types of outliers. Recently,  \cite{herrmann2021geometric} proposed using multidimensional scaling \citep{Cox2008} to reduce functional data to lower dimensional embeddings. Then, an OD method such as the local outlier factors \citep{lof2000} was applied on the embeddings to detect outlying curves.

Fast Massive Unsupervised Outlier Detection (FastMUOD), introduced by \cite{Ojo2021}, belongs to the second group of functional OD methods (outlined above) because it uses three indices, each targeting different outlying behaviours that functional outliers may exhibit. The FastMUOD indices are the \emph{magnitude index}, which targets magnitude outliers; the \emph{shape index}, which targets shape outliers; and the \emph{amplitude index}, which targets amplitude outliers. Because these indices target different outlier types, the outliers identified are also classified as per their types, unsupervised, without the need for inspection or visualisation of the data. The method is fast and simple, making it scalable to (and suitable for) ``big" functional data analysis. 

Nevertheless, despite its advantages, FastMUOD has its limitations. First, its indices are designed for univariate functional data. Second, it is not exactly clear from \cite{Ojo2021} why the FastMUOD indices are suitable for OD from a theoretical perspective, despite the good and scalable performance observed on simulated and real datasets. In this study, we aim to address these two issues by 1) exploring the properties of the FastMUOD indices rigorously and by 2) extending the FastMUOD indices to OD in multivariate functional data.   

The rest of the article is organised as follows. In Section \ref{sec::sec2}, we present definitions and properties of the FastMUOD indices.  Next, in Section \ref{sec::sec3}, we describe several extensions of the FastMUOD indices to outlier detection in multivariate functional data. In Section \ref{sec::sec4}, we evaluate the extensions presented in Section \ref{sec::sec3} in a simulation study. We demonstrate the extensions on two real data applications in Section \ref{sec::sec5}. We end the article with some concluding remarks in Section \ref{sec::sec6}.

\section{Definitions and Properties of the FastMUOD Indices}
\label{sec::sec2}
We present the sample and population definitions of the FastMUOD indices and explore their properties. First, we describe the notations used in this article. We assume that functions are defined on the unit interval $[0, 1]$ and denote by $L^2([0, 1])$, the space of all square-integrable functions defined over $[0, 1]$. We denote by $\langle f, g\rangle$ (unless otherwise specified), the inner product of two functions $f,g\in L^2([0,1])$. The norm of $f\in L^2([0, 1])$ induced by this inner product is denoted by $\Vert f \Vert$.

\subsection{Definitions of the Univariate FastMUOD Indices }
\label{subsec::deffmuod}
\begin{definition}[Definitions of FastMUOD indices]
Let $X$ be a stochastic process in $L^2([0,1])$ with distribution $F_X$ and $\mu(t) = \mathbb{E}[X(t)]$ be its population mean function. We define the \textbf{shape index} of a function $y\in L^2([0,1])$ (which may be a realization of $X$) with respect to (w.r.t.) $F_X$ as 
\begin{equation*}
I_S(y, F_X) \coloneqq 1 - \frac{\int\tilde{y}(t)\tilde{\mu}(t)dt}{\left[\int \tilde{y}(t) ^2 dt \right]^{1/2}\left[ \int \tilde{\mu}(t) ^2 dt \right]^{1/2}} =  1 - \frac{\langle \tilde{y}, \tilde{\mu}\rangle}{\Vert \tilde{y} \Vert  \cdot \Vert \tilde{\mu} \Vert},
\end{equation*}
where $\tilde{y}(t)$ and $\tilde{\mu}(t)$ denote the centered curves given by: $\tilde{y}(t)\coloneqq y(t) - \int y(r) dr, $ and $\tilde{\mu}(t) \coloneqq \mu(t) - \int \mu(r) dr$, respectively. We define the \textbf{amplitude index} of $y$ w.r.t. $F_X$ as
\begin{equation*}
     I_A(y, F_X) \coloneqq \frac{\int\tilde{y}(t)\tilde{\mu}(t)dt}{ \int \tilde{\mu}(t) ^2 dt}  - 1 = \frac{\langle\tilde{y},\tilde{\mu}\rangle }{ \Vert \tilde{\mu} \Vert ^2}  - 1. 
\end{equation*}
Finally, we define the \textbf{magnitude index} of a function $y$ w.r.t. $F_X$ as \begin{equation*}
    I_M(y, F_X) \coloneqq   \int y(t)dt - \beta(y)\int \mu(t)dt,
\end{equation*}
where $\beta(y) =I_A(y, F_X) + 1.$
\end{definition}

In practice, functions are usually observed on a finite number of points in the domain. In this case, an approximation to the FastMUOD indices can be obtained by replacing the integral with a summation, yielding the following trivial definitions, which we include for completeness.

\begin{definition}[Finite-dimensional approximation of FastMUOD indices]
Suppose the function $y$ is observed on the finite points $T = \{t_1 = 0, t_2, \ldots, t_k = 1\} \subset [0, 1]$ with $t_j - t_{j-1} = \Delta$, a constant.  Moreover, let ${\mu}(t_j) = \mathbb{E}[X(t_j)]$  for all $t_j \in T$. Then, we define the finite-dimensional version of the shape index of $y$ w.r.t $F_X$ as:
% Suppose the stochastic process $X(t)$ and the function $y$ are observed on the same finite points $T = \{t_1, t_2, \ldots, t_k \} \in [0, 1]$ with $t_j - t_{j-1} = \Delta$, a constant. 
\begin{equation*}
    I_{S_k}(y, F_X) \coloneqq 1 - \frac{\sum_{j = 1}^k\tilde{y}(t_j)\tilde{\mu}(t_j)}{\left[\sum_{j = 1}^k \tilde{y}(t_j) ^2  \right]^{1/2}\left[ \sum_{j = 1}^k \tilde{\mu}(t_j)^2 \right]^{1/2}},
\end{equation*}
where $\tilde{y}(t_j)$ and $\tilde{\mu}(t_j)$ denote the centered functions: 
$\tilde{y}(t_j) \coloneqq y(t_j) - \frac{1}{k}\sum_{j = 1}^k y(t_j), $
and $\tilde{\mu}(t_j) \coloneqq \mu(t_j) - \frac{1}{k}\sum_{j = 1}^k \mu(t_j)$, respectively. The finite-dimensional versions of the amplitude and magnitude indices are, respectively, defined as:
\begin{equation*}
I_{A_k}(y, F_X) \coloneqq \frac{\sum_{j = 1}^k\tilde{y}(t_j)\tilde{\mu}(t_j)}{\sum_{j = 1}^k \tilde{\mu}(t_j)^2 } - 1, 
\end{equation*}
and 
\begin{equation*}
I_{M_k}(y, F_X) \coloneqq  \left(\frac{1}{k} \sum_{j = 1}^k y(t_j)\right) - \beta_k(y) \left(\frac{1}{k}\sum_{j = 1}^k \mu(t_j)\right), 
\end{equation*}
with $\beta_k(y) =  I_{A_k}(y, F_X) + 1.$
\end{definition}

\begin{proposition}[Convergence of the finite-dimensional approximation of FastMUOD indices]

For a stochastic process $X(t)\in L^2([0,1])$ and a function $y$ observed on the finite points $T = \{t_1, t_2, \ldots, t_k \} \subset [0, 1]$ with $t_j - t_{j-1} = \Delta$, a constant, the indices $I_{S_k}(y, F_X)$, $I_{A_k}(y, F_X)$, and $I_{M_k}(y, F_X)$ converge (in limit) to $I_{S}(y, F_X)$, $I_{A}(y, F_X)$, and $I_{M}(y, F_X)$ respectively, as $k \longrightarrow \infty$ (and $\Delta \longrightarrow 0$).    

% For a function $y$ and stochastic process $X$ observed on the same finite points $T = \{t_1, t_2, \ldots, t_k \} \in [0, 1]$ with $t_j - t_{j-1} = \Delta$, a constant, the indices $I_{S_k}(y, F_X)$, $I_{A_k}(y, F_X)$, and $I_{M_k}(y, F_X)$ converge to $I_{S}(y, F_X)$, $I_{A}(y, F_X)$, and $I_{M}(y, F_X)$ respectively, as $k \longrightarrow \infty$.    
\end{proposition}

\begin{proof}
The proof follows trivially from the definition of the Riemann integral.
\end{proof}

Depending on whether the realizations of $X$ are continuously or discretely sampled in time, the sample versions of the indices can be defined by replacing the mean function, $\mu$, of $X$ with an appropriate empirical estimate. Suppose that $X_1(t), \ldots, X_n(t)$ are independent and identically distributed (iid) realizations from $X(t)$, the (pointwise) sample mean function, which we will denote by $\bar{X}(t)$, and given by $\bar{X}(t) = n^{-1} \sum_{i = 1}^n X_i(t),$ is an estimate of the mean function $\mu(t)$. This leads to a direct proposal of the following definitions for the sample versions of the FastMUOD indices.

\begin{definition}[Sample version of FastMUOD indices]
\label{def:d3}
Let $X_1(t), \ldots, X_n(t)$ be iid realizations of the stochastic process $X\in L^2([0,1])$, with empirical distribution $F_{X_n}$. We define the sample shape index as: 
\begin{equation*}
    I_{S_n}(y, F_{X_n}) \coloneqq 1 - \frac{\int\tilde{y}(t)\tilde{\bar{X}}(t)dt}{\left[\int \tilde{y}(t) ^2 dt \right]^{1/2}\left[ \int \tilde{\bar{X}}(t) ^2 dt \right]^{1/2}} = 1 - \frac{\langle \tilde{y}, \tilde{\bar{X}}\rangle}{\Vert \tilde{y} \Vert \cdot \Vert \tilde{\bar{X}} \Vert},
\end{equation*}
where $\tilde{\bar{X}}(t)$ denotes the centered sample mean function: $\tilde{\bar{X}}(t) \coloneqq \bar{X}(t) -  \int \bar{X}(t)dt.$ The sample amplitude and magnitude indices are then defined as:
\begin{equation*}
    I_{A_n}(y, F_{X_n}) \coloneqq \frac{\int\tilde{y}(t)\tilde{\bar{X}}(t)dt}{ \int \tilde{\bar{X}}(t) ^2 dt}  - 1 = \frac{\langle\tilde{y},\tilde{\bar{X}}\rangle }{ \Vert \tilde{\bar{X}} \Vert ^2}  - 1,
\end{equation*}
and 
\begin{equation*}
    I_{M_n}(y)\coloneqq  \int y(t)dt - \beta_n(y)\int \bar{X}(t)dt, 
\end{equation*}
with $\beta_n(y) = I_{A_n}(y, F_{X_n})+1.$
\end{definition}

\begin{definition}[Finite sample version of FastMUOD indices]
Let $X_1(t), \ldots, X_n(t)$ be iid realizations of $X$ and let each $X_i$ be observed on finite points $T = \{t_1, t_2, \ldots, t_k \} \subset [0, 1]$ where $t_j - t_{j-1} = \Delta$, a constant. For a function $y$ observed on the same set of domain points $T$, we define an approximation to the sample version of the shape index as:
\begin{equation*}
    I_{S_{n,k}}(y, F_{X_{n,k}}) \coloneqq 1 - \frac{\sum\limits_{j = 1}^k \tilde{y}(t_j) \tilde{\bar{X}}(t_j)}{\left[\sum\limits_{j = 1}^k  \tilde{y}(t_j)^2 \right]^{1/2} \left[\sum\limits_{j = 1}^k \tilde{\bar{X}}(t_j)^2\right]^{1/2}} = 1 - \hat{\rho}(y, \bar{X}),
\end{equation*}
where $\hat{\rho}(y, \bar{X})$ denotes the sample Pearson correlation coefficient between the observed points of the sample mean function $\bar{X}$ and $y$. Approximations of the sample amplitude and magnitude indices are defined as:
$$ I_{A_{n,k}}(y, F_{X_{n,k}}) \coloneqq \frac{\sum_{j = 1}^k\tilde{y}(t_j)\tilde{\bar{X}}(t_j)}{\sum_{j = 1}^k \tilde{\bar{X}}(t_j)^2 } - 1, $$
and 
$$I_{M_{n,k}}(y, F_{X_{n,k}}) \coloneqq  \left(\frac{1}{k} \sum_{j = 1}^k y(t_j)\right) - \beta_{n,k}(y) \left(\frac{1}{k}\sum_{j = 1}^k \bar{X}(t_j)\right),$$
where
$\beta_{n,k} (y) = I_{A_{n,k}}(y, F_{X_{n,k}})+ 1.$
\end{definition}

\begin{proposition}[$L^2$-consistency of the sample indices]

Let $X\in L^2([0,1])$ be a stochastic process. Then for another function $y\in L^2([0,1])$, $I_{S_n}(y, F_{X_n})$, $I_{A_n}(y, F_{X_n})$, and $I_{M_n}(y, F_{X_n})$,  are $L^2$-consistent estimators of $I_{S}(y, F_X)$, $I_{A}(y, F_X)$, and $I_{M}(y, F_X)$, respectively.  
\end{proposition}

\begin{proof} See the Supplementary Material.
\end{proof}

\subsection{Properties of the Univariate FastMUOD Indices}
\label{subsec::prop_fastmuod}
In this subsection, we present some results showing how the FastMUOD indices behave under certain simple transformations and why this behaviour makes them ideal for detecting outliers. 

% \begin{proposition}
% Suppose $X \in C[0,1]$ is a stochastic process with mean function $\mu(t) = \mathbb{E}(X(t))$, and $y \in C[0,1]$ is another function which may or may not be a realization of $X$. Consider the new function $y'(t) = y(t) + k$, for some $k\in \mathbb{R}$ and $k \ne 0$. Then $I_M(y, F_X) \ne I_M(y', F_X)$. 

% \end{proposition}

\begin{proposition}[Properties of FastMUOD indices]
\label{prop:p3}
Let $X$ be a stochastic process in $L^2([0,1])$ with distribution $F_X$ and mean function $\mu(t)$. Let $y$ and $z$ be other functions in $L^2([0,1])$ (which may be realizations of $X$) and let $a, b \in \mathbb{R}$. Then, the following statements hold.
\begin{enumerate}
    \item  For a new function $y'(t) = a y(t) + b$ we have $I_M(y', F_X) = a I_M(y, F_X) + b $;  $I_A(y', F_X) = aI_A(y, F_X) +a -1$; and if $a \ne 0$, then $I_S(y, F_X) = I_S(y', F_X)$.
    \item For a new function $y'(t) = y(t) + z(t)$ we have $I_M(y', F_X) =  I_M(y, F_X)+I_M(z, F_X)$;  $I_A(y', F_X) = I_A(y, F_X) \iff \langle \tilde{z}, \tilde{\mu}\rangle = 0$; and $I_S(y, F_X) = I_S(y', F_X)\iff \frac{\langle \tilde{y}, \tilde{\mu}\rangle} {\Vert  \tilde{y} \Vert} = \frac{\langle \tilde{y}, \tilde{\mu}\rangle+ \langle \tilde{z}, \tilde{\mu}\rangle}{{\Vert  \tilde{y}+\tilde{z} \Vert}}$.
%    \item For a new function $y'(t) = ay(t) + b$: $I_A(y', F_X) = aI_A(y, F_X) +a -1$; 
%    \item For a new function $y'(t) = y(t) + z(t)$: $I_A(y', F_X) = I_A(y, F_X)$ iff $\langle \tilde{z}, \tilde{\mu}\rangle = 0$; 
    \item For a new function $y'(t) = z(t)y(t)$, we have $I_A(y, F_X) = I_A(y', F_X) \iff \langle \tilde{y}, \tilde{\mu}\rangle = \langle \tilde{zy}, \tilde{\mu}\rangle$; and $I_S(y, F_X) = I_S(y', F_X) \iff \frac{\langle \tilde{y}, \tilde{\mu}\rangle} {\Vert  \tilde{y} \Vert} = \frac{\langle \tilde{zy},  \tilde{\mu}\rangle}{{\Vert  \tilde{zy} \Vert}}$. 
%    \item For $y'(t) = ay(t) + b$, with $a \ne 0$: $I_S(y, F_X) = I_S(y', F_X)$; 
%    \item For $y'(t) = y(t) + z(t)$: $I_S(y, F_X) = I_S(y', F_X)$ iff $\frac{\langle \tilde{y}, \tilde{\mu}\rangle} {\Vert  \tilde{y} \Vert} = \frac{\langle \tilde{y}, \tilde{\mu}\rangle+ \langle \tilde{z}, \tilde{\mu}\rangle}{{\Vert  \tilde{y}+\tilde{z} \Vert}}$;
    %\item For $y'(t) = z(t)y(t)$: $I_S(y, F_X) = I_S(y', F_X)$ iff $\frac{\langle \tilde{y}, \tilde{\mu}\rangle} {\Vert  \tilde{y} \Vert} = \frac{\langle \tilde{zy},  \tilde{\mu}\rangle}{{\Vert  \tilde{zy} \Vert}}$. 
\end{enumerate}
\begin{proof}See the Supplementary Material.
\end{proof}
\end{proposition}
\begin{figure}[t]
\centering
\includegraphics[scale = .5]{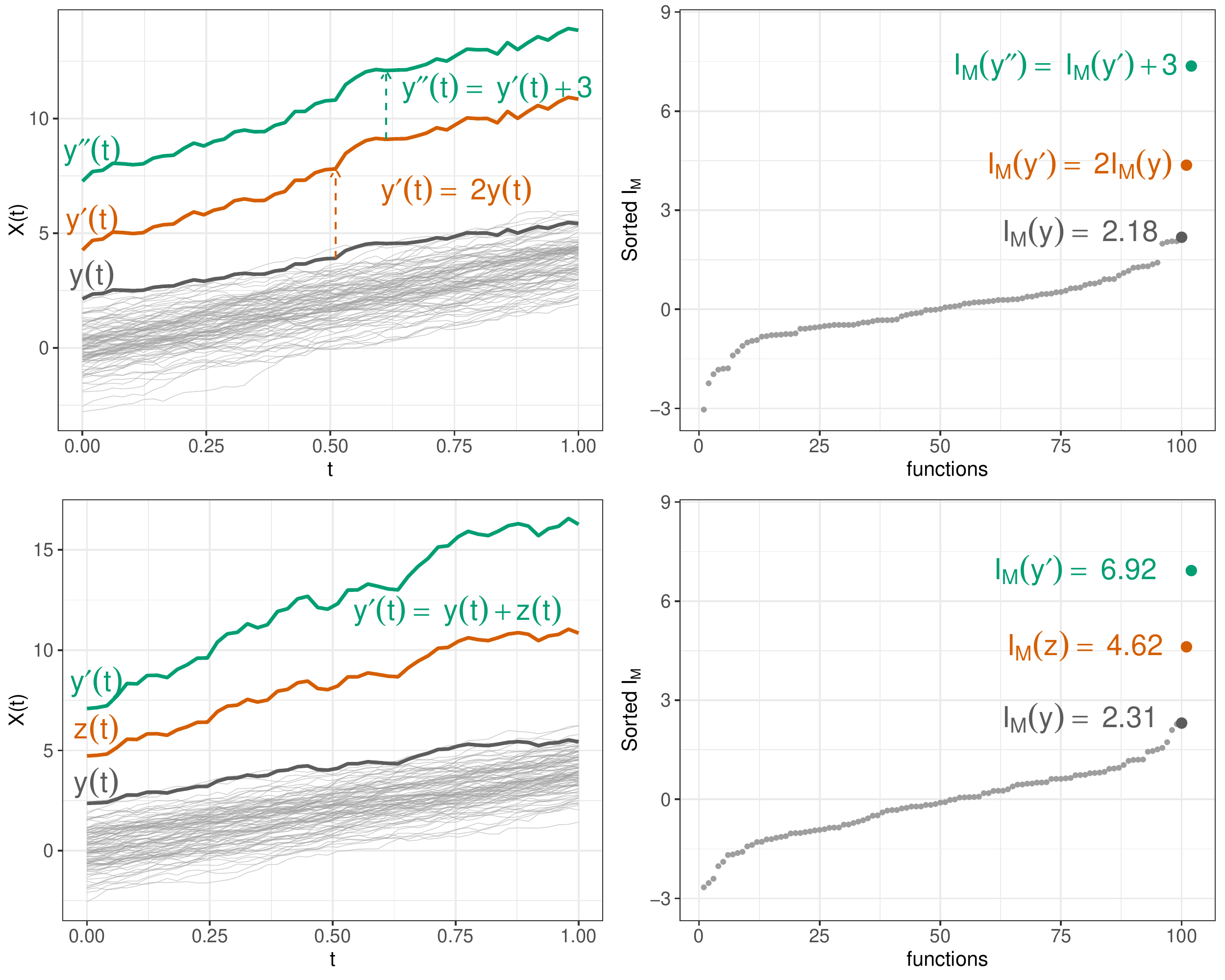}
\caption{Illustration of the magnitude indices under scaling and translation.  Functions and their sorted magnitude indices are shown in the first and second columns, respectively. Functions in grey are the bulk of the data. The function in black is $y(t)$. Functions in orange and green are transformed functions. The same colour code applies to points representing the indices.\label{fig::prop_mag_illus}}
\end{figure}
Proposition \ref{prop:p3} provides insights into how the different FastMUOD indices behave under transformations and hence, why they are useful for targeting the corresponding types of outliers. The first property (Proposition 3.1) demonstrates that $I_M$ is sensitive to the translation and scaling of a function (by real numbers), which is a desirable property, because $I_M$ is intended to be a measure of magnitude outlyingness; it should consequently capture any magnitude shift to be such a worthy measure (of magnitude outlyingness). This property is shown in the first row of Figure \ref{fig::prop_mag_illus} where 100 realizations of the process $X_i(t) = 4t$ are obtained by adding some noise generated from a Gaussian process $e_i(t)$ with zero mean and covariance function $\gamma(s,t) = \exp\{ -0.3|t-s| \}$, for $s, t \in [0,1]$ and $i = 1,\ldots,100$. One of these realizations is then transformed by scaling ($a = 2$) and shifting it ($b = 3$). The second plot on the first row of Figure \ref{fig::prop_mag_illus} shows that the index of the transformed function ($I_M(y', F_X)$) is equal to that of the original realization of $X(t)$ scaled and shifted with the same values ($a I_M(y, F_X) + b$).  

The second property (Proposition 3.2), illustrated in the second row of Figure \ref{fig::prop_mag_illus}, demonstrates that the magnitude index preserves the functional addition operation, which is a desirable property because functional addition causes a shift in magnitude; this shift is captured by the magnitude index. Thus, for a new function $y'(t)$ obtained by adding another function $z(t)$ to $y(t)$, $I_M(y') = I_M(y) \iff I_M(z) = 0$. 

\begin{figure}[t]
	\centering
\includegraphics[scale = .5]{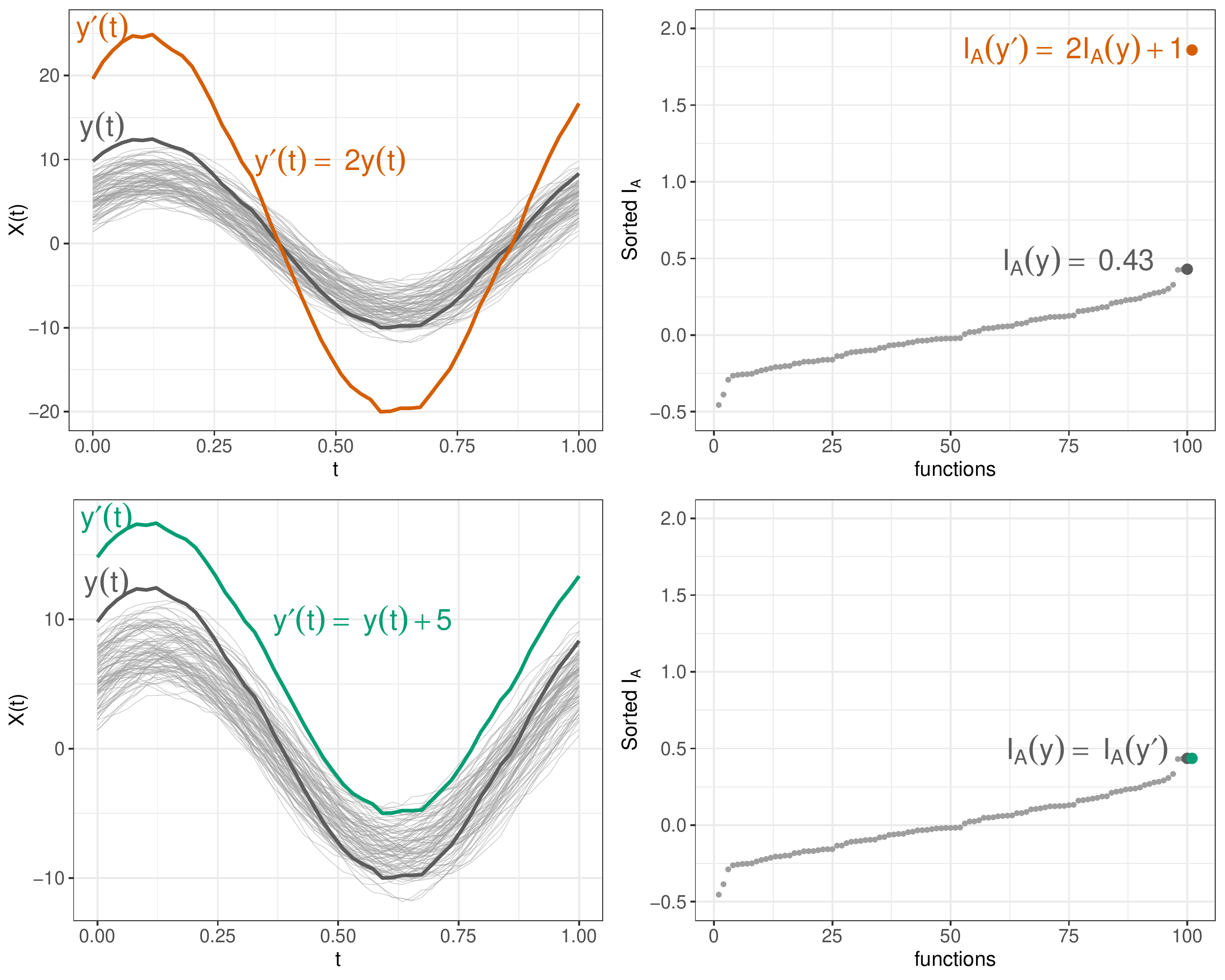}
\caption{Illustration of the amplitude indices under simple transformation.  Functions and their sorted amplitude indices are shown in the first and second columns, respectively. The functions in grey are the bulk of the data. The function in black is $y(t)$. The functions in orange and green are transformed functions. The same colour code applies to the points representing the indices.\label{fig::prop_amp_illus}}
\end{figure}

\begin{figure}[t]
	\centering
\includegraphics[scale = .55]{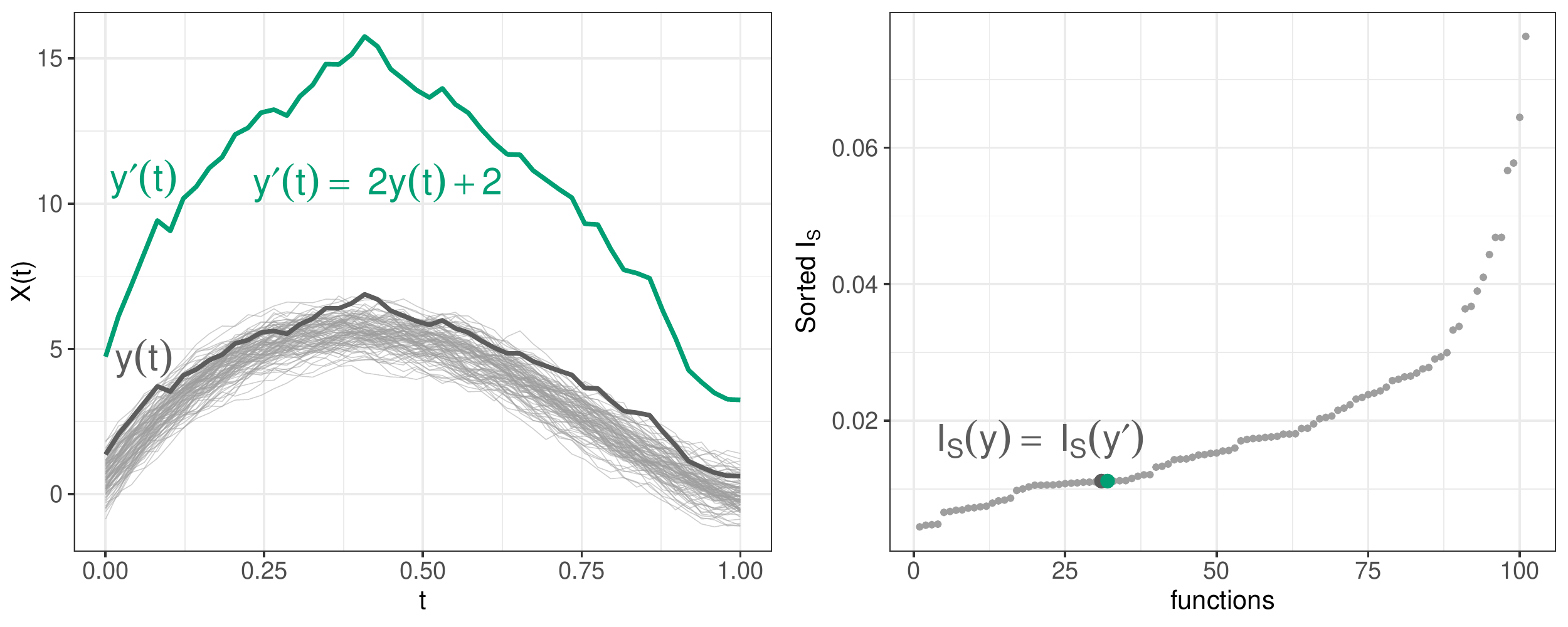}
\caption{Illustration of the shape indices under simple transformation. Functions and their sorted shape indices are shown in the first and second columns, respectively. The functions in grey are the bulk of the data. The function in black is $y(t)$. The function in green is the transformed function $(y'(t))$. The same colour code applies to the points representing the indices.\label{fig::prop_sha_illus}}
\end{figure}

Unlike the magnitude index, the amplitude index is not sensitive to translation by a scalar (Proposition 3.1). Because shifting a (periodic) function in magnitude does not inherently change its amplitude, a good measure of amplitude outlyingness should ignore such a transformation. However, the index $I_A$ is sensitive to scaling by a scalar because this transformation changes the amplitude of a function. In fact, Proposition 3.1 indicates that for a transformed function $y'(t) = ay(t), \ \ a\in \mathbb{R}$, $I_A(y') = I_A(y) \iff a = 1$.  This property of the amplitude index is illustrated in the first and second rows of Figure \ref{fig::prop_amp_illus}. Proposition 3.1 further shows that $I_S$ is neither sensitive to scaling nor translation by scalar values (Figure \ref{fig::prop_sha_illus}). The remaining properties in Proposition \ref{prop:p3} establish conditions under which the amplitude and shape indices of a transformed function remain the same.

The FastMUOD indices defined above slightly differ from those used in \cite{Ojo2021}. The original amplitude and magnitude indices, $I_{A_v}$ and $I_{M_v}$, used by \cite{Ojo2021} had absolute values that guaranteed these indices were positive. These indices also exhibit slightly different properties (because of the use of the absolute value function). For completeness, we have provided definitions and properties of the original indices, $I_{A_v}$ and $I_{M_v}$, used by \cite{Ojo2021} in Section \ref{sec::original_fastmuod} of the Supplementary Material.

\subsection{Implementation and Cutoffs for FastMUOD Indices}
\label{subsec::imp_cutoff}
The sample versions of $I_{A}$ and $I_{M}$ were implemented in R \citep{r2022} using an algorithm similar to the one presented by \cite{Ojo2021}, Algorithm 2. Furthermore, the point-wise median is used in the implementation of $I_{A_n}$ and $I_{M_n}$ (Definition \ref{def:d3}) rather than the point-wise sample mean as the former is more robust to outliers. In \cite{Ojo2021}, where the sample versions of $I_{A_v}$, $I_{M_v}$, and $I_S$ were proposed, a classical boxplot was proposed to determine a cutoff for $I_{A_v}$, $I_{M_v}$, and $I_S$ by considering only the upper whisker of the boxplot (the third quartile extended by 1.5 times the inter-quartile range of $I_{A_v}$, $I_{M_v}$, and $I_S$) as a cutoff.  This worked because $I_{A_v}$, $I_{M_v}$, and $I_S$ are always non-negative and their distributions are right-skewed. For the sample versions of $I_{A}$ and $I_{M}$, both the upper and lower whiskers of the boxplot have to be considered for a cutoff because outliers have indices that occur on both tails of the distribution of these indices. Thus, we propose to consider amplitude outliers as observations with $I_{A_n}$ value greater than $Q_{3{I_{A_n}}}+ 1.5\times IQR_{I_{A_n}}$ or less than $Q_{1{I_{A_n}}}- 1.5\times IQR_{I_{A_n}}$. Moreover, we propose the same rule for flagging magnitude outliers with $I_{M_n}$. 

\section{Extensions to Multivariate Functional Data}
\label{sec::sec3}
FastMUOD was proposed for univariate functional data; however, many real functional data are multivariate in nature. Consequently, we present some techniques for detecting outliers in multivariate functional data using FastMUOD indices. The proposed techniques all involve applying the univariate FastMUOD indices on univariate functional datasets obtained from the multivariate functional data of interest; hence, for our multivariate applications, the definitions and properties presented in Section~\ref{sec::sec2} are relevant. 

\subsection{Marginal Outlier Detection with FastMUOD Indices}
\label{sec::sec3_1}
Suppose $\{\bm{Y}(t), t\in \mathcal{I}\} $ is a stochastic process defined on $\mathcal{I} = [0,1]$ taking values in $\mathbb{R}^d$. Let the distribution of $\bm{Y}(t)$ be $F_{\bm{Y}(t)}$ and let $F_{Y^j(t)}$ be the distribution of the $j^{th}$ marginal component of $\bm{Y}(t)$, with $j = 1,\ldots, d$. Consider a set of $n$ realizations of $\bm{Y}$: $\{\bm{Y}_i(t)\}_{i = 1}^n$. To identify outliers in $\{\bm{Y}_i(t)\}_{i = 1}^n$, a first option is to apply FastMUOD to the $d$ marginals of the observed curves, (i.e.,  $Y_i^j(t)$) and identify $\bm{Y}_i$ as an outlier if it is an outlier (of any type: shape, amplitude, or magnitude) in any of the $d$ margins. However, this technique has the obvious limitation of not detecting ``joint-outliers", i.e.,  observations that are not outliers in any of the marginal distributions but are outlying compared to the joint distribution of the data. It is also quite prone to false positives (FPs) because the final FPs of the procedure is the union of the FPs of the three indices for each margin of the multivariate functional data to which FastMUOD is applied.  

\subsection{Stringing Marginal Functions into Univariate Functional Data}
\label{sec::sec3_2}
For a multivariate functional observation, $\bm{Y}_i(t)$, we can concatenate or ``string" its $d$ univariate dimensions (i.e., $Y_i^1(t), Y_i^2(t), \ldots, Y_i^d(t)$) together into a single univariate function. Thus, we can obtain univariate functional curves $Z_i(t')$ defined on the interval $[0, d]$ from the original multivariate observations $\{\bm{Y}_i(t)\}_{i = 1}^n$ given by $Z_i(t') := Y^j_i(t'-j+1)$, whenever $t' \in (j-1, j]$, for $j = 1, \ldots, d$ (set $Z_i(0) := Y^1_i(0))$. Then, FastMUOD can be applied on $\{Z_i(t')\}_{i = 1}^n$ by estimating for each $Z_i$, the indices $I_{S_n}(Z_i, F_{Z_n})$, $I_{A_n}(Z_i, F_{Z_n})$   and $I_{M_n}(Z_i, F_{Z_n})$ and applying the cutoffs described in Subsection~ \ref{subsec::imp_cutoff}. There are two potential issues with this technique. First, the $d$ univariate functions ($Y_i^1(t), Y_i^2(t), \ldots, Y_i^d(t)$) may have different ranges in which case it might be convenient to scale the dimensions of $\bm{Y}_i(t)$  into the same range (e.g., using a min-max scaling). Furthermore, changing the order of stringing of the marginal functions into univariate functions might have an effect on which observations are detected as outliers or not, e.g., for a set of multivariate functional curves $\{\bm{R}(t)\}_{i = 1}^n$, the set of stringed functions $\{A_i(t') = R^j_i(t'-j+1) \}_{i = 1}^n$ and $\{B_i(t') = Y^{d+1-j}_i(t'-j+1) \}_{i = 1}^n$ might produce different outliers.

% For a multivariate functional observation, $\bm{Y}_i(t)$, we can concatenate or ``string" its $d$ univariate dimensions (i.e., $Y_i^1(t), Y_i^2(t), \ldots, Y_i^d(t)$) together into a single univariate function. Thus, we can obtain univariate functional curves $\{Z_i(t) =[Y_i^1(t), Y_i^2(t), \ldots, Y_i^d(t)]^\top\}_{i = 1}^n$ from the original multivariate observations $\{\bm{Y}_i(t)\}_{i = 1}^n$ (the domain can be rescaled, for example using: $Z_i(t) =Y_i^j\left(\frac{t+j-1}{d}\right)$, for $t\in [0, 1]$, $\left(\frac{t+j-1}{d}\right) \in [0,1]$, and $j = 1, \ldots, d$). FastMUOD can then be applied on $\{Z_i(t)\}_{i = 1}^n$ by estimating for each $Z_i$, the indices $I_{S_n}(Z_i, F_{Z_n})$, $I_{A_n}(Z_i, F_{Z_n})$   and $I_{M_n}(Z_i, F_{Z_n})$ and applying the cutoffs described in Section \ref{subsec::imp_cutoff}. There are a couple of potential issues with this method. First, the $d$ univariate functions ($Y_i^1(t), Y_i^2(t), \ldots, Y_i^d(t)$) may have different range in which case it might be convenient to scale the dimensions of $\bm{Y}_i(t)$  into the same range (e.g., using a min-max scaling). Furthermore, changing the order of stringing of the marginal functions into univariate functions might have an effect on which observations are detected as outliers or not, e.g., for a set of bivariate functional curves $\{\bm{R}(t)\}_{i = 1}^n$, the set of stringed functions $\{Z_i(t) =[R_i^1(t), R_i^2(t)]^\top\}_{i = 1}^n$ and $\{Z_i(t) =[R_i^2(t), R_i^1(t)]^\top\}_{i = 1}^n$ might produce different outliers. 

\subsection{Random Projections}
\label{sec::sec3_3}
Owing to the potential limitations of the two techniques presented in Subsections \ref{sec::sec3_1} and \ref{sec::sec3_2}, we introduce a new technique based on random projections. For a multivariate functional dataset $\{\bm{Y}_i(t)\}_{i = 1}^n$ taking values in $\mathbb{R}^d$, we generate $L$ random unit vectors $\{\bm{\hat{a}}_l \in \mathbb{R}^d: l = 1, \dots, L\}$. Then, we compute the projection of $\bm{Y}_i(t)$ in the direction of $\bm{\hat a}_l$:
 $$
 Y_{i,l}(t) = \bm{\hat a}_l^\top \bm{Y}_i(t) = \sum\limits_{j = 1}^d a^j_l Y^j_i(t)\in \mathbb{R},
 $$
 where $ Y_i^j(t)$ is the $j^{th}$ component of $\bm{Y}_i$ evaluated at $t$ and $a^j_l$ is the $j^{th}$ component of $\bm{\hat a}_l$. Then, FastMUOD can be applied on the univariate functional data $\{Y_{i,l}(t)\}_{i = 1}^n$ by estimating for each $Y_{i,l}(t)$, the indices $I_{S_{n}}(Y_{i,l}, F_{Y_{n,l}})$, $I_{A_n}(Y_{i,l}, F_{Y_{n,l}})$, and $I_{M_n}(Y_{i,l}, F_{Y_{n,l}})$, where $F_{Y_{n,l}}$ is the empirical distribution of $\{\bm{Y}_i(t)\}_{i = 1}^n$ projected on $\bm{\hat a}_l$. Applying the cutoff described in Subsection \ref{subsec::imp_cutoff} on the sets $\{I_{S_{n}}(Y_{i,l}, F_{Y_{n,l}})\}_{i = 1}^n$, $\{I_{A_{n}}(Y_{i,l}, F_{Y_{n,l}})\}_{i = 1}^n$, and $\{I_{M_{n}}(Y_{i,l}, F_{Y_{n,l}})\}_{i = 1}^n$ reveals whether $\bm{Y}_{i}(t)$ is a shape, amplitude, or magnitude outlier, respectively, when projected in the direction of $\bm{\hat a}_l$. 
 \subsubsection{Threshold for the Random Projections}
 To combine all information from the $L$ projections, we adopt a ``voting system" in which a multivariate function is flagged as an outlier of a specific type if it is an outlier of that type in more than a fixed proportion of the projection directions. To this end, we define the following indicator functions: 
\begin{equation}
\label{eqn::indicator_o_t}
\begin{aligned}
    O_{S,l}(\bm{Y}_i) &\coloneqq \mathbbm{1}\{\text{if } Y_{i,l}(t) \text{ is a shape outlier}\}, \\
    O_{A,l}(\bm{Y}_i) &\coloneqq \mathbbm{1}\{ \text{if } Y_{i,l}(t) \text{ is an amplitude outlier}\}, \\
    O_{M,l}(\bm{Y}_i) &\coloneqq \mathbbm{1} \{ \text{if } Y_{i,l}(t) \text{ is a magnitude outlier}\}. \\
   \end{aligned}
\end{equation}   
These indicator functions indicate whether $\bm{Y}_i$ is a shape, amplitude, or magnitude outlier when $\bm{Y}_i(t)$ is projected in the direction of $\bm{\hat a}_l$. Then, we fix the threshold triple $Q = (\tau_S, \tau_A, \tau_M)$, where $\tau_S, \tau_A, \tau_M \in [0,1]$, and declare $\bm{Y}_i(t)$ a ``shape" outlier if $\mathbb{E}_l[O_{S,l}(\bm{Y}_i)] \ge \tau_S$. Similarly, we declare $\bm{Y}_i(t)$ an ``amplitude" outlier if $\mathbb{E}_l[O_{A,l}(\bm{Y}_i)]  \ge \tau_A$ and a ``magnitude" outlier if $\mathbb{E}_l[O_{M,l}(\bm{Y}_i)] \ge \tau_M.$ Note that the classification of a multivariate functional outlier into a specific type (``amplitude", ``magnitude" or ``shape") now indicates that the function is an outlier of that type in at least $\tau$ proportion of the projections (for $\tau \in Q$). This classification is also not necessarily disjoint since an observation can be flagged as an outlier of more than one type in each projection (for examples, see Subsections \ref{subsec::simres} and \ref{subsec::chardata}).  

The threshold triple $Q$ helps to control the false positive rate (FPR) of the procedure. The lower the value of $\tau\in Q$, the more aggressive the procedure is in flagging an observation as an outlier (because flagging an outlier requires less number of ``votes" from the random projections). Higher values of $\tau$, on the other hand, make the procedure more conservative in detecting outliers. For the amplitude and magnitude indices, we find (in our simulation tests) that limiting the value of both $\tau_A$ and $\tau_M$ to the interval $[0.3, 0.7]$ works well for most applications (see Section \ref{subsec::q_exp_res} in the Supplementary Material). In the case when there are magnitude (or amplitude) outliers in the projected data, $\tau_M$ (or $\tau_A$) should be close to the lower bound of $0.3$, which is sufficiently low to allow for flagging the outliers without introducing many FPs. When there are no magnitude (amplitude) outliers, $\tau_M$ (or $\tau_A$) should be close to $0.7$, which is a sufficiently high proportion to prevent FPs. When there are magnitude (or amplitude) outliers, some random projections of the data will not detect the true outliers, and therefore it is imperative not to set $\tau_M$ (or $\tau_A$) to a high proportion in this case. However, setting $\tau_M$ (or $\tau_A$) to a very low proportion, even when there are magnitude (amplitude) outliers, will yield many FPs because some non-outliers will be erroneously flagged as outliers in some of the projections.  For the shape index, we suggest limiting $\tau_S$ to the interval $[0.4, 0.7]$ because we know from previous studies that the shape index is more prone to FPs than the magnitude and amplitude indices (partly because of its skewed distribution and sometimes because of random noise in the data, see \cite{Ojo2021}). 
\subsubsection{Selecting the Thresholds Q}
It is possible to select the threshold values in $Q$ (within the suggested intervals $[0.3, 0.7]$ for $\tau_A$/$\tau_M$ and $[0.4, 0.7]$ for $\tau_S$) in a data-driven way if the distribution or model from which the functional data come from is known. Consider, for example, the following model for $T\in \{S, A, M\}$ (where $S$, $A$, and $T$ denote shape, amplitude, and magnitude, respectively): 
\begin{equation}
\label{eqn::tau_t}
    \tau_T \coloneqq \begin{cases} 
      \gamma_T - \eta_T \frac{\Delta_{PT}}{\Delta_{C}} & \text{if } \frac{\Delta_{PT}}{\Delta_{C}} \in [0,1],\\
      \gamma_T - \eta_T & \text{if } \frac{\Delta_{PT}}{\Delta_{C}} > 1,\\
      \gamma_T & \text{otherwise,} 
\end{cases}
\end{equation}
 where $\gamma_T$, $\eta_T \in [0,1]$. The term $\Delta_{PT}$ is an estimate of the proportion of outliers of type $T$ present in the data computed by subtracting the expected proportion of FP of type $T$ under the null model (a model where there are no outliers) from the average proportions of outliers of type $T$ found over all $L$ projections: 
 \begin{equation}
\label{eqn::delta_pt}
\Delta_{PT} = \frac{\sum\limits_{l = 1}^L \sum\limits_{i = 1}^n \hat{O}_{T,l} (\bm{Y}_i)}{n\times L} - \hat{B}_T,
\end{equation}
 where $\hat{B}_T$ is an estimate of the ``baseline" expected proportion of  FP of type $T$ under the null model and $\hat{O}_{T,l}$ is an estimate of the indicator functions in Equation (\ref{eqn::indicator_o_t}). Likewise, $\Delta_{C}$ is an estimate of the proportion of all unique outliers (regardless of their type) present in the data computed by subtracting the expected proportion of total FPs under the null model from the average proportions of total unique outliers found over all $L$ projections: 
\begin{equation}
\label{eqn::delta_c}\Delta_{C} = \frac{\sum\limits_{l = 1}^L \sum\limits_{i = 1}^n \hat{O}_l (\bm{Y}_i)}{n\times L} - \hat{B}_C,
\end{equation}
where $\hat{O}_l$ is an estimate of the indicator function $O_l(\bm{Y}_i) \coloneqq \mathbbm{1} \{ \text{if } \text{ any } O_{T,l}(Y_i) = 1 \text{ for } T\in \{S, A, M\}\} $ and $\hat{B}_C$ is an estimate of the ``baseline" expected proportion of total FPs (of any type) under the null model.  To ensure that $\tau_T$ is within an interval $[a, b]\subset [0,1]$ of interest in Equation (\ref{eqn::tau_t}), it suffices to set $\gamma_T = b$ and $\eta_T = b-a$. For example, to ensure that $\tau_S \in [0.4, 0.7]$, we can set $\gamma_S = 0.7$ and $\eta_S = 0.3$ in Equation (\ref{eqn::tau_t}). % gives the model for $\tau_S$: 
% \begin{equation}
% \label{eqn::tau_t_shape}
%     \tau_S \coloneqq \begin{cases} 
%       0.7 - 0.3 \frac{\Delta_{PS}}{\Delta_{C}} & \text{if } \frac{\Delta_{PS}}{\Delta_{C}} \in [0,1]\\
%       0.7 - 0.3 & \text{if } \frac{\Delta_{PS}}{\Delta_{C}} > 1\\
%       0.7 & \text{otherwise,} 
% \end{cases}
% \end{equation}
The intuition is that if there are only shape outliers in the data, the proportion $\frac{\Delta_{PS}}{\Delta_{C}}$ will be close to 1, resulting in $\tau_S \approx 0.4$, which is the lower bound of the suggested interval $[0.4, 0.7]$ for shape outliers. On the other hand, if there are no shape outliers, $\frac{\Delta_{PS}}{\Delta_{C}}$ will be close to 0 so that $\tau_S \approx 0.7$, which is the upper bound of the suggested interval $[0.4, 0.7]$, thereby controlling for FPs. However, to estimate the proportion $\frac{\Delta_{PS}}{\Delta_{C}}$, it is necessary to have an estimate of the baseline values $\hat{B}_S$ and $\hat{B}_C$ in Equations (\ref{eqn::delta_pt}) and (\ref{eqn::delta_c}), respectively. If the model or distribution of the data is known, it is possible to estimate these baseline values by simulating the null model (observation from the model without outliers) and estimating the proportion of FP of type $S$ ($\hat{B}_S$) and the proportion of all FPs ($\hat{B}_C$). 

However, for real applications, the distribution or model which the data come from is unknown, and therefore it is impossible to estimate the baselines $B_T$ and $B_C$; hence, the model in Equation (\ref{eqn::tau_t}) cannot be used to fix the threshold values in $Q$. An obvious option is to consider as an outlier of type $T$ any observation that is flagged as an outlier of type $T$ in at least one projection, i.e., flag $\bm{Y}_i(t)$ as an outlier of type $T$ if $\mathbb{E}_l[O_{T,l}(\bm{Y}_i)] > 0$. This has the downside of being prone to FPs since it does not control for any FP due to the projection directions and the FastMUOD indices. Another option, which we recommend, is to use the threshold triple $Q = (\tau_S, \tau_A, \tau_M) = (0.4, 0.3, 0.3)$ that we have found to have a well-balanced performance across various scenarios in our simulation studies (see Section \ref{subsec::q_exp_res} of the Supplementary Material).

\section{Simulation Study}
\label{sec::sec4}
We performed a simulation study to compare the various techniques discussed in Section \ref{sec::sec3} with state-of-the-art methods. 
\subsection{Simulation Models}
\label{sec::sec4_1}
We simulated trivariate ($d = 3$) functional datasets from models based on the truncated Karhunen--Lo\`eve expansion for multivariate functional data \citep{Happ2018}: 
\newcommand\myfollow{\mathrel{\overset{\makebox[0pt]{\mbox{\normalfont\tiny\sffamily iid}}}{\sim}}}
$$\bm{Y}_i(t) = \bm{\mu}(t) + \sum\limits_{m = 1}^M\rho_{i,m}\bm{\psi}_m(t) + \bm{\epsilon}(t), \ \ \ i =1, \ldots, n,\ M\in \mathbb{N},$$
where $\bm{Y}_i(t) \in \mathbb{R}^3$, $\bm{\mu}(t)\in \mathbb{R}^3$ is the multivariate mean function, and $\bm{\psi}_m(t) \in \mathbb{R}^3$, for $m = 1, \ldots, M$ are multivariate eigenfunctions. The scores $\rho_{i,m}\myfollow \mathcal{N}(0, \nu_m)$ for eigenvalues $\nu_m$ that are linearly decreasing $(\nu_m = \frac{M+1-m}{M})$. The errors $\bm{\epsilon}(t)\in \mathbb{R}^3$, and $\bm{\epsilon}(t) \myfollow \mathcal{N}_3(\bm{0, \Sigma})$, with $\bm{\Sigma} = \text{diag}(\sigma_1, \sigma_2, \sigma_3)$, where $\sigma_i \myfollow U[0.1, 0.3]$, for $i = 1,2,3$. The eigenfunctions $\bm{\psi}_m(t)$ were constructed by splitting orthonormal Fourier functions into $d = 3$ pieces and shifting them to the required domain \citep{HappKurz2020}.  We set $M = 9$ basis functions. The sample size for each dataset is $n = 100$ and we considered a contamination rate of $10\%$. The simulated functions are evaluated at $50$ equidistant points in $[0,1]$.  For each simulation model considered, the non-outliers were generated from a main model while the outliers were generated from a contaminated model, both listed below:

\begin{enumerate}
    \item Simulation Model 0 (No outliers): 
    
    \textbf{Main Model}: $\bm{Y}_i(t) = \bm{\mu}(t) + \sum\limits_{m = 1}^M\rho_{i,m}\bm{\psi}_m(t) + \bm{\epsilon}(t);$ where $\bm{\mu}(t) = (4t, 30t(1 - t)^{\frac{3}{2}}, 5(t-1)^2 )^\top.$
    \item Simulation Model 1 (Persistent magnitude outliers): 
    
    \textbf{Main Model}: The same as Model 0. 
    
    \textbf{Contamination Model}:  
     $\bm{Y}_i(t) = \bm{\mu}(t) + \bm{u}(t) + \sum\limits_{m = 1}^M\rho_{i,m}\bm{\psi}_m(t) + \bm{\epsilon}(t);$ where $\bm{\mu}(t)$ is the same as in Model 0 and $\bm{u}(t)$ is given by: $u^{j}_i (t) = 8W_j$ for $j = 1,2,3$. $W_j$ is sampled from $\{-1, 1\}$ with equal probability.
    
    \item Simulation Model 2 (Non-persistent magnitude outliers): %model3
    
    \textbf{Main Model}: The same as Model 0 but with $\bm{\mu}(t) = (5\sin(2\pi t), 5\cos(2\pi t), 5(t-1)^2)^\top.$ 
    
    \textbf{Contamination Model}: $\bm{Y}_i(t) = \bm{\mu}(t) + \bm{u}(t) + \sum\limits_{m = 1}^M\rho_{i,m}\bm{\psi}_m(t) + \bm{\epsilon}(t)$ where $\bm{\mu}(t)$ is the same as the Main Model above, and $\bm{u}(t)$ is given by: $$u^{j}_i (t) = 8W_j\cdot \mathbbm{1}\{\text{if } t \in [T_q, T_q +0.1]\}.$$    
    $W_j$ is the same as in Model 1 and $T_q \sim U[0, 0.9]$.

    \item Simulation Model 3 (Shape outlier I): %model 5
    
    \textbf{Main Model}: The same as Model 0 but with $\bm{\mu}(t) = (5\sin(2\pi t), 5\cos(2\pi t), 5(t-1)^2)^\top.$ 
    
    \textbf{Contamination Model}: The same as Model 0 but with $\bm{\mu}(t)$ changed to: 
    $$\bm{\mu}(t) = (5\sin(2\pi (t - 0.3)), 5\cos(2\pi (t - 0.2)), 5(0.1 - t)^2)^\top.$$
    
    \item Simulation Model 4 (Shape outlier II): %model 6
    
    \textbf{Main Model}: $\bm{Y}_i(t) = \bm{\mu}(t) + \bm{u}(t) + \sum\limits_{m = 1}^M\rho_{i,m}\bm{\psi}_m(t) + \bm{\epsilon}(t);$
       where $\bm{\mu}(t) = (5\sin(2\pi t), 5\cos(2\pi t), 5(t-1)^2)^\top$, and $\bm{u}(t)$ is given by $u^{j}_i (t) = \varrho_j$, with $\varrho_j \myfollow U[-2.1, 2.1]$. 
    
   \textbf{Contamination Model}: Same as the Main Model above but with $\bm{u}(t)$ changed to: 
    $$\bm{u}(t) = (2\sin(4\pi t), 2\cos(4\pi t), 2\cos(8\pi t))^\top.$$
    
    \item Simulation Model 5 (Amplitude outliers): %model 8
    
    \textbf{Main Model}: The same as Model 0 but with $\bm{\mu}(t) = (5\sin(2\pi t), 5\cos(2\pi t), 5(t-1)^2)^\top$. 
    
    \textbf{Contamination Model}: $\bm{Y}_i(t) = \bm{\mu}(t) + \bm{u}(t) + \sum\limits_{m = 1}^M\rho_{i,m}\bm{\psi}_m(t) + \bm{\epsilon}(t);$
       where $\bm{\mu}(t)$ is the same as the Main Model above and $\bm{u}(t)$ is given by:
       $$\bm{u}_i (t) =   ((2+R_i^{1})\mu^1(t),\ (2+R_i^{2})\mu^2(t), (2+R_i^{3})\mu^3(t) - 6   )^\top.$$
    $\mu^j(t)$ are the components of $\bm{\mu}(t)$ in the Main Model above and $R_i^j \myfollow \text{Exp}(2)$, for $j = 1,2,3.$
    
    \item Simulation Model 6 (Shape outlier III): %model 9
    
    \textbf{Main Model}: $\bm{Y}_i(t) = \bm{\mu}(t) + \bm{u}(t) + \sum\limits_{m = 1}^M\rho_{i,m}\bm{\psi}_m(t) + \bm{\epsilon}(t);$ where $\bm{\mu}(t) = (5\sin(2\pi t), 5\cos(2\pi t), 5(t-1)^2)^\top$, and $\bm{u}(t) = (8t\sin(\pi t), t\cos(\pi t), 6\sin(2\pi t) - 3)^\top$. 
    
    \textbf{Contamination Model}: Same as the Main Model above but with $\bm{u}(t)$ changed to: 
    $$\bm{u}(t) = (10t\sin(\pi t), 11t\cos(\pi t), 10\sin(2\pi t) - 6)^\top.$$
\end{enumerate}

Some sample data from these models are shown in Figures \ref{fig::sim_model_1} and \ref{fig::sim_model_2} of the Supplementary Material. 

\subsection{Outlier Detection Methods}
\label{sec::sec4_2}
Data were simulated from the seven models presented in Subsection \ref{sec::sec4_1}. For each model, we compared the OD performance of the proposed extensions in Section \ref{sec::sec3}. We also compared our proposals to other multivariate OD methods such as MS-plot \citep{Dai2018}, FOM,  and functional adjusted outlyingness (FAO) \citep{Rousseeuw2018}. Since our FastMUOD-based proposals use indices that target different types of outliers, we can consider the OD performance of each index or consider a union of outliers flagged by the three indices. Thus, we considered the following methods in our comparison.

\begin{itemize}
\item[-] FST-MAR: This is the union of all outlier types detected by applying FastMUOD to the marginal distributions of the multivariate functional data (described in Subsection \ref{sec::sec3_1}). Consequently, an observation is an outlier if it is flagged as an outlier, of any type, in any of the three dimensions of the simulated multivariate functional data.

\item[-] FST-STR: This is the union of all outlier types detected by applying FastMUOD to the univariate functional data obtained by stringing marginal functions together (described in Subsection \ref{sec::sec3_2}). 

\item[-] FST-PRJ: This is the union of all outlier types detected by applying FastMUOD using random projections (described in Subsection \ref{sec::sec3_3}).  For our simulation tests, 60 random directions were used for projection. To generate each random direction, the components of the vector were simulated from $U[-1,1]$ and the resulting vector was then normalised to have a unit norm. Generating the random directions in this manner is straightforward and fast with limited computational burden. 

We used Equation (\ref{eqn::tau_t}) to determine the threshold triple $Q = (\tau_S, \tau_A, \tau_M)$. Because we knew the base model (Model 0) from which the simulated data were generated, we could estimate the values of $B_T$ and $B_C$ used in Equations (\ref{eqn::delta_pt}) and (\ref{eqn::delta_c}), respectively. For this purpose, we simulated data from Model 0 (null model without outliers) and computed the total FPR (of all outlier types) and the FPR of each type of outliers (shape, magnitude, and amplitude). Then, we used the computed FPR of all outliers as an estimate of $B_C$ and the computed FPRs of outliers of each type as an estimate of $B_T$, for $T\in \{S, A, M\}$.  Our simulation results yielded the following baseline values: $B_A = B_M = 0.009$, $B_S = 0.075$, and $B_C = 0.09$, which we then used in the estimation of $\frac{\Delta_{PT}}{\Delta_{C}}$ in Equation (\ref{eqn::tau_t}). Finally, we set the parameters $\gamma_T$ and $\eta_T$ in Equation (\ref{eqn::tau_t}) to $\gamma_T = 0.7$ for $T \in \{S, A, M\}$, $\eta_S = 0.3$, and $\eta_A, \eta_M = 0.4$ so that $\tau_S \in [0.4, 0.7]$ and $\tau_A, \tau_M \in [0.3, 0.7]$, as reported in Subsection \ref{sec::sec3_3}.
\item[-] FST-PRJ-MG: This is the set of ONLY the ``magnitude" outliers detected using FST-PRJ. 

\item[-] FST-PRJ-AM: This is the set of ONLY the ``amplitude" outliers detected using FST-PRJ. 

\item[-] FST-PRJ-SH: This is the set of ONLY the ``shape" outliers detected using FST-PRJ. 

\item[-] FST-PRJ1: This is similar to FST-PRJ, but uses the threshold triple $Q = (\tau_S, \tau_A, \tau_M) = (0.4, 0.3, 0.3)$, which we recommend in real application when it impossible to use Equation (\ref{eqn::tau_t}) to determine the values of $Q$. 

\item[-] FST-PRJ1-MG: This is the set of ONLY the ``magnitude" outliers detected using FST-PRJ1. 

\item[-] FST-PRJ1-AM: This is the set of ONLY the ``amplitude" outliers detected using FST-PRJ1. 

\item[-] FST-PRJ1-SH: This is the set of ONLY the ``shape" outliers detected using FST-PRJ1. 

\item[-] FST-PRJ2: This is similar to FST-PRJ, but rather than using Equation (\ref{eqn::tau_t}) to select the threshold $Q$, we consider an observation as an outlier of type $T$ if it is flagged as an outlier of that type in ANY projection, i.e., an observation is an outlier of type $T$ if $\mathbb{E}_l[O_{T,l}(\bm{Y}_i)] > 0$ for $T \in \{S, A, M\}$.    

\item[-] FST-PRJ2-MG: This is the set of ONLY the ``magnitude" outliers detected using FST-PRJ2. 

\item[-] FST-PRJ2-AM: This is the set of ONLY the ``amplitude" outliers detected using FST-PRJ2. 

\item[-] FST-PRJ2-SH: This is the set of ONLY the ``shape" outliers detected using FST-PRJ2. 

\item[-] MSPLOT: This is a multivariate functional outlier detection and visualisation method based on the directional outlyingness proposed in \cite{Dai2018}. 

%It is a scatter plot of the mean directional outlyingness ``\textbf{MO}" against the variation of directional outlyingness ``VO". For a multivariate function $\bm{Y}(t)$ defined on the domain $\mathcal{I}$ with distribution $F_{\bm{Y}}$, the mean directional outlyingness is given by: 
%$$\bm{MO}(\bm{Y}, F_{\bm{Y}}) = \int_\mathcal{I} %\bm{O}(\bm{Y}(t), F_{\bm{Y}(t)}) w(t) dt.$$
%and the variation of directional outlyingness is: 
%$$VO(\bm{Y}, F_Y) = \int_\mathcal{I} \lVert %\bm{O}(\bm{Y}(t), F_{\bm{Y}(t)}) - \bm{MO}(\bm{Y}, %F_{\bm{Y}})\rVert^2 w(t)dt,$$
%where $w(t)$ is a weight function. The term $\bm{O}(\bm{Y}(t), F_{\bm{Y}(t)})$ is the directional outlyingness of $\bm{Y}$ at point $t$ given by: 
%$$\bm{O}(\bm{Y}(t), F_{\bm{Y}(t)}) = \frac{1}{D(\bm{Y}(t), F_{\bm{Y}(t)} )}\cdot \bm{v}(t), $$
%where $D$ is multivariate depth notion and $\bm{v}$ is a unit vector pointing from the direction of the median of $F_{\bm{Y}(t)}$ to $\bm{Y}(t)$.  In our simulations, the depth notion $D$ is taken to be the projection depth.  

\item[-] FOM: This is another multivarate functional outlier detection technique. It was proposed in \cite{Rousseeuw2018} and is based on another type of functional directional outlyingness (fDO).

\item[-] FAO: The functional adjusted outlyingness (FAO) is based on the adjusted outlyingness (AO) and its functional extension (fAO) proposed by \cite{Brys2005} and \cite{Hubert2015}, respectively.% (instead of the DO and fDO) in a functional outlier map. %The functional adjusted outlyingness (fAO) of function $\bm{Y}(t)$ w.r.t. $F_{\bm{Y}(t)}$ is the (weighted) integral of its pointwise AO values over the domain: 
%$$\text{fAO}(\bm{Y}, F_{\bm{Y}}) = \int_I AO(\bm{Y}(t), F_{\bm{Y}(t)})w(t)dt.$$
%The fAO and its variation (vAO) can then be used in a functional outlier map as in FOM above. 

\end{itemize}

\subsection{Simulation Results}
\label{subsec::simres}
% latex table generated in R 4.1.2 by xtable 1.8-4 package
% Thu Feb 24 16:38:24 2022
For each of the models  in Subsection \ref{sec::sec4_1}, we tested the methods described Section \ref{sec::sec4_2}. We set the contamination rate to $10\%$ and performed 200 repetitions for each possible model. Table \ref{tab:sim-results-100-50} shows the performance of the proposed techniques on the different simulation models. Because Model 0 is a null model without outliers, we only show the FPRs of the techniques. For Models 1-6, we show the true positive rate (TPR) and the FPR together with their respective standard deviations in parentheses. 
% *** Highlight values of FPR at most 1 and TPR at least 95 *** Add discussion on this table results ***

\begin{table*}[htbp!]
\caption{\label{tab:sim-results-100-50}Mean and Standard Deviation (in parentheses) of the true positive rate (TPR) and the false positive rate (FPR) (in percentage) over 200 repetitions for each model. Sample size $n=100$, evaluation points $t_j=50$, and contamination rate is $10\%$. Comparatively high TPRs ($\ge$ 95\%) and low FPRs ($\le$ 1\%) are marked in bold. The proposed techniques are in italics.}
\centering
{\renewcommand{\arraystretch}{.8}
  \footnotesize
\begin{tabular}{@{}lccccccc@{}}  \toprule
  \multirow{2}{*}{Method} & Model 0  & \multicolumn{2}{c}{Model 1} & \multicolumn{2}{c}{Model 2} & \multicolumn{2}{c}{Model 3} \\
  \cmidrule{2-2} \cmidrule{3-4} \cmidrule{5-6} \cmidrule{7-8}
& FPR & TPR & FPR & TPR & FPR & TPR & FPR \\ 
  \midrule
 \textit{FST-MAR} & 26.2(4.0) & \textbf{100.0(0.0)} & 25.4(4.0) & \textbf{99.9(1.2)} & 14.0(3.6) & \textbf{100.0(0.0)} & 13.4(3.5) \\ 
\textit{FST-STR} & 4.6(2.4) & \textbf{100(0.7)} & 2.6(1.8) & 89.7(12.6) & 2.3(1.8) & \textbf{100.0(0.0)} & 1.8(1.6) \\  \hline
\textit{FST-PRJ} & 1.7(2.6) & \textbf{100.0(0.0)} & \textbf{0.2(0.5)} & \textbf{99.0(3.4)} & \textbf{0.8(1.0)} & \textbf{100.0(0.0)} & \textbf{0.7(0.8)} \\
\textit{FST-PRJ-SH} &1.7(2.5) & 0.4(2.2) & \textbf{0.1(0.4)} & \textbf{98.9(3.4)} & \textbf{0.8(1.0)} & \textbf{100.0(0.0)} & \textbf{0.7(0.8)} \\ 
\textit{FST-PRJ-AM} & \textbf{0.1(0.3)} & 0.0(0.0) & \textbf{0.0(0.0)} & 0.1(1.0) & \textbf{0.0(0.0)} & \textbf{100.0(0.0)} & \textbf{0.0(0.1)} \\ 
\textit{FST-PRJ-MG} & \textbf{0.0(0.0)} & \textbf{100.0(0.0)} & \textbf{0.0(0.2)} & 0.0(0.0) & \textbf{0.0(0.0)} & 0.0(0.0) & \textbf{0.0(0.0)} \\ \hline
\textit{FST-PRJ1} & 3.6(2.1) & \textbf{100.0(0.0)} & 3.5(2.1) & \textbf{99.1(3.0)} & \textbf{0.9(1.0)} & \textbf{100.0(0.0)} & \textbf{0.9(0.9)} \\ 
\textit{FST-PRJ1-SH} & 3.4(2.0) & 4.4(6.4) & 3.4(2.1) & \textbf{98.9(3.4)} & \textbf{0.7(0.8)} & \textbf{100.0(0.0)} & \textbf{0.7(0.8)} \\ 
\textit{FST-PRJ1-AM} & \textbf{0.2(0.4)} & 0.5(2.3) & \textbf{0.2(0.5)} & 4.6(6.6) & \textbf{0.1(0.2)} & \textbf{100.0(0.0)} & \textbf{0.0(0.1)} \\ 
\textit{FST-PRJ1-MG} & \textbf{0.1(0.3)} & \textbf{100.0(0.0)} & \textbf{0.1(0.2)} & 2.7(5.2) & \textbf{0.2(0.6)} & 1.7(3.9) & \textbf{0.2(0.5)} \\ \hline
\textit{FST-PRJ2} & 52.1(3.3) & \textbf{100.0(0.0)} & 50.8(4.1) & \textbf{100.0(0.0)} & 35.5(3.8) & \textbf{100.0(0.0)} & 34.2(4.0) \\ 
\textit{FST-PRJ2-SH} & 46.8(3.6) & 48.9(15.3) & 46.8(4.0) & \textbf{100.0(0.0)} & 27.9(3.1) & \textbf{100.0(0.0)} & 26.8(3.3) \\ 
 \textit{FST-PRJ2-AM} & 12.3(3.7) & 14.0(10.7) & 12.8(4.1) & 65.3(16.9) & 10.4(3.6) & \textbf{100.0(0.0)} & 6.1(2.8) \\ 
  \textit{FST-PRJ2-MG} & 13.7(4.1) & \textbf{100.0(0.0)} & 7.8(3.0) & 35.3(15.8) & 8.7(3.0) & 43.9(19.1) & 9.1(3.5) \\ \hline
 MSPLOT & 1.4(1.6) & \textbf{100.0(0.0)} & \textbf{0.6(1.0)} & \textbf{100.0(0.0)} & \textbf{1.0(1.1)} & \textbf{100.0(0.0)} & \textbf{0.1(1.2)} \\ 
 FOM & \textbf{0.4(0.8)} & \textbf{100.0(0.0)} & \textbf{0.1(0.3)} & \textbf{98.0(5.9)} & \textbf{0.1(0.3)} & 71.3(29.9) & \textbf{0.0(0.2)} \\ 
  FAO & \textbf{0.3(0.6)} & \textbf{100.0(0.0)} & \textbf{0.0(0.1)} & 86.8(17.6) & \textbf{0.1(0.2)} & 40.5(35.8) & \textbf{0.0(0.2)} \\ 
   
\end{tabular}}
\setlength\tabcolsep{11.5pt} % default value: 6pt
{\renewcommand{\arraystretch}{.8}
  \footnotesize
\begin{tabular}{@{}lcccccc@{}}  \toprule
\multirow{2}{*}{Method} & \multicolumn{2}{c}{Model 4} & \multicolumn{2}{c}{Model 5} & \multicolumn{2}{c}{Model 6}  \\
  \cmidrule{2-3} \cmidrule{4-5} \cmidrule{6-7} 
& TPR & FPR & TPR & FPR & TPR & FPR \\ 
 \midrule
 \textit{FST-MAR} & 76.7(16.5) & 15.7(3.6) & \textbf{100.0(0.0)} & 25.3(4.0) & \textbf{100.0(0.0)} & 14.6(3.9) \\ 
\textit{FST-STR} & 47.0(25.9) & 2.6(1.8) & \textbf{100.0(0.0)} & 4.4(2.3) & \textbf{99.8(1.4)} & 2.5(1.6) \\ \hline
\textit{FST-PRJ} & 14.5(20.7) & \textbf{0.2(0.6)} & \textbf{100.0(0.0)} & \textbf{0.1(0.3)} & \textbf{99.8(1.6)} & \textbf{0.6(0.9)} \\ 
\textit{FST-PRJ-SH} & 13.0(19.5) & \textbf{0.1(0.6)} & 0.0(0.0) & \textbf{0.0(0.2)} & \textbf{95.8(9.2)} & \textbf{0.5(0.8)} \\ 
\textit{FST-PRJ-AM} & 0.0(0.0) & \textbf{0.0(0.1)} & \textbf{100.0(0.0)} & \textbf{0.0(0.2)} & 3.5(10.2) & \textbf{0.0(0.0)} \\ 
\textit{FST-PRJ-MG} & 1.9(6.0) & \textbf{0.1(0.2)} & 56.7(26.8) & \textbf{0.0(0.0)} & 89.2(15.3) & \textbf{0.1(0.3)} \\ \hline
\textit{FST-PRJ1} & 43.0(18.3) & 1.1(1.2) & \textbf{100.0(0.0)} & 3.6(1.8) & \textbf{100.0(0.0)} & \textbf{0.9(0.9)} \\ 
\textit{FST-PRJ1-SH} & 40.2(18.3) & \textbf{0.9(1.1)} & 0.0(0.0) & 3.5(1.8) & \textbf{99.0(3.8)} & \textbf{0.8(0.9)} \\ 
\textit{FST-PRJ1-AM} & 0.2(1.4) & \textbf{0.1(0.4)} & \textbf{100.0(0.0)} & \textbf{0.0(0.1)} & 23.6(19.9) & \textbf{0.1(0.2)} \\ 
\textit{FST-PRJ1-MG} & 4.3(7.0) & \textbf{0.2(0.5)} & 87.4(12.9) & \textbf{0.1(0.3)} & 91.0(11.8) & \textbf{0.1(0.3)} \\ \hline
\textit{FST-PRJ2} & \textbf{96.4(6.6)} & 36.7(4.0) & \textbf{100.0(0.0)} & 51.6(3.8) & \textbf{100.0(0.0)} & 38.5(4.2) \\ 
\textit{FST-PRJ2-SH} & 93.8(8.3) & 29.1(3.3) & 0.6(2.6) & 49.4(3.6) & \textbf{100.0(0.0)} & 33.7(4.0) \\ 
 \textit{FST-PRJ2-AM} & 13.6(11.2) & 13.1(4.0) & \textbf{100.0(0.0)} & 5.6(2.6) & \textbf{95.2(8.0)} & 9.5(3.4) \\ 
  \textit{FST-PRJ2-MG} & 39.8(21.2) & 8.2(3.5) & \textbf{99.5(2.3)} & 5.7(2.9) & \textbf{99.9(1.2)} & 7.0(3.1) \\ \hline
 MSPLOT & 33.9(22.3) & 1.0(1.3) & \textbf{100.0(0.0)} & \textbf{0.9(1.3)} & 93.5(9.6) & \textbf{0.9(1.2)} \\ 
 FOM & 1.4(4.4) & \textbf{0.1(0.3)} & \textbf{100.0(0.0)} & \textbf{0.1(0.3)} & 48.2(31.5) & \textbf{0.1(0.3)} \\ 
  FAO & 0.8(2.9) & \textbf{0.1(0.4)} & \textbf{99.8(1.7)} & \textbf{0.0(0.2)} & 25.4(30.1) & \textbf{0.0(0.3)} \\ 
   \bottomrule
\end{tabular}}
\end{table*}

The results of Model 0 show that FST-MAR and FST-PRJ2 both have very high FPRs. For FST-MAR, FastMUOD was independently applied on each of the three dimensions of the simulated trivariate functional data. For each dimension, the three FastMUOD indices contributed some FPs and the union of all these FPs over the three dimensions of the dataset yielded the overall FPR of about 26\% for the null model in Model 0. Moreover, the extremely high FPR of FST-PRJ2 justifies the need to impose the threshold $Q = (\tau_S, \tau_A, \tau_M)$ used for determining if an observation is an outlier. Simply flagging an observation as an outlier if it is detected as an outlier in any random projection does not work for the FastMUOD indices. First, because a non-outlier might sometimes appear to be an outlier in the projected direction, and second because the FastMUOD indices (especially the shape index) and the boxplot cutoff procedure described in Subsection \ref{subsec::imp_cutoff} also produce some FPs. These high FPRs can be observed for both methods (FST-MAR and FST-PRJ2) across all tested models. 

For Model 1 where the outliers are clear magnitude outliers in all dimensions, all methods performed well, except for FST-MAR and FST-PRJ2 because of their high FPRs. The high TPRs of FST-PRJ1-MG and FST-PRJ-MG demonstrate that the magnitude indices detect the magnitude outliers while the other indices (FST-PRJ-AM, FST-PRJ1-AM, FST-PRJ-SH, and FST-PRJ1-SH) do not contribute significantly to the FPs, thus yielding good overall results for FST-PRJ and FST-PRJ1. This also shows that the multivariate magnitude outliers in Model 1 remained magnitude outliers after the projection procedure since only the magnitude indices were activated. FST-STR, which used the stringing procedure described in Subsection \ref{sec::sec3_2}, also showed a very high TPR with low FPR on this magnitude model. In Model 2, which contained non-persistent magnitude outliers, FST-PRJ and FST-PRJ1 show very good OD performance but this time powered by their shape indices (FST-PRJ-SH and FST-PRJ1-SH). FAO and FST-STR however struggled with this model with less than $90\%$ TPR and high standard deviations.  Both FOM and FAO did not perform well in Model 3, while FST-PRJ and FST-PRJ1 showed excellent OD performance on this model, helped by their amplitude and shape indices (FST-PRJ-SH, FST-PRJ-AM, FST-PRJ1-SH, and FST-PRJ1-AM). This reiterates that outlier classification is not necessarily disjoint because on the average, all the outliers in Model 3 were flagged as both amplitude and shape outliers. The ``non-disjoint" classification of outliers can also be observed in the results of FST-PRJ and FST-PRJ1 on Models 5 and 6. All the methods showed poor TPRs (and FPRs) in Model 4, because Model 4 contains pure shape outliers that follow the overall trend of the data and are hidden within the bulk of the data. Apart from Model 4, MSPLOT maintains an excellent OD performance across all other models, except for Model 6 where it did not perform quite as well with a TPR of $93.5\%$ compared to $100\%$ for FST-PRJ1 and FST-PRJ. 

Most of the simulation models used in this study have outliers outlying in all three dimensions (except for Model 6). In the Supplementary Material (Section \ref{subsec::more-sim-res}), we show the performance of the presented methods with similar simulation models but with the outliers only outlying in one or two of the three dimensions of the functional data. 

To summarise, FST-PRJ and FST-PRJ1 showed the best performance across all tested simulation models. We recommend FST-PRJ1 in most usual applications because the underlying data distribution will be unknown, and it will consequently be impossible to compute the threshold $Q$ for FST-PRJ.

\section{Data Examples}
\label{sec::sec5}
In this section, we apply the FastMUOD extension using random projections described in Subsection~\ref{sec::sec3_3} to detect outliers in two multivariate functional datasets: character and video datasets.

\subsection{Character Dataset}
\label{subsec::chardata}
\begin{figure}[t]
	\centering
\includegraphics[scale = .40]{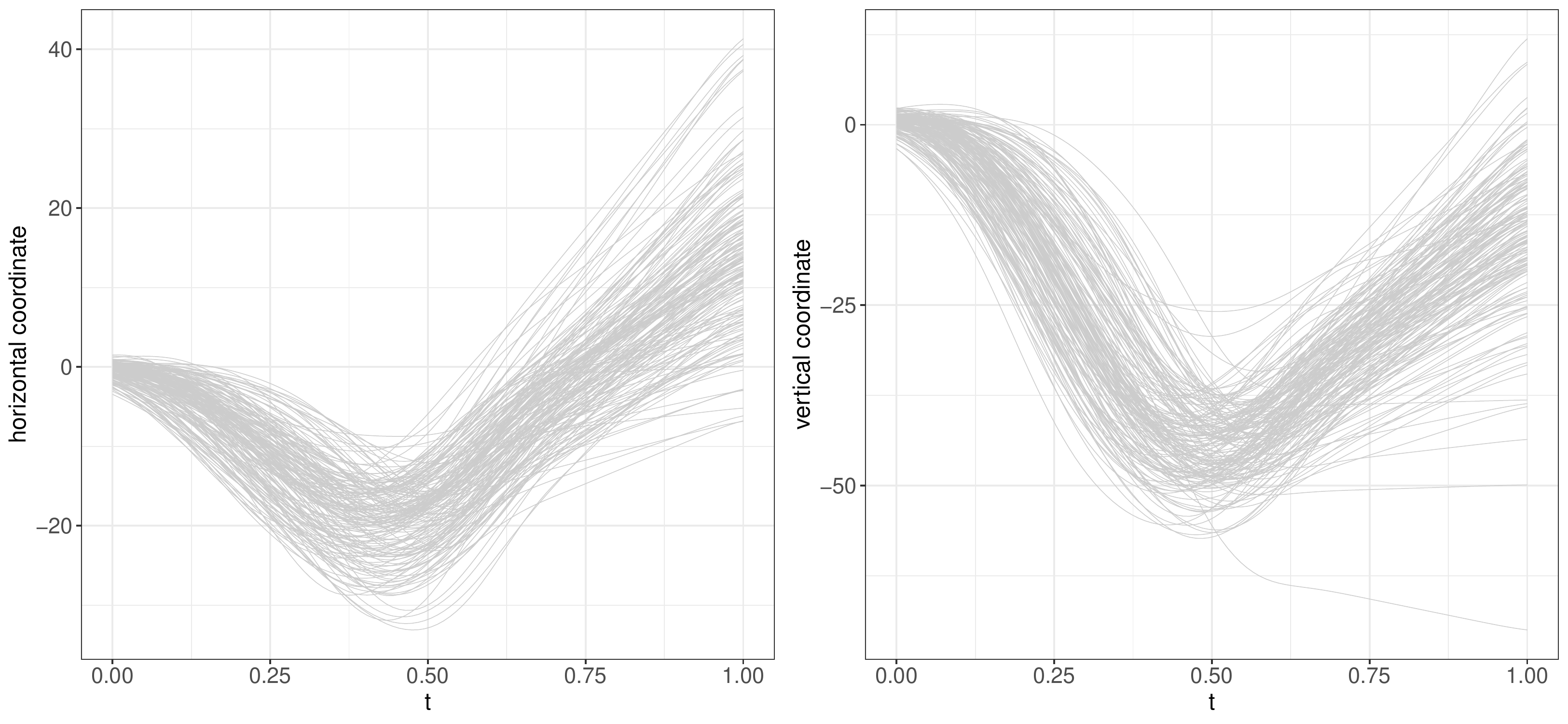}
\includegraphics[scale = .40]{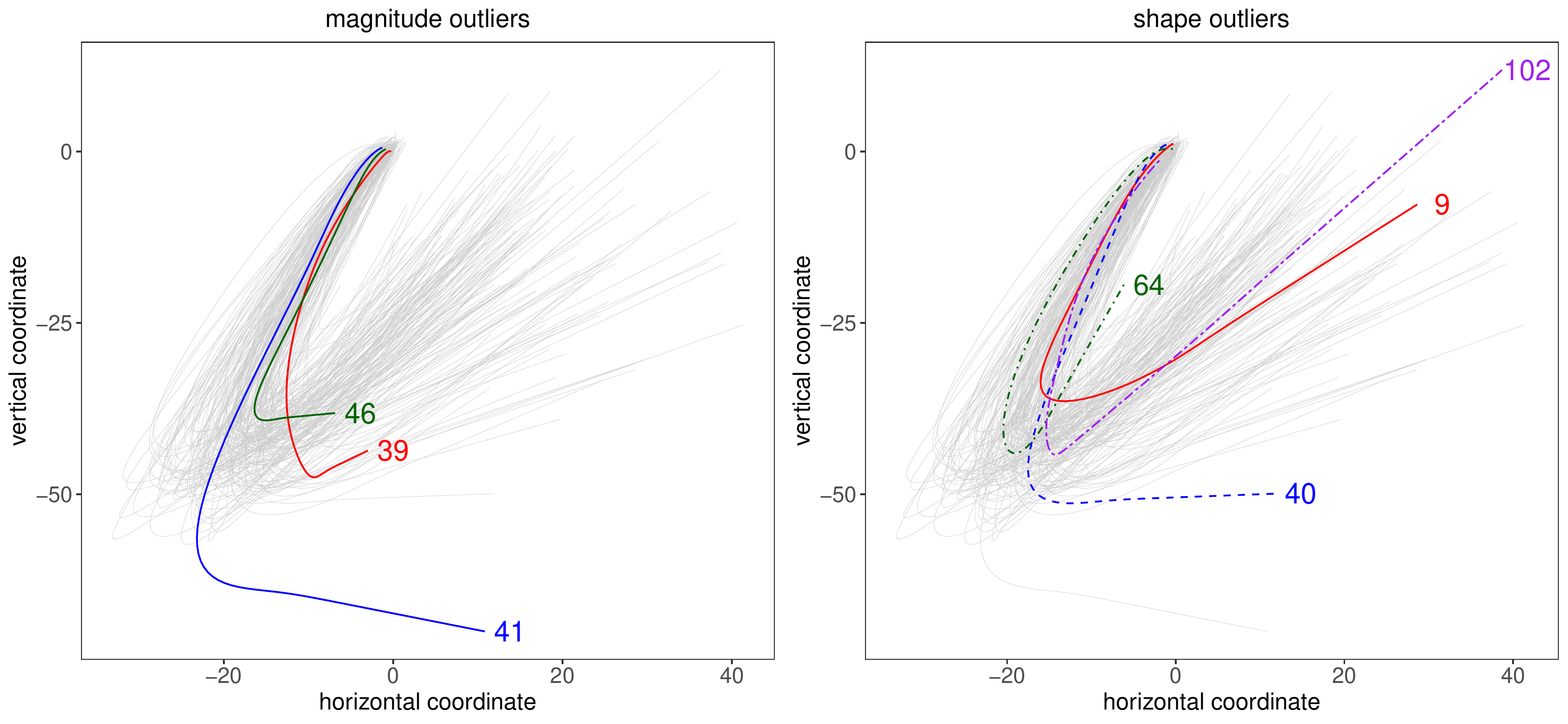}
\caption{First Row: Horizontal and vertical trajectories for letter ``i" data. Second Row: All magnitude outliers and some shape outliers detected in letter ``i" data.\label{fig::charI_mag}}
\end{figure}

\begin{figure}[htbp!]
	\centering
\includegraphics[scale = .40]{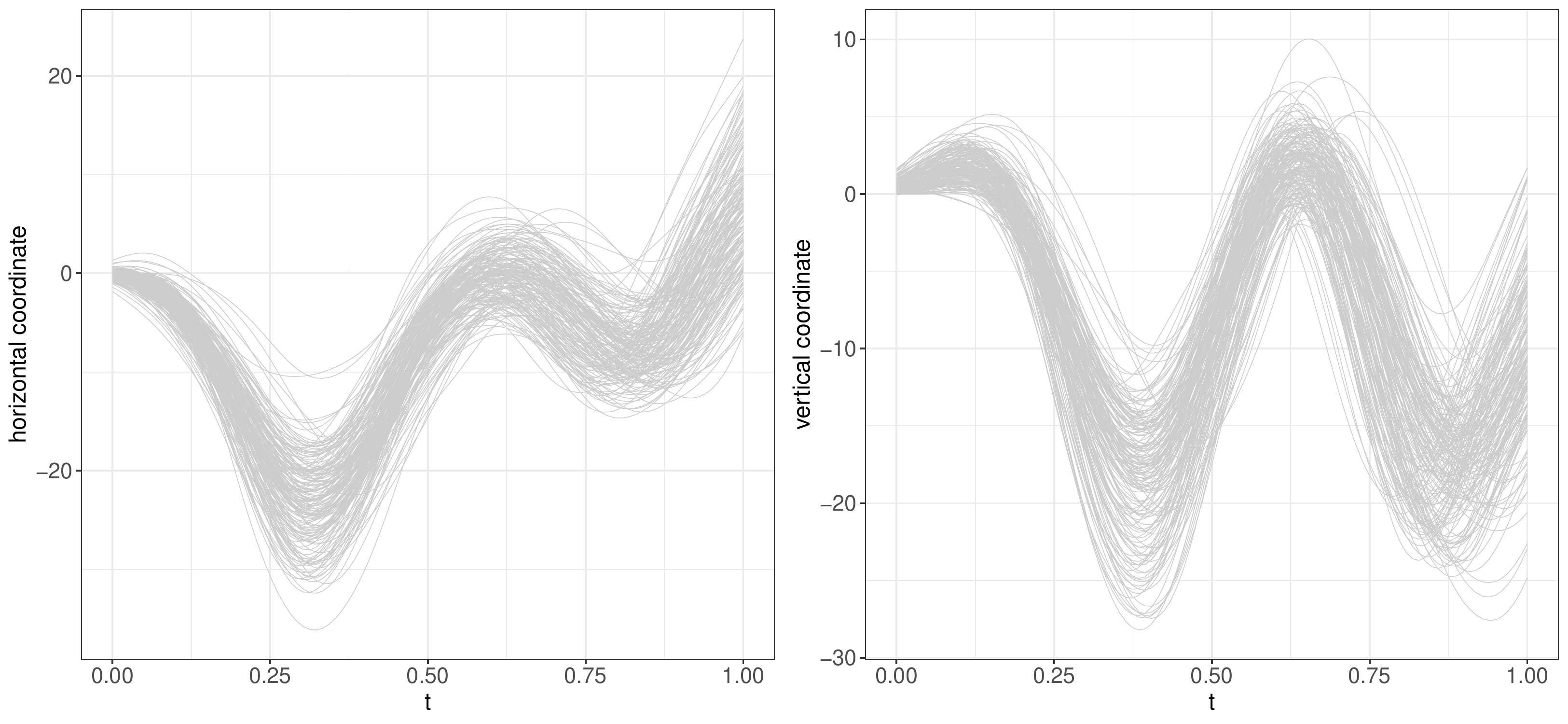}
\includegraphics[scale = .40]{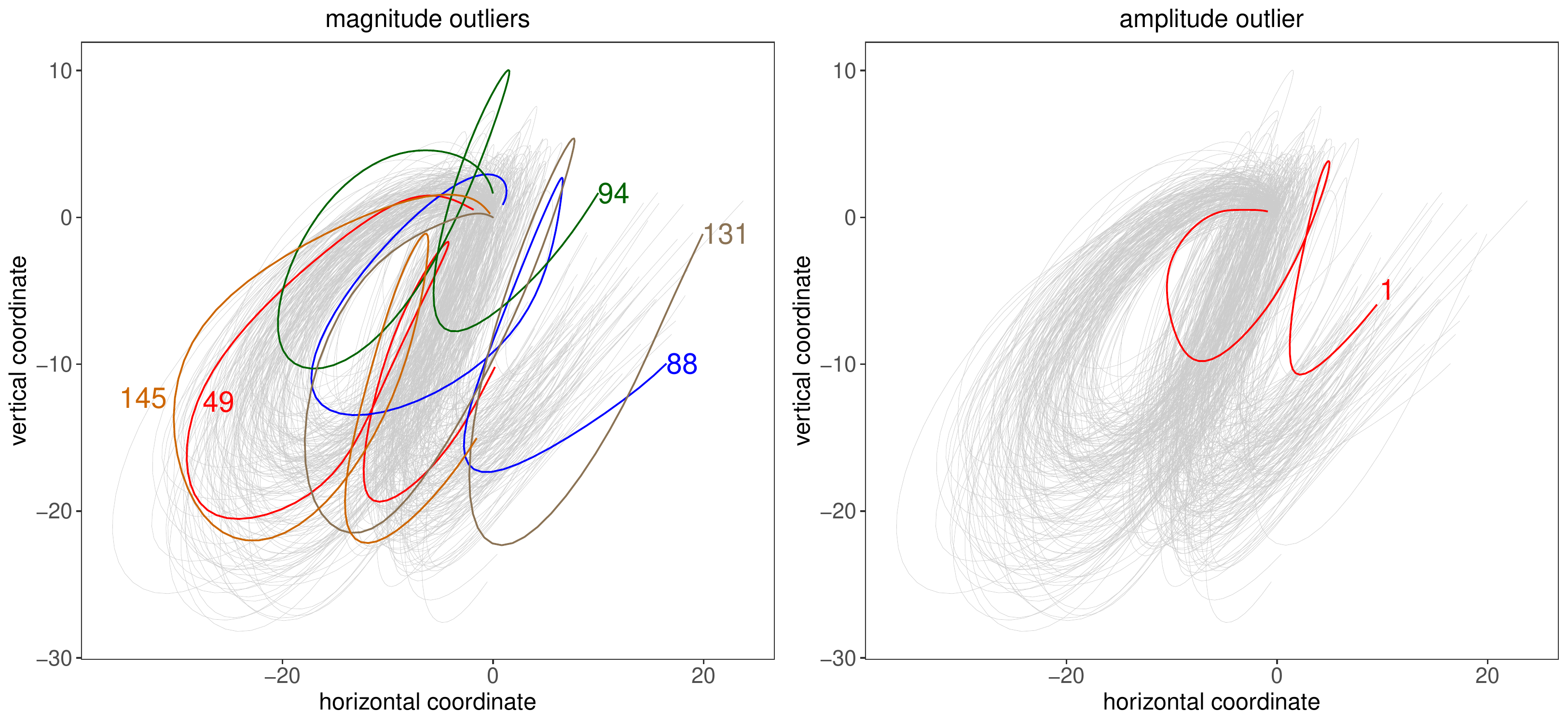}
\includegraphics[scale = .40]{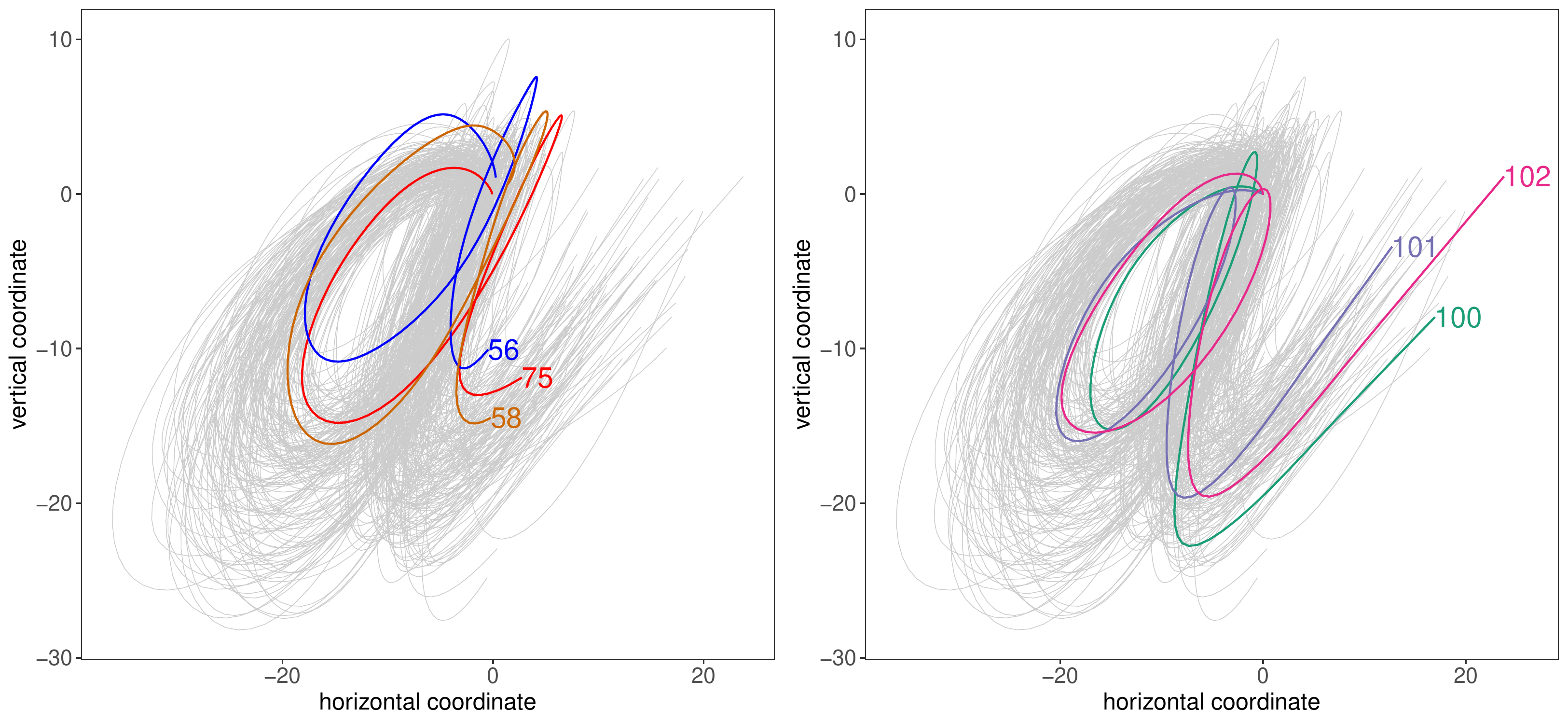}
\caption{First Row: Horizontal and vertical trajectories for letter ``a" data. Second Row: Magnitude and amplitude outliers detected by FastMUOD (FST-PRJ1). Third Row: Shape outliers with short (left) and long (right) ``follow-throughs" respectively.\label{fig::mag_amp_charA}}
\end{figure}

The character dataset comprises bivariate functional data of trajectories of a pen tip along the $x$ and $y$ axes while a subject repeatedly writes various letters of the English alphabet. The original data were provided as part of the Character Trajectories dataset on the UCI machine learning repository \citep{williams2006extracting}. The versions of the dataset used in this study are for the letters ``i" (without the dot) and ``a" provided in the $mrfDepth$ R package \citep{segaert2017mrfdepth}. For the letter ``i", the dataset consists of $n_i = 174$ bivariate functions, observed at 100 times points, whereas for letter the ``a", there are $n_a = 171$ bivariate functions observed at the same number of time points. The bivariate functions in both datasets are the vertical and horizontal coordinates of the pen tip while the subject wrote $n_i$ or $n_a$ copies of each corresponding letter. The first row of Figures \ref{fig::charI_mag} and \ref{fig::mag_amp_charA} show the bivariate functions for letters ``i" and ``a", respectively.

%\begin{figure}[t]
%	\centering
%\includegraphics[scale = .40]{gfx/characterIA_combined.pdf}
%\caption{The plot of the trajectories for the %letters ``i" and ``a".\label{fig::charI_comb}}
%\end{figure}

Plotting the vertical coordinates against the horizontal coordinates in both datasets reveals the handwritten characters (Figures \ref{fig::charI_mag} and \ref{fig::mag_amp_charA}). The aim is to use the FastMUOD via projections (FST-PRJ1) to detect outliers in both datasets. For each dataset, we generated $60$ random unit vectors in $\mathbb{R}^2$ (entries of the vectors follow $U[-1, 1]$ and each vector is normalised to have a unit norm) and projected the data. FastMUOD was then applied on the projections and we used a threshold triple of $Q = (\tau_S, \tau_A, \tau_M) = (0.4, 0.3, 0.3)$ to determine which observations were flagged as outliers of the various types. 

\subsubsection{Letter ``i"}
For the letter ``i", curves $39, 41$, and $46$ were flagged as magnitude outliers; curves $41$ and $46$ were flagged as amplitude outliers and curves $3, 5, 6, 9, 35, 39, 40, 41, 46, 64, 90$, and $102$ were flagged as shape outliers. Thus, 12 unique outliers were flagged in total. All magnitude outliers are also shape outliers, and curves $41$ and $46$ are outliers of all types. The observations marked as ``magnitude outliers"  are shown in the bottom left plot of Figure \ref{fig::charI_mag}. Curve 41 deviates from the overall trend of the data while curves 46 and 39 do not have enough ``follow through" compared to other curves. Some of the shape outliers are shown on the bottom right plot of Figure \ref{fig::charI_mag}. Like curve 41, curves 40 and 64 deviate from the overall trend of the data, while the curve 102 looks rather similar to a ``v" instead of an ``i". Although curve 9 seems to follow the overall trend of the data and does not appear to be an outlier, a closer look at its horizontal and vertical coordinate curves (see Figure \ref{fig::charI_sha_shift} of the Supplementary Material) reveals that the minimum points of both curves are horizontally shifted (to the right) compared to other curves; hence, it was flagged as a shape outlier. Curves 3, 5, 6, and 90 (shown in Figure \ref{fig::charI_sha_shift} of the Supplementary Material) were also flagged as outliers for this same reason. 

For comparison, we applied MSPLOT on the character dataset for letter ``i". MSPLOT discovered a total of 18 unique outliers. The curves flagged by MSPLOT were: $3$, $5$, $6$, $9$, $11$, $12$, $14$, $39$, $40$, $41$, $67$, $73$, $90$, $102$, $109$, $110$, $111$, and $141$. Among the 12 unique outliers flagged by FastMUOD, 9 were flagged by MSPLOT. The functions flagged as outliers by only FastMUOD are curves: 35, 46, and 64; while those flagged by only MSPLOT are curves: 11, 12, 14, 67, 73, 109, 110, 111, and 141.  These curves are shown in Figure \ref{fig::ex_outliers} of the Supplementary Material. 

Considering the curves detected by only FastMUOD and MSPLOT reveals certain interesting features in the data. Curves 35, 46 and 64, detected by only FastMUOD, clearly show some deviation from the trend of the data, especially in their follow-through. On the other hand, curves 67 and 111, flagged by only MSPLOT, seem to resemble a slanted ``v" rather than an ``i". Curve 67 also looks vertically shifted compared to the other curves. To summarise, MSPLOT seems to be more aggressive in declaring curves as outliers compared to FastMUOD.

Finally, we compared the results obtained to those of FOM, which only flagged curve 41 (shown in the bottom left plot of Figure \ref{fig::charI_mag}) as an outlier, probably because FOM is more suited to detecting magnitude outliers rather than shape outliers. Curve 41 demonstrates a clear magnitude deviation in its vertical axis (Figure \ref{fig::curve_41_fom} of the Supplementary Material).  

\subsubsection{Letter ``a"}
For the letter ``a", FST-PRJ1 flagged 16 curves as outliers. Curves: 49, 88, 94, 131, and 145 were flagged as magnitude outliers. Curve 1 is the only curve flagged as an amplitude outlier, while curves 21, 56, 58 75, 100, 101, 102, 110, 114, 117, 125,  and 136 were flagged as shape outliers. The curves flagged as magnitude and amplitude outliers (shown in the middle row of Figure \ref{fig::mag_amp_charA}) show a shift, either in the vertical, horizontal, or both axes.  

Some of the flagged shape outliers have their peaks (or turning point) shifted to the right, particularly in the vertical axis, compared to the bulk of the data, resulting in letter ``a"s with very small follow-through (when both axes are plotted) compared to the bulk of the data. Some of these functions are shown in the bottom row of Figure \ref{fig::mag_amp_charA}. On the other hand, some of the shape outliers have their peaks shifted to the left, which results in letter ``a"s with a long follow through compared to the bulk of the data, thereby making the corresponding letters look like a ``q" rather than an ``a". These functions are shown in Figure \ref{fig::mag_amp_charA} (see Figures \ref{fig::charA_sha_grp1} and \ref{fig::charA_sha_grp2} of the Supplementary Material for the plots of the horizontal and vertical coordinates of these curves). 

We applied MSPLOT on the data for letter ``a" and MSPLOT flagged 13 unique outliers compared to the 16 outliers flagged by FastMUOD (FST-PRJ1). The outliers flagged by MSPLOT are the curves: 1, 21, 23, 49, 56, 58, 75, 102, 114, 125, 131, 158, and 166. Among these 13 unique outliers, 10 were also flagged by FST-PRJ1, indicating a good overlap between the outliers flagged by MSPLOT and FastMUOD. Curves 23, 158, and 166 were flagged as outliers by MSPLOT but not by FastMUOD, while the curves 94, 100, 101, 117, 136, and 145 were flagged by FastMUOD but not by MSPLOT. Some of these curves are shown in Figure \ref{fig::ex_outliers_carA} of the Supplementary Material. FOM detected only 3 unique outliers. These are curves 23, 56, and 58, which can be seen in Figure \ref{fig::mag_amp_charA} (and Figures \ref{fig::charA_sha_grp1} and \ref{fig::ex_outliers_carA} of the Supplementary Material).  

\subsection{Video Data}
In the second application, we applied FST-PRJ1 on a surveillance video data named ``WalkByShop1front" (made available by the the EC Funded CAVIAR project/IST 2001 37540 at: \href{http://homepages.inf.ed.ac.uk/rbf/CAVIAR/}{homepages.inf.ed.ac.uk/rbf/CAVIAR/}). The video consists of a 94 seconds long recording of a surveillance camera in front of a clothing shop in a shopping mall in Lisbon. At various time stamps in the course of the video clip, people passed by the front of the shop; sometimes they entered the shop to explore the products too. The aim is to identify time frames during which people passed by or entered the shop.  This video dataset was analysed in \cite{Ojo2021}; they converted the video to greyscale, and used the original FastMUOD to analyse the resulting univariate functional data (with each frame represented as a function and each pixel being an evaluation point on the curve). Because the original video is coloured, some information is lost in the conversion to grayscale.  We represent the coloured video as trivariate functional data with each dimension being the RBG values of each pixel. We aim to apply FST-PRJ1 to the trivariate functional data and compare the performance to the unviariate analysis of the greyscale values done in \cite{Ojo2021}. 

The video clip is provided at 25 frames/seconds and there are a total of 2,359 frames. The resolution of the video is $384\times 288$, and therefore each frame contains $384\times 288 = 110{,}592$ pixels. For each frame, we arranged the RBG pixel values into an array of size $110{,}592 \times 3$. Thus, the trivariate functional dataset is of dimension $2{,}359\times 110{,}592 \times 3$ representing 2,359 functions (the frames) evaluated at 110,592 points (the pixels) where the value of each point is a vector in $\mathbb{R}^3$ (the RGB pixels intensity). Then, we projected the constructed trivariate functional data on 30 random unit vectors in $\mathbb{R}^3$ and applied FastMUOD (FST-PRJ1) on the 30 univariate functional data of size  $2{,}359\times 110{,}592$. We then set the threshold triple to $Q = (0.4, 0.3, 0.3)$. 

\begin{figure}[t]
	\centering
\includegraphics[scale = .39]{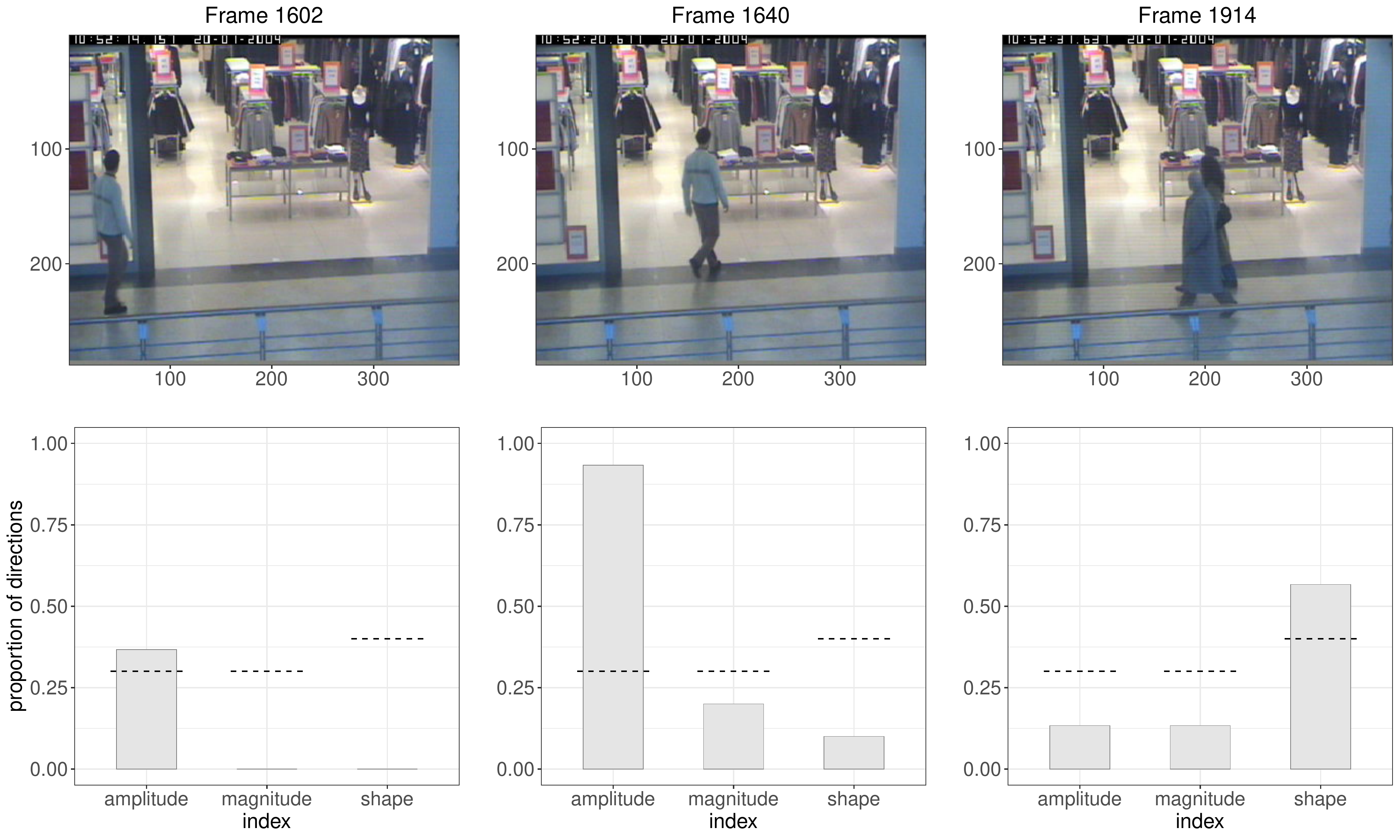}
\caption{Some selected frames detected as outliers by FST-PRJ1. The bar chart below each frame shows the proportion of projections in which they are flagged as an outlier of a particular type. The dotted lines indicate threshold values in $Q$. \label{fig::grouped_frames1}}
\end{figure}

In total, 363 unique frames were flagged as outliers with 191, 312, and 238 frames flagged as shape, amplitude, and magnitude outliers, respectively (Figure \ref{fig::grouped_frames1}). A total of 143 frames were flagged as outliers of all types. The 363 unique outliers flagged are an improvement over the 294 unique frames detected as outliers in the previous analysis of the greyscale pixel values performed by \cite{Ojo2021}. Similar to the analysis in \cite{Ojo2021}, all the frames flagged as outliers correspond to frames during which people pass by or enter the shop. This improvement underlines the advantage of using the multivariate data of the video data compared to performing a univariate analysis of the greyscale values as done in  \cite{Ojo2021}.

\begin{figure}[t]
	\centering
\includegraphics[scale = .39]{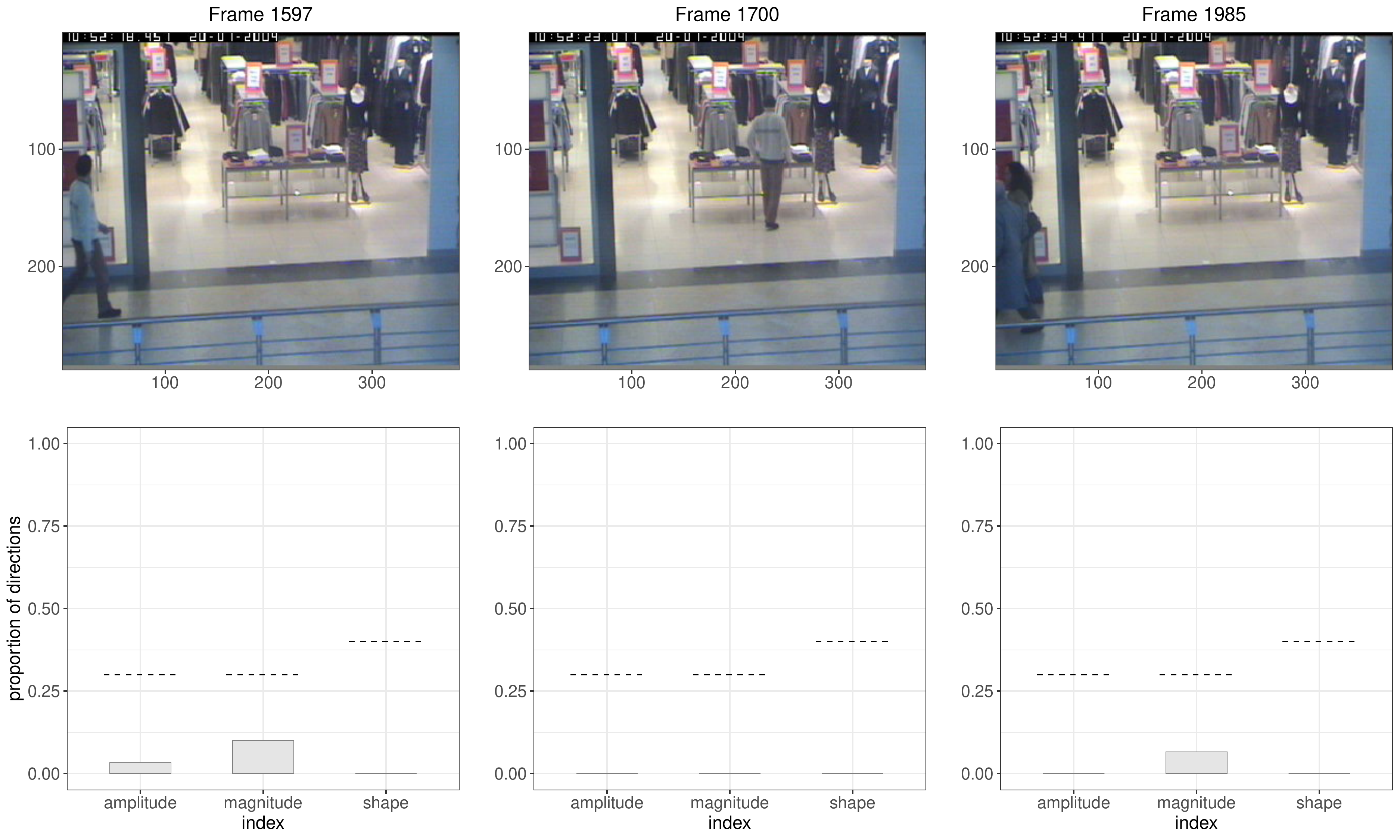}
\caption{Some selected frames not detected as outliers by FST-PRJ1. The bar charts below each frame show the proportion of projections in which that corresponding frame was flagged as an outlier of a particular type. The dotted lines indicate threshold values in $Q$. \label{fig::grouped_frames2}}
\end{figure}

To evaluate the performance of FastMUOD (with projections) in detecting the video frames of interest, it is necessary to understand the distribution of the outlying video frames. The video itself contains three major segments during which various people passed by or entered the front of the shop. The first segment contains frames 803--908, during which a woman passed by the front of the shop. The second segment contains frames 1,588--2,000, when a man entered the shop (to check the products on sale) and two other women passed by the shop. The third segment contains frames 2,073--2,359, which show another man entering the shop. All frames detected as outliers are within the frames of the three main segments, and so there are no FPs. However, similar to the results obtained by \cite{Ojo2021}, there are pockets of timestamps in these segments not flagged as outliers. For instance, the first frames detected as outliers are in the frames of the first segment (frames 803-908) with FastMUOD flagging frames 823, 829--846, and 882--902 as outliers while missing some frames at the beginning (frames 803--828), middle (frames 847--881), and end (frames 903--908) of this segment. We observed the same behaviour for the second and third segments, with certain pockets of a few frames in the beginning, middle and end of the segments not flagged as outliers. Usually, the pocket of frames not flagged as outliers in the middle of the segments correspond to timestamps when someone enters the shop and stands beside the products on display, yielding insufficient contrast in the pixel values of the person and the products on display in the shop. This is shown in Figure \ref{fig::grouped_frames2}, which shows some frames in the second segment of outlying frames that were not flagged as outliers. The first frame, frame 1,597, shows when a man just entered the camera view. The second frame, frame 1,700, shows the same man in the store checking out the products.  Figure \ref{fig::grouped_frames1}, on the other hand, shows some selected frames in the second segment that were flagged as outliers. In addition to frames shown in Figures \ref{fig::grouped_frames1} and \ref{fig::grouped_frames2}, the proportion of directions in which the frames are detected as outliers of each type are also shown. 

Since the frames detected as outliers depend on the threshold triple $Q$, it is useful to visualise the frames of the video together with an animation of the proportion of directions in which the frames are flagged as outliers of different types. Such an animation can be seen by clicking on this \href{https://drive.google.com/file/d/1kcXDT3Ms5myxYbCergyW3ly5AuSazcx5/view?usp=sharing}{[link]} and it shows the variation in the proportion of directions in which frames are outlying as people pass by or enter the shop. 

For comparison, we applied MSPLOT on the same video data. MSPLOT flagged 1,001 frames (out of the 2,359 frames in the data) as outliers; although most frames in the three segments of interest were flagged as outliers, about 200 additional frames that are clearly out of the outlying segments were detected as outliers. However, FOM performed very well on the data, flagging 774 outliers with all the flagged frames coming from the three outlying segments in the video data. FOM excels in the analysis of image and video data because it computes a directional outlyingness at each grid point of the functional data, and image and video data usually consist of thousands of gridpoints (or pixels) per observation (image or frame). Only a few frames from the beginning and end of the outlying segments were not flagged by FOM.

To briefly examine the computational burden of the three methods, Table \ref{tab::pc_time} shows the computational time for each method to analyse the video data. FOM is the fastest requiring about 47 minutes to complete the analysis. FastMUOD with 30 projections used about 51 minutes, while MSPLOT required over 679 minutes ($>$11 hours) to complete the analysis. Although FOM had the fastest running time, it also required the largest amount of random access memory (RAM) for the analysis. MSPLOT and FastMUOD ran in a computer with 64 GB memory, while FOM required $\ge$ 64 GB memory to complete the analysis. The running time experiment was performed on an Ubuntu Server containing an AMD Opteron CPU containing 64 Cores (each running at 2.3 GHz) with 512 GB of RAM. The codes used in this experiment are those provided with the published paper on the methods or those provided on the website of the authors without any prior optimization.

\begin{table}[!htbp]
    \centering
    \caption{Computational time in minutes for video data}
\label{tab::pc_time}
    \begin{tabular}{l r}
\toprule
Method & Times (Minutes)\\
\hline
FOM & 47.4\\
%\hline
FastMUOD & 51.1\\
%\hline
MSPLOT & 679.7\\
\bottomrule
\end{tabular}

\end{table}

\section{Conclusion}
\label{sec::sec6}
The FastMUOD indices, introduced by \cite{Ojo2021} are useful and scalable tools for OD for functional data. However, their use presented in \cite{Ojo2021} was limited to univariate functional data. We sought to address that in this work by presenting some techniques for using these indices for outlier detection in multivariate functional data settings. 

To do that, we presented definitions, sample estimates, and the corresponding finite dimensional versions of the FastMUOD indices. Next, we presented and illustrated certain properties of these indices that make them useful for OD in functional data. We then proposed various techniques for applying the FastMUOD indices to multivariate functional data. Among the various proposed techniques, using random projections showed the most effective results. This involves projecting the multivariate functional data of interest on different unit vectors and then applying FastMUOD indices on the resulting projected univariate functional data. Then, an observation is flagged as an outlier if it is detected as an outlier in at least a fixed proportion of the projection directions.     

We demonstrated the proposed methods on various simulated and real datasets and compared their performance to other multivariate functional OD methods. Our simulation results show the need for adequate selection of a threshold triple $Q = (\tau_S, \tau_A, \tau_M)$ of the proportion of projection directions used in determining whether an observation is an outlier (of a particular type). Instead of declaring an observation as an outlier if it is detected as an outlier in any direction (possibly resulting in a high FPR), carefully selecting the threshold helps to control the FPR. A possible direction for improvement is to develop a method to select these threshold values even when the distribution and base model of the data are unknown. With the proposed techniques, the FastMUOD indices add to the available options of OD tools for multivariate functional data.

\section*{Acknowledgments}
This work has been partially supported by the Comunidad de Madrid grant EdgeData-CM \newline (P2018/TCS4499, cofunded by FSE \& FEDER) and the Ministerio de Econom\'ia, Industria y competitividad, Gobierno de España, grant number PID2019-104901RB-I00. 
The work is also part of the agreements of the Community of Madrid  (Ministry of Education, Universities, Science and Spokesperson) with the Carlos III University of Madrid and the IMDEA Networks Institute for the funding of research projects on SARS-CoV-2 and COVID-19 disease, project names ``Multi-source and multi-method prediction to support COVID-19 policy decision making" and ``COVID-19 Monitoring via Data-Intensive Analysis", which are supported with REACT-EU funds from the European regional development fund ``a way of making Europe". Marc G. Genton and Oluwasegun Ojo's research was supported by the King Abdullah University of Science and Technology (KAUST).

\bibliographystyle{ECA_jasa}
\bibliography{bib}

\clearpage

\section*{Supplementary Material}

%%%%%%%%%% Prefix a "S" to all equations, figures, tables and reset the counter %%%%%%%%%%
%
%\setcounter{figure}{0}
%\setcounter{table}{0}
%\setcounter{page}{1}
% \setcounter{subsection}{0}
% \makeatletter
% \renewcommand{\thesection}{S-\Roman{section}}
\numberwithin{equation}{section}
\setcounter{equation}{0}
\renewcommand{\thesubsection}{S.\arabic{subsection}}
\renewcommand{\theequation}{S\arabic{equation}}
\renewcommand{\thefigure}{S\arabic{figure}}
\renewcommand{\thetable}{S\arabic{table}}
\renewcommand{\thepage}{S\arabic{page}}
\renewcommand{\citenumfont}[1]{S#1}
%%%%%%%%%% Prefix a "S" to all equations, figures, tables and reset the counter %%%%%%%%%%
\subsection{Proof of Proposition 2}

The pointwise sample mean function $\bar{X}(t)$ is an $L^2$-consistent estimator of the mean function $\mu = \mathbb{E}[X(t)]$, i.e., $\Vert\mu - \bar{X} \Vert \overset{P}{\rightarrow} 0$, where $\lVert X \rVert = \left(\int X(t)^2dt\right)^{1/2}$ is the $L^2$-norm, and $\overset{P}{\rightarrow}$ indicates convergence in probability \citep{kokoszka2017introduction}. Thus, $\bar{X}(t) \overset{P}{\rightarrow} \mu(t)$,
as $n \rightarrow \infty$. By the continuous mapping theorem, then: $$\int_{0}^1 \bar{X}(t) dt \overset{P}{\to}  \int_{0}^1 \mu(t)dt $$ and 
$$\left[\bar{X}(t) -  \int_{0}^1 \bar{X}(s)ds\right] \overset{P}{\longrightarrow}  \left[\mu(t) -   \int_{0}^1 \mu(s)ds\right].$$
Now for a given $y\in L^2([0,1])$, which may or may not be a realization of $X$,  define: $$\tilde{y}(t) \coloneqq y(t) - \int y(s)ds.$$
%Note that $\tilde{y}(t)$ is not a random function in $L^2([0,1])$, hence [NEEDED??]: 
By the continuous mapping theorem again, 
$$\tilde{y}(t) \left[\bar{X}(t) -   \int_{0}^1 \bar{X}(s)ds\right] \overset{P}{\longrightarrow}  \tilde{y}(t) \left[\mu(t) -   \int_{0}^1 \mu(s)ds\right],$$
%and by the continuous mapping theorem again, 

\begin{equation}
    \int \tilde{y}(t) \left[\bar{X}(t) -   \int_{0}^1 \bar{X}(s)ds\right]dt \overset{P}{\longrightarrow}   \int  \tilde{y}(t) \left[\mu(t) -  \int_{0}^1 \mu(s)ds\right] dt. \label{eqn:: PRF2_1}
\end{equation}
Using similar arguments, since $\bar{X}(t) \overset{P}{\rightarrow}  \mu(t)$
as $n\to \infty$, 

$$\left[\bar{X}(t) -  \int_{0}^1 \bar{X}(s)ds\right]^2 \overset{P}{\longrightarrow}  \left[\mu(t) -  \int_{0}^1 \mu(s)ds\right]^2,$$
and thus, 
$$\left( \int \left[\bar{X}(t) -  \int_{0}^1 \bar{X}(s)ds\right]^2 dt \right)^{1/2}  \overset{P}{\longrightarrow}  \left(  \int \left[\mu(t) -  \int_{0}^1 \mu(s)ds\right]^2dt \right)^{1/2}.$$
So, 
\begin{equation}
\left(\int \tilde{y}(t)^2 dt \right)^{1/2} \left( \int \tilde{\bar{X}}(t)^2 dt \right)^{1/2} \overset{P}{\longrightarrow}  \left(\int \tilde{y}(t)^2 dt \right)^{1/2} \left(  \int \tilde{\mu}(t) ^2 dt \right)^{1/2}, \label{eqn:: PRF2_2}
\end{equation}
where: 
$$\tilde{\bar{X}}(t) = \bar{X}(t) -   \int_{0}^1 \bar{X}(s)ds \ \ \ \  \text{and}\ \ \ \ \   \tilde{\mu}(t) = \mu(t) -   \int_{0}^1 \mu(s)ds.$$ 
If $\tilde{y}(t) \ne 0$ and $\tilde{\mu}(t) \ne 0$ almost surely, for all $t \in [0,1]$, then from Equation (\ref{eqn:: PRF2_1}) and (\ref{eqn:: PRF2_2}): 

$$I_{S_n}(y) = 1 -  \frac{\int \tilde{y}(t) \tilde{\bar{X}}(t) dt}{\left( \int \tilde{y}(t)^2 dt \right)^{1/2} \left(  \int \tilde{\bar{X}}(t)^2dt \right)^{1/2} } \overset{P}{\longrightarrow} 1 -\frac{ \int \tilde{y}(t) \tilde{\mu}(t) dt}{\left( \int \tilde{y}(t)^2 dt \right)^{1/2} \left(  \int \tilde{\mu}(t)^2dt \right)^{1/2} }  = I_S(y) $$
Using similar arguments, we see that 
$$I_{A_n}(y)  \overset{P}{\longrightarrow} I_{A}(y), $$
and 
$$I_{M_n}(y) \overset{P}{\longrightarrow} I_M(y). $$
\qed
\subsection{Proof of Proposition 3} 

\begin{enumerate}
    \item By definition,  
\begin{align}
\label{eq:q4}
I_M(y', F_X) &=  \int y'(t) dt - \beta(y')\int \mu(r)dr\\\nonumber
&= \int (ay(t)+b) dt - \beta(ay(t)+b)\int \mu(r)dr. 
\end{align}

But,
\begin{align*}
  \beta(ay(t)+b) &= \frac{\int\left[ay(t)+b - \int (ay(r)+b)dr \right]\tilde{\mu}(t)dt}{\int \tilde{\mu}(t)^2 dt}\\ 
  &= \frac{\int\left[ay(t) - \int ay(r)dr \right]\tilde{\mu}(t)dt}{\int \tilde{\mu}(t)^2 dt}\\
  &= \frac{a\int\tilde{y}(t)\tilde{\mu}(t)dt}{\int \tilde{\mu}(t)^2 dt} = a\beta(y).
\end{align*}

So Equation (\ref{eq:q4}) becomes 
\begin{align*}
I_M(y', F_X) &= b+ a\int y(t) dt - a\beta(y)\int \mu(r)dr\\ &= aI_M(y, F_X)+b.
\end{align*}

For the amplitude index, by definition, we have that
\begin{align*}
    I_A(y', F_X) &= \frac{\int \left[ay(t) + b - \int (ay(r) + b) dr\right] \left[\mu(t)  -  \int \mu(r) dr\right] dt}{\int \left[\mu(t) - \int \mu(r) dr\right]^2dt } - 1\\
    &=  \frac{a\int \left[y(t) - \int y(r)dr\right] \left[\mu(t)  -  \int \mu(r) dr\right] dt}{\int \left[\mu(t) - \int \mu(r) dr\right]^2dt } - 1\\
    &= \frac{a\int \tilde{y}(t) \tilde{\mu}(t)dt}{\int \tilde{\mu}(t)^2 dt } - 1\\ 
    &= a (I_A(y, F_X)+1) - 1 \\
    &= aI_A(y, F_X) +a -1.
\end{align*} 

For the shape index, since $a\ne 0$, we have by definition: 
\begin{align*}
    I_{S}(y', F_X) &= 1 -  \frac{\int \left[ay(t) + b - \int (ay(r) +b )dr\right] \left[\mu(t)  -  \int \mu(r) dr\right] dt}{\left( \int \left[ay(t) + b - \int (ay(r) +b )dr\right]^2 dt\right)^{1/2}\left( \int \left[\mu(t) - \int \mu(r) dr\right]^2dt \right)^{1/2}}\\
    &= 1 -  \frac{a\int \left[y(t) - \int y(r) dr\right] \left[\mu(t)  -  \int \mu(r) dr\right] dt}{a\left( \int \left[y(t) - \int y(r)dr\right]^2 dt\right)^{1/2}\left( \int \left[\mu(t) - \int \mu(r) dr\right]^2dt \right)^{1/2}}\\
    &= 1 -  \frac{\int \tilde{y}(t) \tilde{\mu}(t) dt}{\left( \int \tilde{y}(t)^2 dt\right)^{1/2}\left( \int \tilde{\mu}(t)^2dt \right)^{1/2}} = I_S(y, F_X).
\end{align*}
Thus, for any $a,b \in \mathbb{R}$, $a\ne 0$, we have that  $I_S(y, F_X) = I_S(y', F_X)$. 

\item Since 
\begin{align*}
    \beta(y') &= \beta(y(t)+z(t))\\
    &= \frac{\int\left[y(t)+z(t) - \int (y(r)+z(r))dr \right]\tilde{\mu}(t)dt}{\int \tilde{\mu}(t)^2 dt} \\ 
    &= \frac{\int\left[y(t) - \int y(r)dr + z(t) - \int z(r)dr\right]\tilde{\mu}(t)dt}{\int \tilde{\mu}(t)^2 dt}\\
    &= \frac{\int\left[\tilde{y}(t) + \tilde{z}(t)\right] \tilde{\mu}(t)dt}{\int \tilde{\mu}(t)^2 dt} \\
    &= \beta(y) + \beta(z),
\end{align*}
we have (by definition) that 
\begin{align*}
I_M(y', F_X) &=  \int y'(t) dt - \beta(y')\int \mu(r)dr\\
&= \int (y(t)+ z(t)) dt - [\beta(y) + \beta(z)]\int \mu(r)dr\\
&= \int (y(t)+ z(t)) dt - \beta(y)\int \mu(r)dr - \beta(z)\int \mu(r)dr\\
&= I_M(y, F_X) + I_M(z, F_X).
\end{align*}

For the amplitude index, assume that for some $z \in L^2([0,1])$, $\langle \tilde{z}, \tilde{\mu}\rangle = 0$, we have that

\begin{align*}
    I_A(y', F_X) &= \frac{\int \tilde{y}'(t)\tilde{\mu}(t)}{\int \tilde{\mu}(t)^2 dt} - 1 \\
    &=  \frac{\int \left[y(t) + z(t) -\int (y(r) + z(r))  dr\right]\tilde{\mu}(t) dt}{\int \tilde{\mu}(t)^2 dt} - 1\\ 
    &= \frac{\int \left[y(t) - \int y(r)dr + z(t) -\int z(r)  dr\right]\tilde{\mu}(t) dt}{\int \tilde{\mu}(t)^2 dt} - 1\\
    &= \frac{\int \left[\tilde{y}(t) + \tilde{z}(t) \right]\tilde{\mu}(t) dt}{\int \tilde{\mu}(t)^2 dt} - 1 \\ 
    &= \frac{\int \tilde{y}(t)\tilde{\mu}(t) dt}{\int \tilde{\mu}(t)^2 dt} + \frac{\int \tilde{z}(t)\tilde{\mu}(t) dt}{\int \tilde{\mu}(t)^2 dt} - 1 \\ 
    &= \frac{\langle\tilde{y},\tilde{\mu}\rangle }{ \Vert \tilde{\mu} \Vert ^2} + \frac{\langle\tilde{z},\tilde{\mu}\rangle }{ \Vert \tilde{\mu} \Vert ^2} - 1 \\ 
    &= \frac{\langle\tilde{y},\tilde{\mu}\rangle }{ \Vert \tilde{\mu} \Vert ^2}  - 1 \\
    &= I_A(y, F_X).
\end{align*}

Now, assume that $I_A(y, F_X)  =  I_A(y', F_X)$, we have (by definition of $I_A$) that
\begin{align*}
    \int \tilde{y}(t)\tilde{\mu}(t) dt &=  \int \tilde{y}'(t)\tilde{\mu}(t) dt\\
    &= \int \left[y(t) + z(t) -\int (y(r) + z(r))  dr\right]\tilde{\mu}(t) dt\\ \nonumber
    &= \int \left[y(t) - \int y(r)dr + z(t) -\int z(r)  dr\right]\tilde{\mu}(t) dt\\ \nonumber
    &= \int \left[\tilde{y}(t) + \tilde{z}(t) \right]\tilde{\mu}(t) dt\\ \nonumber
    &= \int \tilde{y}(t)\tilde{\mu}(t) dt + \int \tilde{z}(t)\tilde{\mu}(t) dt, 
\end{align*}
which implies that:

$$\int \tilde{z}(t)\tilde{\mu}(t) dt = \langle \tilde{z}, \tilde{\mu}\rangle = 0.$$

For the shape index, assume that $I_S(y, F_X) = I_S(y', F_X)$. By definition, $I_S(y, F_X) = I_S(y', F_X)$ implies that 
\begin{align*}
    \frac{\int \tilde{y}(t)\tilde{\mu}(t) dt}{[\int \tilde{y}(t)^2 dt]^{1/2}}  &= \frac{\int \tilde{y}'(t)\tilde{\mu}(t) dt}{[\int \tilde{y}'(t)^2 dt]^{1/2}}\\
    &= \frac{\int \tilde{y}(t)\tilde{\mu}(t) dt+\int \tilde{z}(t)\tilde{\mu}(t) dt}{[\int (\tilde{y}(t)+\tilde{z}(t))^2 dt]^{1/2}} \\
    &=  \frac{\langle \tilde{y}, \tilde{\mu}\rangle + \langle \tilde{b}, \tilde{\mu}\rangle}{\Vert  \tilde{y} + \tilde{z} \Vert}. 
\end{align*}

Now suppose that 
$$\frac{\langle \tilde{y}, \tilde{\mu}\rangle} {\Vert  \tilde{y} \Vert} = \frac{\langle \tilde{y}, \tilde{\mu}\rangle + \langle \tilde{z}, \tilde{\mu}\rangle}{\Vert  \tilde{y} + \tilde{z} \Vert},$$ then

\begin{align*}
    I_S(y', F_X) &= 1 - \frac{\langle \tilde{y}', \tilde{\mu}\rangle}{\Vert  \tilde{y}' \Vert \cdot \Vert  \tilde{\mu} \Vert} \\
    &= 1 - \frac{\langle \tilde{y}, \tilde{\mu}\rangle + \langle \tilde{z}, \tilde{\mu}\rangle}{\Vert  \tilde{y} + \tilde{z} \Vert \cdot \Vert  \tilde{\mu} \Vert} \\
    &= 1- \frac{\langle \tilde{y}, \tilde{\mu}\rangle} {\Vert  \tilde{y} \Vert \cdot \Vert  \tilde{\mu} \Vert} \\
    &= I_S(y, F_X).
\end{align*}

\item Proof follows from the definitions of $I_A(y, F_X)$, $I_A(y', F_X)$, $I_S(y, F_X)$ and $I_S(y', F_X)$.
\end{enumerate}
\qed

\newpage
\subsection{Original FastMUOD Magnitude and Amplitude Indices}
\label{sec::original_fastmuod}
The FastMUOD indices in this study slightly differ from the ones used by \cite{Ojo2021}. For completeness, we provide definitions of indices used by \cite{Ojo2021}.   
\begin{definition}[Original FastMUOD indices] 
Let $X$ be a stochastic process in $L^2([0,1])$ with distribution $F_X$ and let $\mu(t) = \mathbb{E}(X(t))$ be its population mean function.  An alternative definition of the amplitude index of a function $y \in L^2([0,1])$ w.r.t. $F_X$ is:
\begin{equation}
     I_{A_v}(y, F_X) \coloneqq \left| \frac{\int\tilde{y}(t)\tilde{\mu}(t)dt}{ \int \tilde{\mu}(t) ^2 dt}  - 1\right| = \left|\frac{\langle\tilde{y},\tilde{\mu}\rangle }{ \Vert \tilde{\mu} \Vert ^2}  - 1\right| =  \left|I_A(y, F_X)\right|.
\end{equation}
Furthermore, an alternative definition of the magnitude index of $y$ w.r.t. $F_X$ is: 
\begin{equation}
    I_{M_v}(y, F_X) \coloneqq  \left|\int y(t)dt - \beta(y)\int \mu(t)dt\right| = \left|I_M(y, F_X)\right|.
\end{equation}

\end{definition}

\begin{figure}[htbp!]
	\centering
\includegraphics[scale = .40]{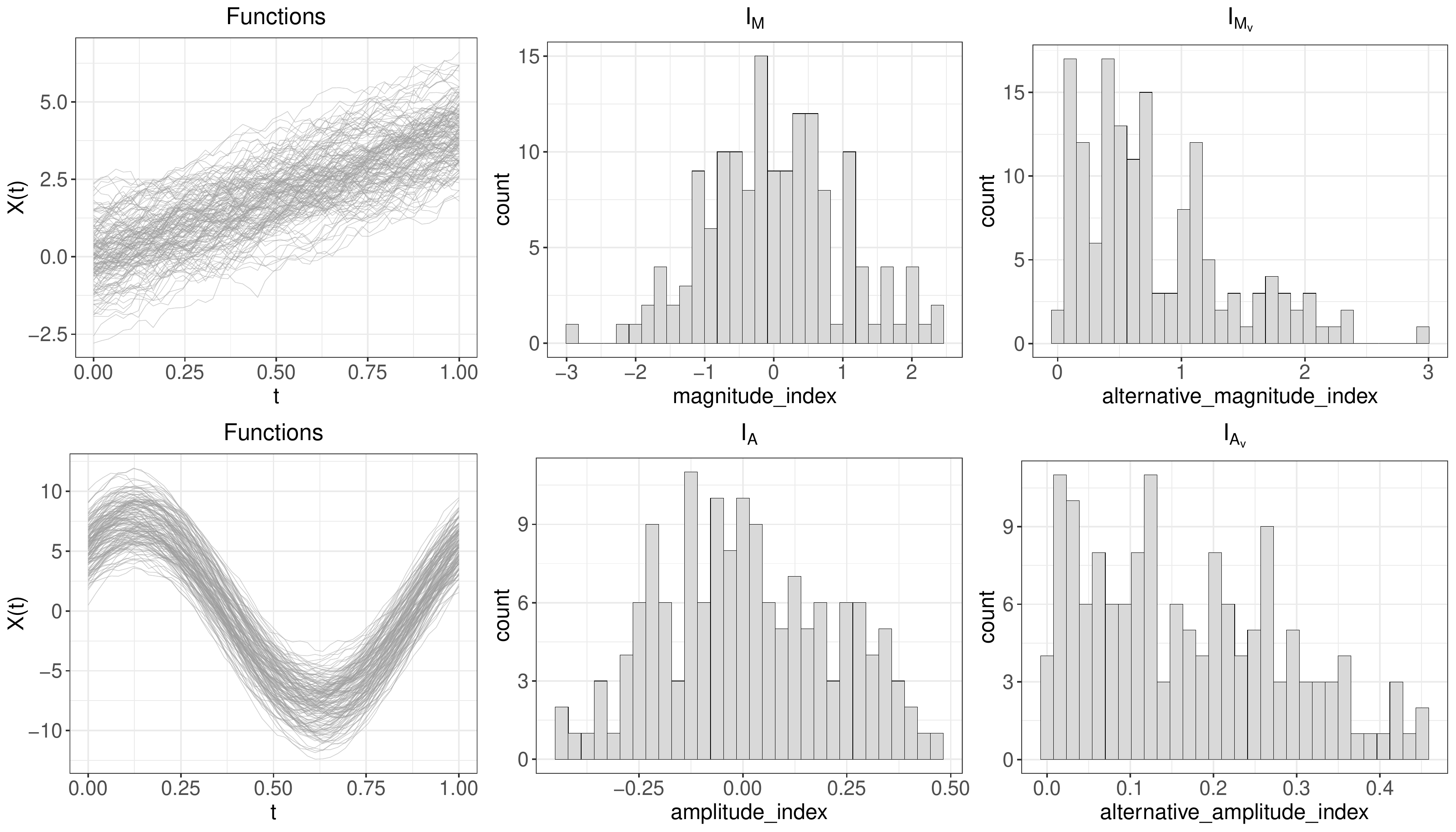}
\caption{FastMUOD and alternative FastMUOD amplitude and magnitude indices. The first row shows $I_M$ and $I_{M_{v}}$ with an approximately normal and right-skewed distribution respectively. The second row shows the same for $I_A$ and $I_{A_{v}}$.\label{fig::fast_abs_nabs}}
\end{figure}

The primary difference between $I_{A}(y, F_X)$ and $I_{A_v}(y, F_X)$ is the use of the  absolute value function in the latter, guaranteeing that $I_{A_v}(y, F_X)$ is always positive (the same applies to $I_{M}(y, F_X)$ and $I_{M_v}(y, F_X)$). However, this makes the distributions of $I_{A_v}(y, F_X)$ and $I_{M_v}(y, F_X)$ right-skewed, compared to normal distribution for $I_{A}(y, F_X)$ and $I_{M}(y, F_X)$ (Figure \ref{fig::fast_abs_nabs}). Note that additional information about the nature of magnitude and amplitude outliers is lost because $I_{M}(y, F_X)$ ($I_{A}(y, F_X)$) assigns larger index values to higher magnitude (amplitude) outliers and smaller index values to lower magnitude (amplitude) outliers, which is not the case with $I_{A_v}(y, F_X)$ and $I_{M_v}(y, F_X)$. For these reasons, we recommend to use $I_{A}(y, F_X)$ and $I_{M}(y, F_X)$ whenever possible. 

It is straightforward to constructs the respective sample and finite dimensional versions of $I_{A_v}(y, F_X)$ and $I_{M_v}(y, F_X)$ following the ideas in Subsection \ref{subsec::deffmuod} of the Main Text. However, their properties slightly differ under simple transformations.

\begin{proposition} [Properties of original FastMUOD indices]

Suppose that $X$ is a stochastic process in $L^2([0,1])$ with distribution $F_X$ and mean function $\mu(t)$. Let $y$ and $z$ other functions in $L^2([0,1])$ (which may be realizations of $X$) and let $a, b \in \mathbb{R}$. Then the following statements hold.

% Suppose that $X$ is a stochastic process in $L^2([0,1])$ with distribution $F_X$ and mean function $\mu(t)$. Let $y$ be another function in $L^2([0,1])$ which may or may not be a realization of $X$. Then the following statements holds true.
\begin{enumerate}
    \item  For a new function $y'(t) = a y(t) + b$: $I_{M_v}(y', F_X) = \lvert  aI_{M}(y, F_X) + b\rvert$.
    
    % \item  For a new function $y'(t) = ay(t)$; $I_{M_v}(y', F_X) = \lvert a I_{M}(y, F_X)\rvert = |a|\lvert  I_{M}(y, F_X)\rvert$.
    
    \item For a new function $y'(t) = y(t) + z(t)$: $I_{M_v}(y', F_X) = | I_{M}(y, F_X) + I_{M}(z, F_X)|$.
    
    \item For a new function $y'(t) = ay(t) + b$: $I_{A_v}(y', F_X) = |a I_{A}(y, F_X) +a-1|$.
    
    % \item For a new function $y'(t) = ay(t)$, 
    
    %\item For a new function $y'(t) = y(t) + z(t)$, if $\langle \tilde{z}, \tilde{\mu}\rangle = 0$ then $I_{A_v}(y', F_X) =I_{A_v}(y, F_X) = \left|I_A(y, F_X)\right|$. 
    \item For a new function $y'(t) = y(t) + z(t)$: if $\langle \tilde{z}, \tilde{\mu}\rangle = 0$ then $I_{A_v}(y', F_X) =I_{A_v}(y, F_X)$.
    %\item For a new function $y'(t) = z(t) y(t)$, if $\langle \tilde{zy}, \tilde{\mu}\rangle = \langle \tilde{y}, \tilde{\mu}\rangle$ then $I_{A_v}(y', F_X) = I_{A_v}(y, F_X) = \left|I_A(y, F_X)\right|$. 
    \item For a new function $y'(t) = z(t) y(t)$: if $\langle \tilde{zy}, \tilde{\mu}\rangle = \langle \tilde{y}, \tilde{\mu}\rangle$ then $I_{A_v}(y', F_X) = I_{A_v}(y, F_X)$. 
   
\end{enumerate}

\begin{proof} 
Proofs of the statements follows directly from the definition of $I_{M_v}$ and $I_{A_v}$, and application of Proposition \ref{prop:p3}.
\end{proof}
\end{proposition}

The following corollary establishes conditions under which the original magnitude and amplitude indices of a function remain the same after they have been scaled and/or translated. 

\begin{corollary}
Suppose that $X$ is a stochastic process in $L^2([0,1])$ with distribution $F_X$ and mean function $\mu(t)$. Let $y$ be another function in $L^2([0,1])$ and let $a, b \in \mathbb{R}$. Then the following statements hold,

\begin{enumerate}
    \item  For a new function $y'(t) = a y(t) + b$: $I_{M_v}(y', F_X) = I_{M_v}(y, F_X)$ iff $b = (-a \pm 1) I_M(y, F_X)$
    
    \item  For a new function $y'(t) = a y(t) + b$: $I_{A_v}(y', F_X) = I_{A_v}(y, F_X)$ iff $a = 1$ or $a = \frac{1 - I_{A}(y, F_X)}{1 + I_{A}(y, F_X)}$
   
\end{enumerate}
\begin{proof} 
The proofs of the statements follow from the definition. 
\begin{enumerate}
    \item Suppose that $I_{M_v}(y', F_X) = I_{M_v}(y, F_X)$, then by definition (of $I_{M_v}$) and Proposition \ref{prop:p3} we have that:
    \begin{equation*}
        |aI_M(y, F_X) +b | = |I_M(y, F_X)|.
    \end{equation*}
    If both $aI_M(y, F_X) +b$ and $I_M(y, F_X)$ have the same sign, then 
    \begin{equation*}
        aI_M(y, F_X) +b  = I_M(y, F_X),
    \end{equation*}
    which implies that $b = (-a+1)I_M(y, F_X)$. However, if $aI_M(y, F_X) +b$ and $I_M(y, F_X)$ have different signs, 
    \begin{equation*}
        aI_M(y, F_X) +b  = -I_M(y, F_X),
    \end{equation*}
    which implies that $b = (-a-1)I_M(y, F_X)$.
    
    To prove the reverse direction, we have to show that whenever $b = (-a \pm 1) I_M(y, F_X)$, $I_{M_v}(y', F_X) = I_{M_v}(y, F_X)$. For the first case, we assume that $b = (-a+1)I_M(y, F_X)$, then 
    \begin{align*}
        I_{M_v}(y', F_X) &= |I_{M}(y', F_X)|\\
        &= |aI_{M}(y, F_X) + b| \\ 
        &= |aI_{M}(y, F_X) + (1-a)I_M(y, F_X)| \\
        &= |I_M(y, F_X)|\\
        &= I_{M_v}(y, F_X).
    \end{align*}
    For the second case, suppose that $b = (-a-1)I_M(y, F_X)$, then
    \begin{align*}
        I_{M_v}(y', F_X) &= |I_{M}(y', F_X)|\\
        &= |aI_{M}(y, F_X) + b| \\ 
        &= |aI_{M}(y, F_X) + (-a-1)I_M(y, F_X)| \\
        &= |-I_M(y, F_X)|\\
        &= |I_M(y, F_X)|\\
        &= I_{M_v}(y, F_X).
    \end{align*}
    So in both cases, we have that $I_{M_v}(y', F_X) = I_{M_v}(y, F_X)$, which completes the proof.
    
    \item Suppose that $I_{A_v}(y', F_X) = I_{A_v}(y, F_X)$, by definition (of $I_{A_v}$) and Proposition \ref{prop:p3} we have
    $$|aI_A(y, F_X) +a-1 | = |I_A(y, F_X)|.$$
    
    If both $aI_A(y, F_X) +a-1$ and $I_A(y, F_X)$ have the same sign,
    $$aI_A(y, F_X) +a-1 = I_A(y, F_X),$$ which indicates that
    $$a = \frac{I_A(y, F_X) + 1}{I_A(y, F_X) + 1} = 1.$$
    Nevertheless, if $aI_A(y, F_X) +a-1$ and $I_A(y, F_X)$ have different signs,$$aI_A(y, F_X) +a-1 = - I_A(y, F_X),$$ which indicates that
    $$a = \frac{1 - I_A(y, F_X) }{1 + I_A(y, F_X)}.$$
    
    To prove the reverse case, we have to show that whenever $a = 1$ or $a = \frac{1 - I_A(y, F_X) }{1 + I_A(y, F_X)}$, $I_{A_v}(y', F_X) = I_{A_v}(y, F_X)$. For the first case, assume that $a = 1$, then: 
    \begin{align*}
        I_{A_v}(y', F_X) &= |I_{A}(y', F_X)|\\
        &= |aI_{A}(y, F_X) + a - 1| \\ 
        &= |I_{A}(y, F_X)| \\
        &= I_{A_v}(y, F_X). 
    \end{align*}
    For the second case, assume that $a = \frac{1 - I_A(y, F_X) }{1 + I_A(y, F_X)}$, then: 
    \begin{align*}
        I_{A_v}(y', F_X) &= |I_{A}(y', F_X)|\\
        &= |aI_{A}(y, F_X) + a - 1| \\ 
        &= \left|\frac{(1 + I_A(y, F_X) (1 - I_A(y, F_X))}{1 + I_A(y, F_X)} -1\right| \\
        &= |-I_{A}(y, F_X)|\\
        &= |I_{A}(y, F_X)|\\
        &= I_{A_v}(y, F_X). 
    \end{align*}
    Thus, in both cases, $I_{A_v}(y', F_X) = I_{A_v}(y, F_X)$, which completes the proof.

\end{enumerate}
\end{proof}
\end{corollary}
\subsection{Plot of Simulation Models}
\begin{figure}[htbp!]
	\centering
\includegraphics[scale = .35]{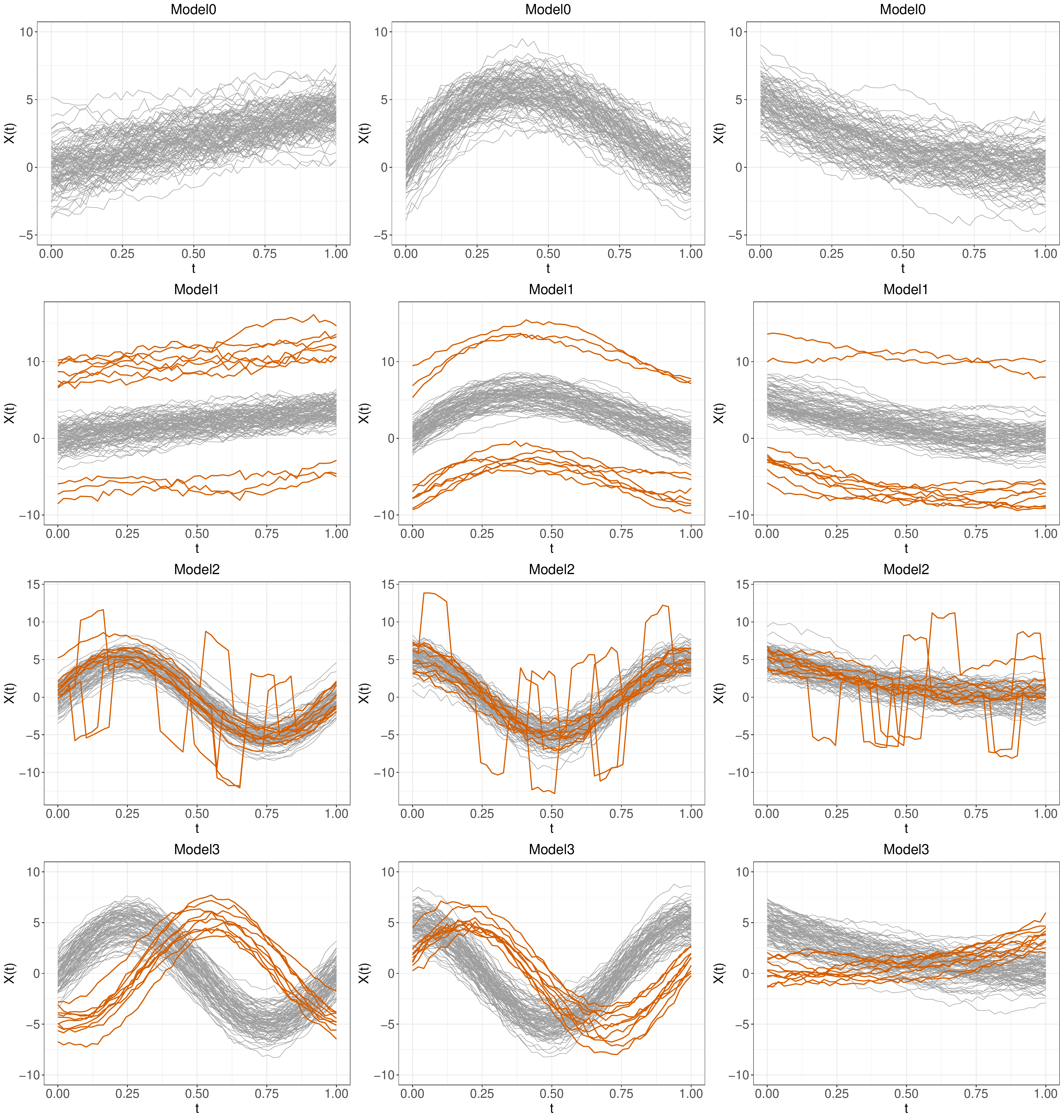}
\caption{Sample data generated by Models 0 -- 3  with contamination rate $\alpha= 0.10$, sample size $n = 100$, and evaluation point $d = 50$. Each row corresponds to a simulation model, and each column corresponds to the margins of the multivariate functional data. Outliers are shown in colour.\label{fig::sim_model_1}}
\end{figure}
\begin{figure}[htbp!]
	\centering
\includegraphics[scale = .35]{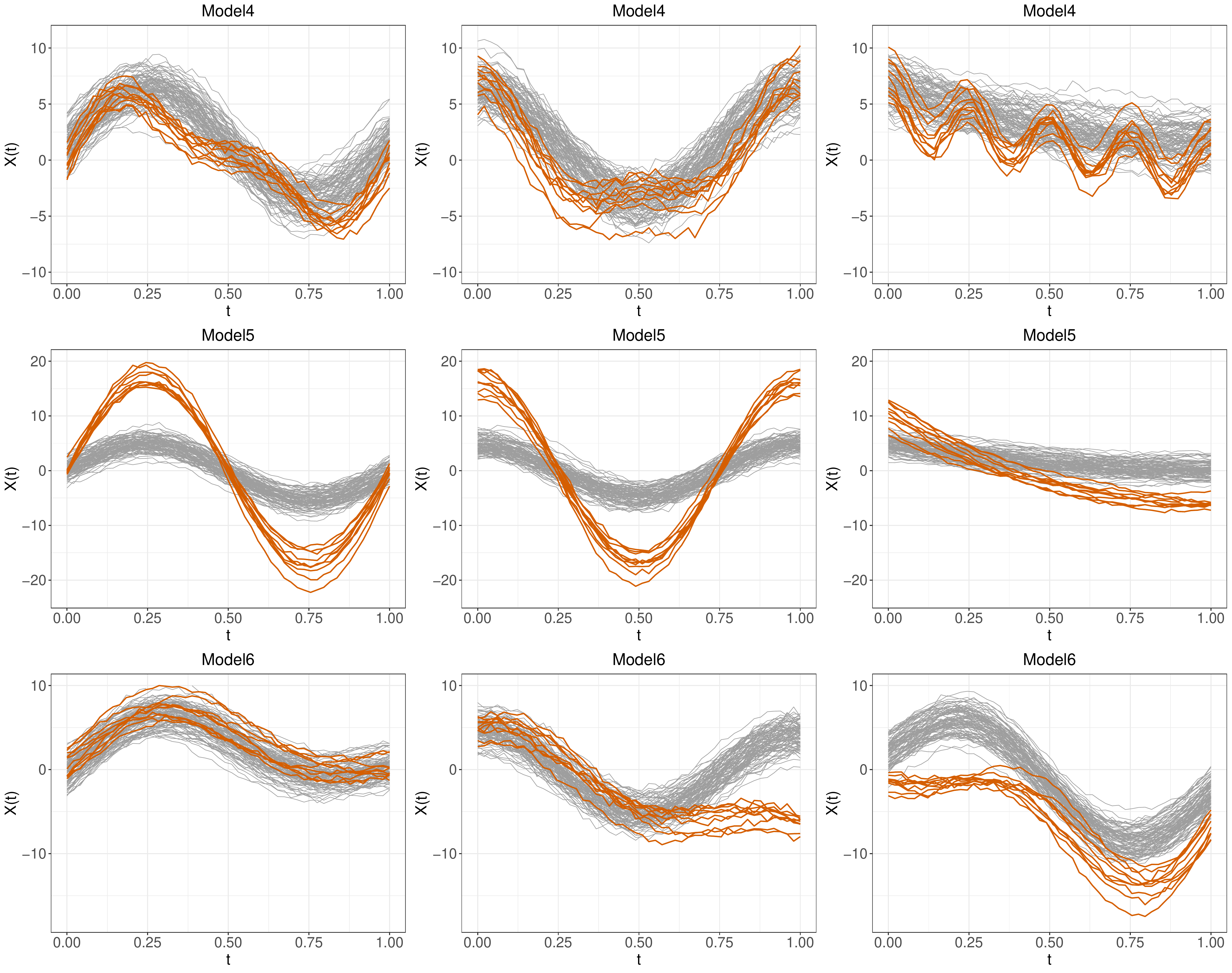}
\caption{Sample data generated by Models 4 -- 6  with contamination rate $\alpha= 0.10$, sample size $n = 100$, and evaluation point $d = 50$. Each row corresponds to a simulation model and each column corresponds to the margins of the multivariate functional data. Outliers are shown in colour.\label{fig::sim_model_2}}
\end{figure}
\newpage
\subsection{More Simulation Results}
\label{subsec::more-sim-res}
\begin{figure}[htbp!]
	\centering
\includegraphics[scale = .35]{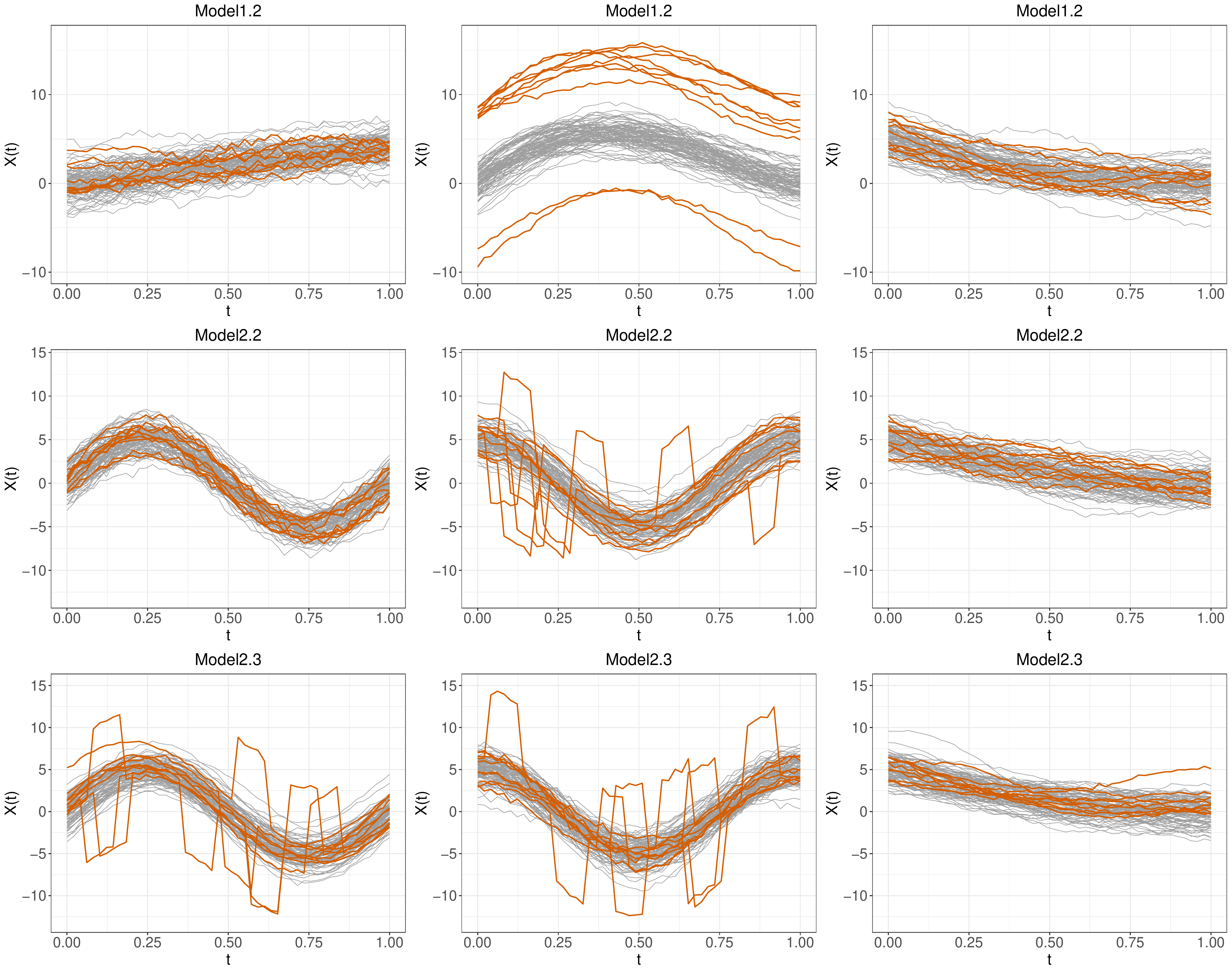}
\caption{Sample data generated by variants of Models 1 and 2  with contamination rate $\alpha= 0.10$, sample size $n = 100$, and evaluation point $d = 50$. Each row corresponds to a simulation model and each column corresponds to the margins of the multivariate functional data. Outliers are shown in colour.\label{fig::sim_model_1_ex}}
\end{figure}

\begin{figure}[htbp!]
	\centering
\includegraphics[scale = .35]{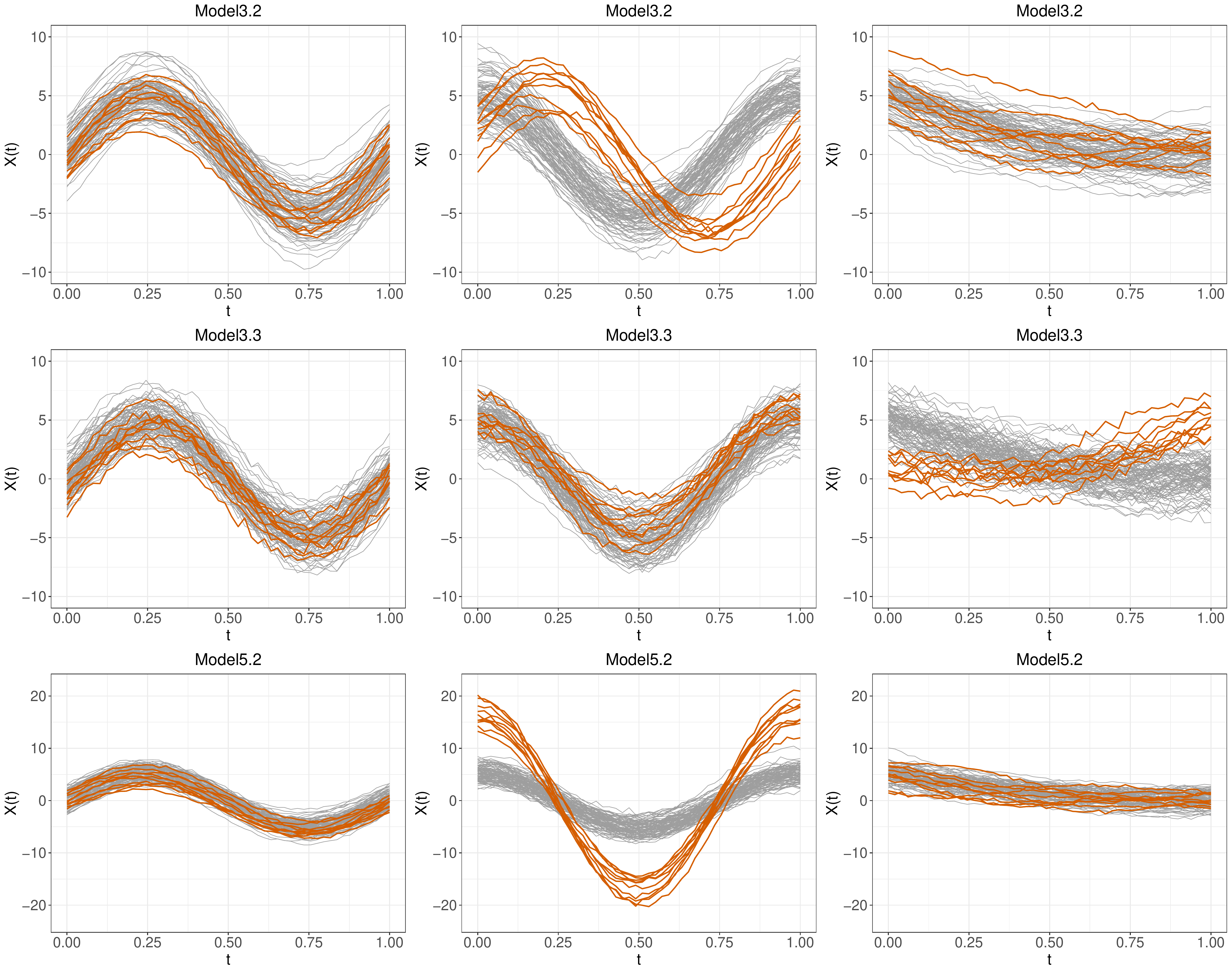}
\caption{Sample data generated by variants of Models 3 and 5  with contamination rate $\alpha= 0.10$, sample size $n = 100$, and evaluation point $d = 50$. Each row corresponds to a simulation model and each column corresponds to the margins of the multivariate functional data. Outliers are shown in colour.\label{fig::sim_model_2_ex}}
\end{figure}

In this section we show the results of the methods outlined in Subsection \ref{sec::sec4_2}  (of the Main Text) on more simulation models. The models considered are variants of the simulation models in Subsection \ref{sec::sec4_1}. The outliers are outlying only in one or two dimensions of the trivariate functional dataset. Figures \ref{fig::sim_model_1_ex} and \ref{fig::sim_model_2_ex} show the simulation models, and the results are shown in Table \ref{tab:sim-results-100-50-extra}.

\begin{table*}[htbp!]
\caption{\label{tab:sim-results-100-50-extra}Mean and Standard Deviation (in parentheses) of the TPR and FPR (in percentage) over 200 repetitions for each model. Sample size $n=100$, evaluation points $t_j=50$, and contamination rate is $10\%$. The proposed methods are in italics.}
\centering
\setlength\tabcolsep{11pt} % default value: 6pt
{\renewcommand{\arraystretch}{.8}
  \footnotesize
\begin{tabular}{@{}lcccccc@{}}  \toprule
  \multirow{2}{*}{Method}  & \multicolumn{2}{c}{Model 1.2} & \multicolumn{2}{c}{Model 2.2} & \multicolumn{2}{c}{Model 2.3} \\
  \cmidrule{2-3} \cmidrule{4-5} \cmidrule{6-7} 
&  TPR & FPR & TPR & FPR & TPR & FPR \\ 
  \midrule
 \textit{FST-MAR} & 100.0(0.7) & 26.0(4.2) & 95.0(8.7) & 20.9(4.0) & 99.8(1.6) & 17.9(3.8) \\ 
\textit{FST-STR} & 87.8(12.7) & 2.1(1.5) & 26.9(16.2) & 2.9(2.0) & 64.9(18.9) & 2.4(1.8) \\ \hline
\textit{FST-PRJ} & 99.6(2.1) & 0.3(0.7) & 34.0(28.1) & 0.9(1.2) & 88.5(13.4) & 0.9(1.0) \\ 
\textit{FST-PRJ-SH} & 0.4(1.8) & 0.2(0.6) & 34.0(28.0) & 0.9(1.2) & 88.5(13.4) & 0.8(1.0) \\ 
\textit{FST-PRJ-AM} & 0.0(0.0) & 0.0(0.0) & 0.1(0.7) & 0.0(0.1) & 0.0(0.0) & 0.0(0.0) \\ 
\textit{FST-PRJ-MG} & 99.6(2.1) & 0.1(0.3) & 0.1(1.0) & 0.1(0.3) & 0.0(0.0) & 0.0(0.2) \\ \hline
\textit{FST-PRJ1} & 99.5(2.4) & 3.8(2.2) & 48.4(18.8) & 1.6(1.4) & 91.5(10.3) & 1.3(1.1) \\ 
\textit{FST-PRJ1-SH} & 3.0(5.6) & 3.6(2.1) & 47.9(18.6) & 1.3(1.2) & 90.4(10.8) & 1.0(1.0) \\ 
\textit{FST-PRJ1-AM} & 0.3(1.6) & 0.2(0.4) & 1.1(3.1) & 0.1(0.3) & 2.7(5.3) & 0.1(0.3) \\ 
\textit{FST-PRJ1-MG} & 99.4(2.7) & 0.1(0.3) & 0.4(2.0) & 0.3(0.6) & 2.0(4.6) & 0.2(0.5) \\ \hline
\textit{FST-PRJ2} & 100.0(0.0) & 50.6(4.2) & 99.7(1.7) & 43.1(4.4) & 100.0(0.0) & 38.3(4.2) \\ 
\textit{FST-PRJ2-SH} & 48.3(15.2) & 46.7(4.0) & 99.0(3.3) & 35.9(3.8) & 100.0(0.7) & 30.6(3.4) \\ 
 \textit{FST-PRJ2-AM} & 11.4(11.0) & 12.0(4.0) & 31.7(14.2) & 12.2(3.8) & 47.2(17.0) & 10.7(3.5) \\ 
   \textit{FST-PRJ2-MG} & 100.0(0.0) & 8.4(3.4) & 16.7(12.0) & 9.9(3.7) & 25.7(13.9) & 9.4(3.7) \\ \hline
 MSPLOT & 100.0(0.7) & 0.6(1.1) & 78.4(17.6) & 1.1(1.3) & 99.0(3.5) & 1.2(1.4) \\ 
 FOM & 87.3(18.9) & 0.1(0.3) & 57.0(22.0) & 0.1(0.4) & 89.7(13.6) & 0.0(0.2) \\ 
  FAO & 90.6(18.2) & 0.0(0.3) & 30.3(19.8) & 0.0(0.2) & 66.3(23.5) & 0.0(0.2) \\ 
   
\end{tabular}}
\setlength\tabcolsep{11pt} % default value: 6pt
{\renewcommand{\arraystretch}{.8}
  \footnotesize
\begin{tabular}{@{}lcccccc@{}}  \toprule
\multirow{2}{*}{Method} & \multicolumn{2}{c}{Model 3.2} & \multicolumn{2}{c}{Model 3.3} & \multicolumn{2}{c}{Model 5.2}  \\
  \cmidrule{2-3} \cmidrule{4-5} \cmidrule{6-7} 
& TPR & FPR & TPR & FPR & TPR & FPR \\ 
 \midrule
\textit{FST-MAR} & 100.0(0.0) & 20.6(3.9) & 98.7(3.4) & 20.8(4.0) & 100.0(0.0) & 24.6(4.1) \\ 
\textit{FST-STR} & 98.0(4.9) & 2.3(1.7) & 48.1(18.5) & 2.9(2.1) & 100.0(0.0) & 2.2(1.7) \\ \hline
\textit{FST-PRJ} & 96.5(6.6) & 1.1(1.1) & 28.5(26.8) & 1.0(1.3) & 100.0(0.0) & 0.1(0.4) \\ 
\textit{FST-PRJ-SH} & 96.4(6.8) & 1.1(1.1) & 27.9(27.0) & 0.9(1.2) & 1.0(3.9) & 0.1(0.3) \\ 
\textit{FST-PRJ-AM} & 30.1(28.2) & 0.0(0.2) & 1.8(5.7) & 0.0(0.2) & 100.0(0.0) & 0.1(0.3) \\ 
\textit{FST-PRJ-MG} & 0.0(0.0) & 0.0(0.0) & 0.2(1.4) & 0.0(0.2) & 0.0(0.0) & 0.0(0.0) \\ \hline
\textit{FST-PRJ1} & 96.8(7.5) & 1.6(1.3) & 49.6(20.8) & 1.9(1.4) & 100.0(0.0) & 1.8(1.3) \\ 
\textit{FST-PRJ1-SH} & 96.7(7.5) & 1.3(1.1) & 49.3(20.9) & 1.5(1.1) & 31.8(20.3) & 1.5(1.3) \\ 
\textit{FST-PRJ1-AM} & 43.8(25.8) & 0.1(0.2) & 3.5(8.1) & 0.1(0.4) & 100.0(0.0) & 0.0(0.2) \\ 
\textit{FST-PRJ1-MG} & 0.8(2.7) & 0.3(0.6) & 0.5(2.2) & 0.4(0.6) & 4.4(6.2) & 0.2(0.5) \\ \hline
\textit{FST-PRJ2} & 100.0(0.0) & 41.9(3.7) & 99.6(2.1) & 43.6(3.9) & 100.0(0.0) & 44.8(4.5) \\ 
\textit{FST-PRJ2-SH} & 100.0(0.0) & 34.8(3.5) & 99.6(2.1) & 36.2(3.6) & 99.1(3.2) & 40.3(4.2) \\ 
\textit{FST-PRJ2-AM} & 99.4(2.8) & 9.7(3.2) & 77.6(16.5) & 10.5(3.8) & 100.0(0.0) & 7.9(3.3) \\ 
\textit{FST-PRJ2-MG} & 18.3(12.9) & 9.6(3.4) & 27.4(18.9) & 10.1(3.5) & 65.3(20.6) & 8.5(3.4) \\ \hline
MSPLOT & 92.5(8.4) & 1.1(1.5) & 41.3(19.0) & 1.0(1.2) & 100.0(0.0) & 0.9(1.2) \\ 
FOM & 2.3(5.7) & 0.1(0.3) & 1.6(4.2) & 0.1(0.5) & 92.4(15.6) & 0.1(0.2) \\ 
FAO & 1.7(5.9) & 0.1(0.3) & 0.8(2.8) & 0.1(0.4) & 77.8(23.6) & 0.0(0.2) \\ 
   \bottomrule
\end{tabular}}
\end{table*}

\newpage

\subsection{Comparison of Various Thresholds $Q$}
\label{subsec::q_exp_res}
Now, we show the distribution of the F1 scores when FastMUOD with projections (FST-PRJ1) is used on the models presented in Section \ref{sec::sec4_1} with different threshold values of $Q = (\tau_S, \tau_A, \tau_M)$ ranging from $Q = (0.2, 0.2, 0.2)$ to  $Q = (0.7, 0.7, 0.7)$. For Model 0 with no outliers, we show only the FPRs.

\begin{figure}[htbp!]
	\centering
\includegraphics[scale = .8]{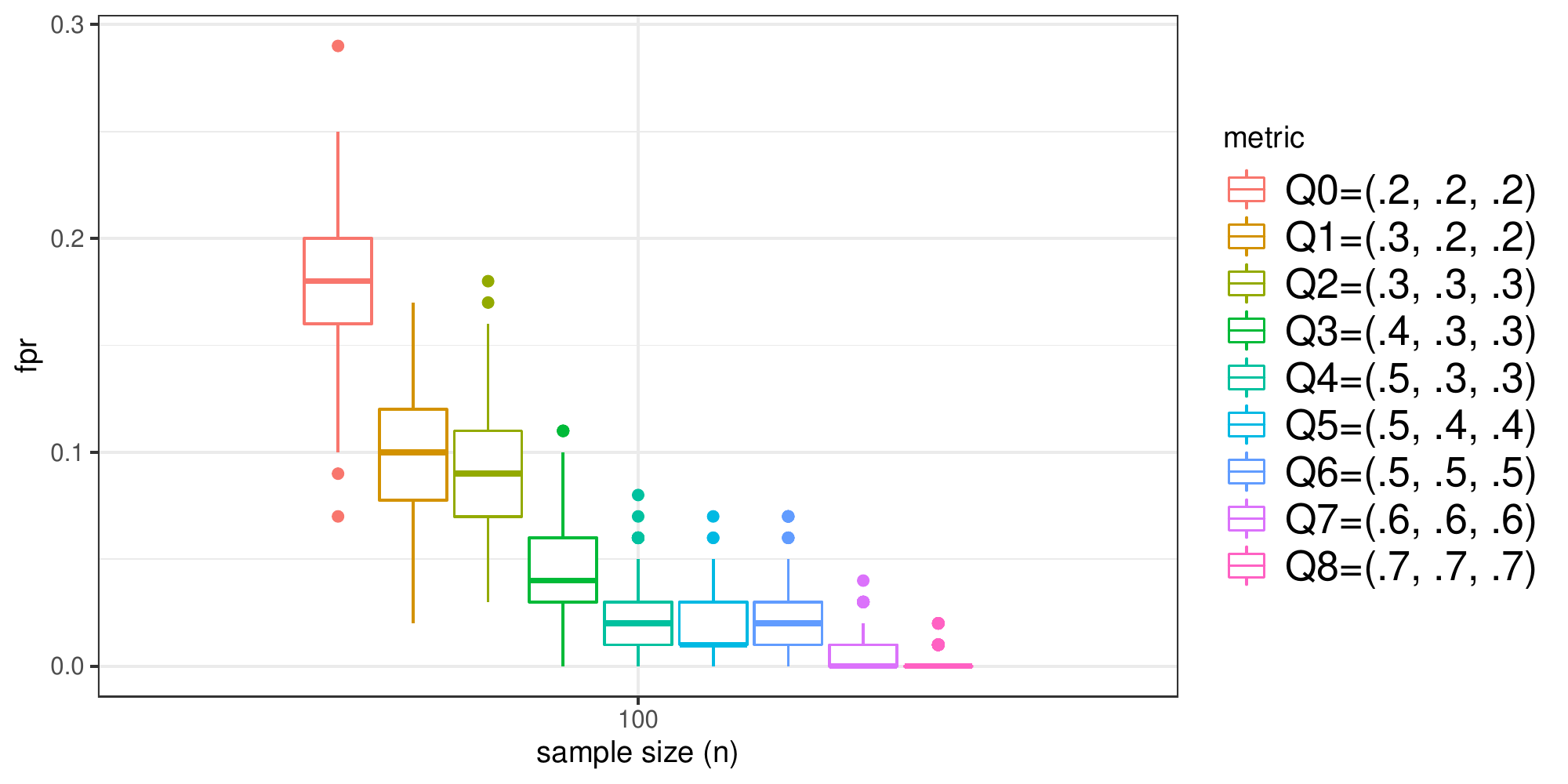}
\caption{The FPRs of FastMUOD with projections (FST-PRJ1) using different threshold values $Q = (\tau_S, \tau_A, \tau_M)$ on Model 0.\label{fig::q_exp_model0}}
\end{figure}

\begin{figure}[htbp!]
	\centering
\includegraphics[scale = .75]{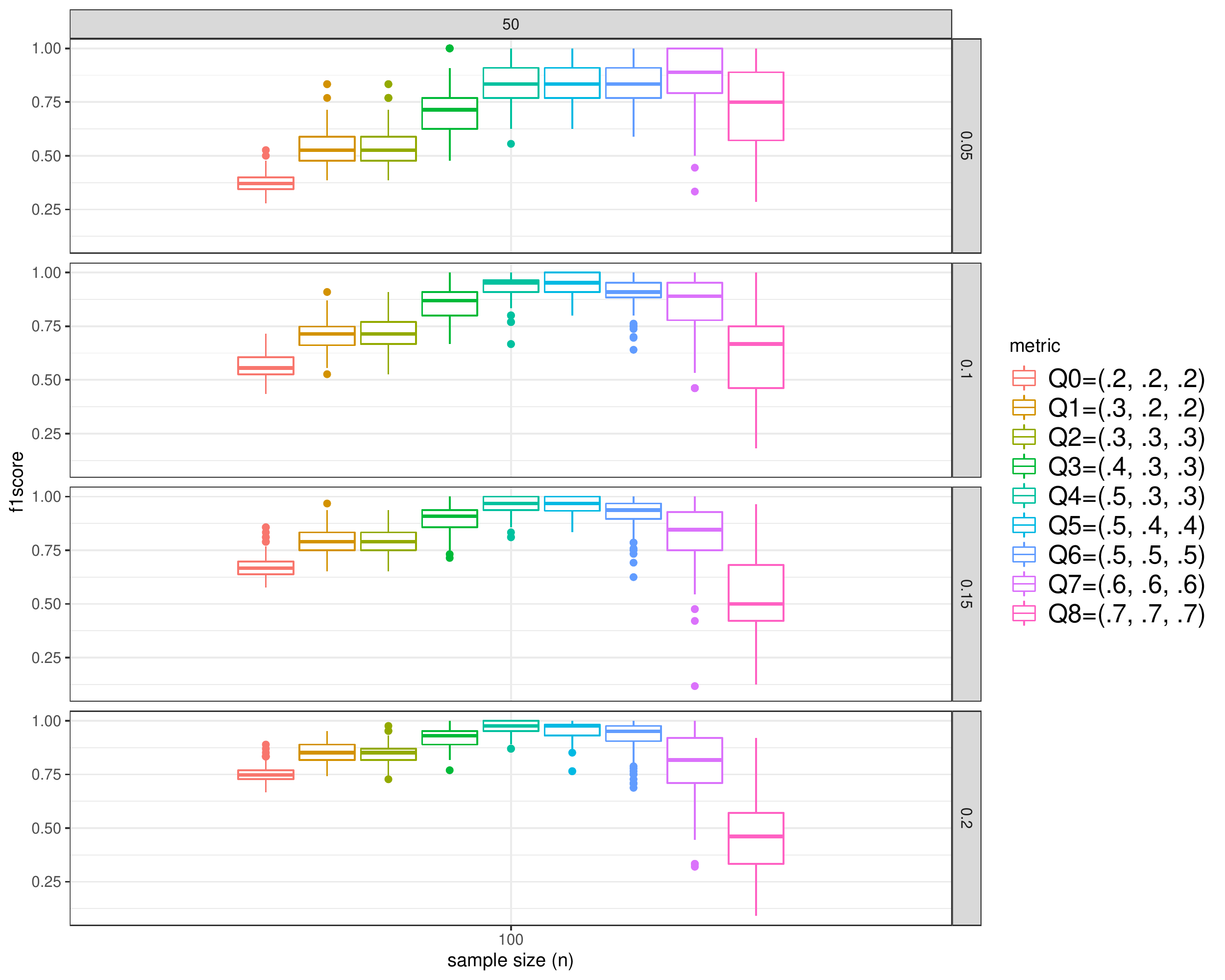}
\caption{F1 scores of FastMUOD with projections (FST-PRJ1) using different threshold values $Q = (\tau_S, \tau_A, \tau_M)$ on Model 1. The horizontal facets indicate the different contamination rates considered ($0.05, 0.1, 0.15, 0.2$).\label{fig::q_exp_model1}}
\end{figure}

\begin{figure}[htbp!]
	\centering
\includegraphics[scale = .75]{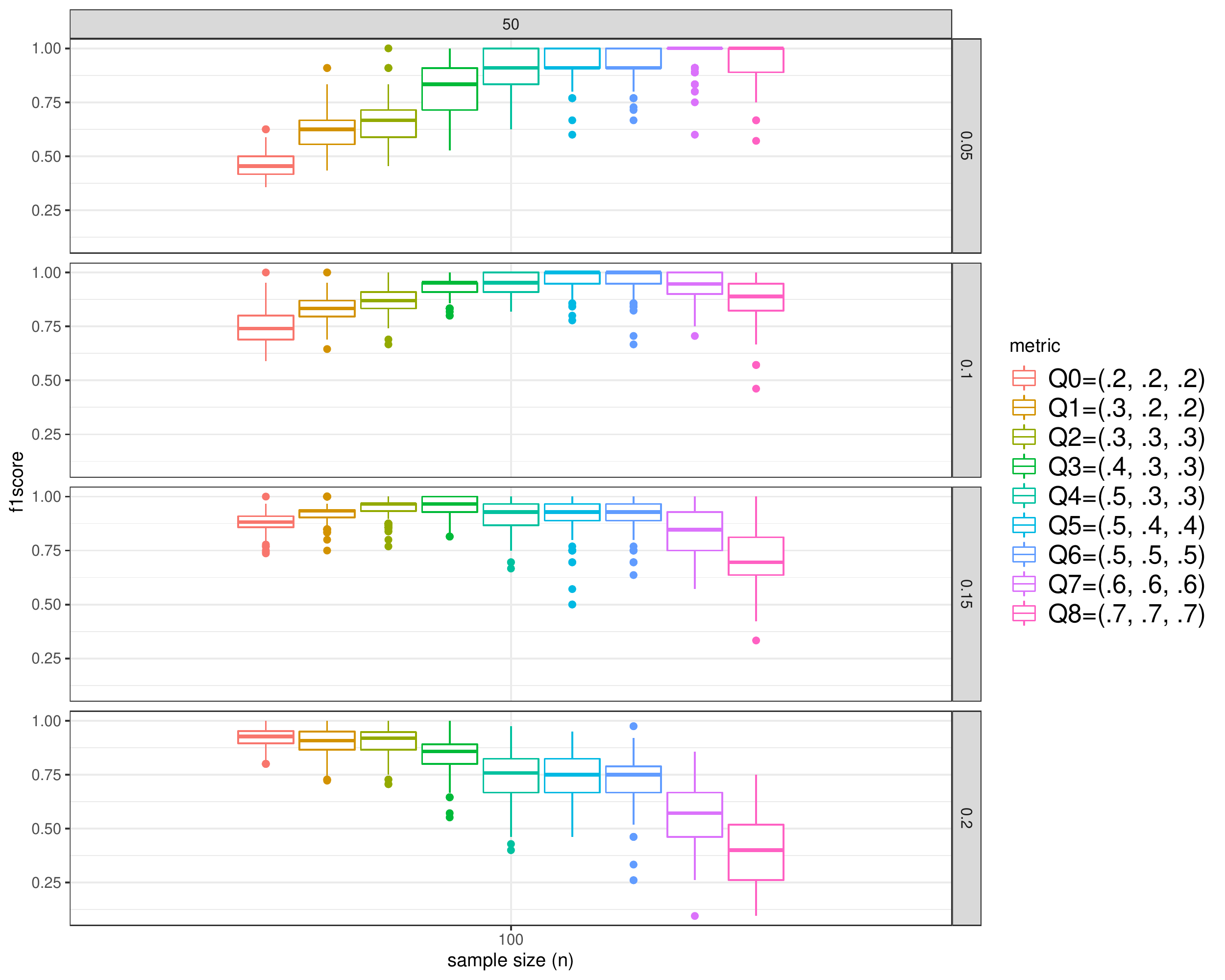}
\caption{F1 scores of FastMUOD with projections (FST-PRJ1) using different threshold values $Q = (\tau_S, \tau_A, \tau_M)$ on Model 2. The horizontal facets indicate the different contamination rates considered ($0.05, 0.1, 0.15, 0.2$).\label{fig::q_exp_model2}}
\end{figure}

\begin{figure}[htbp!]
	\centering
\includegraphics[scale = .75]{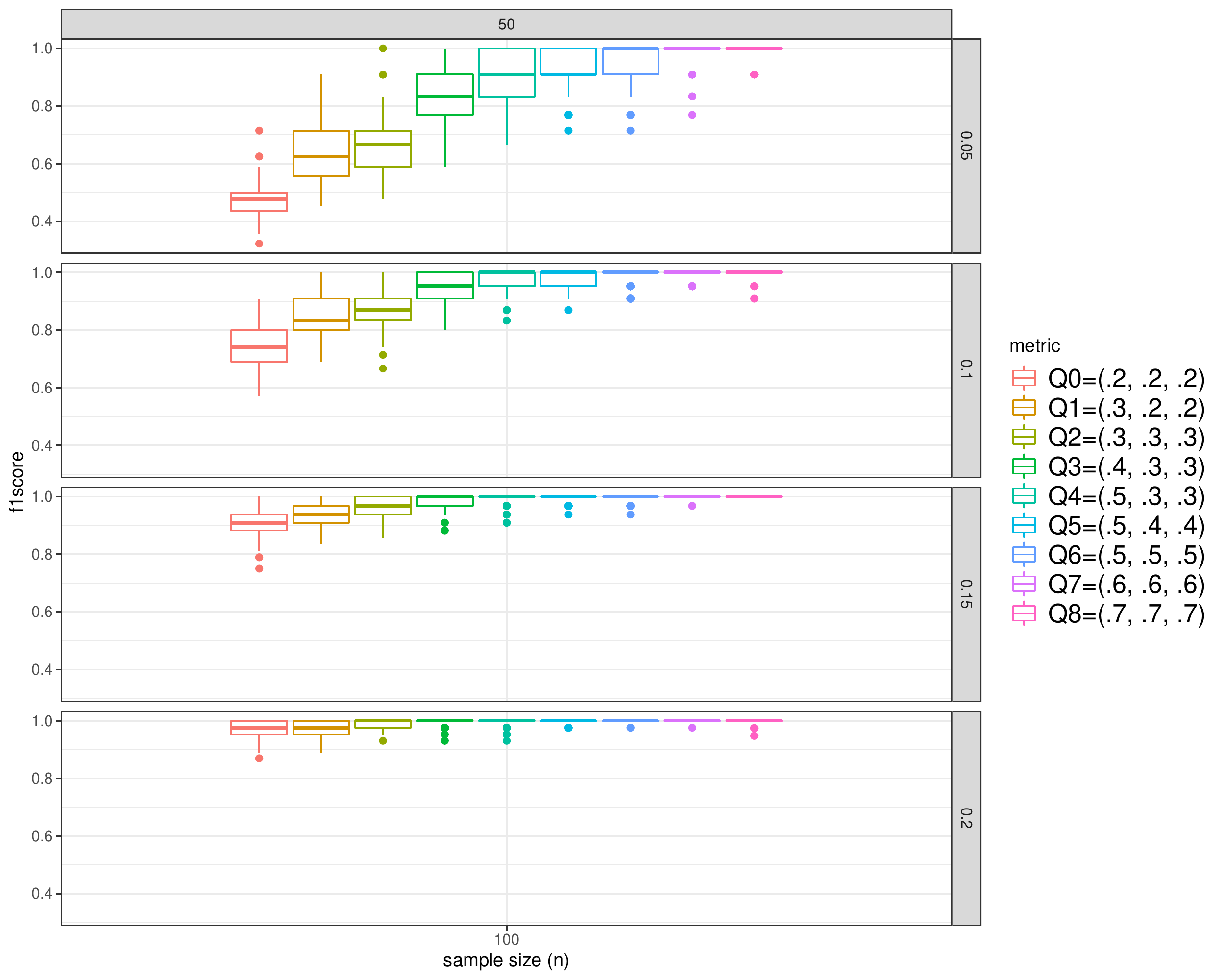}
\caption{F1 scores of FastMUOD with projections (FST-PRJ1) using different threshold values $Q = (\tau_S, \tau_A, \tau_M)$ on Model 3. The horizontal facets indicate the different contamination rates considered ($0.05, 0.1, 0.15, 0.2$).\label{fig::q_exp_model3}}
\end{figure}

\begin{figure}[htbp!]
	\centering
\includegraphics[scale = .75]{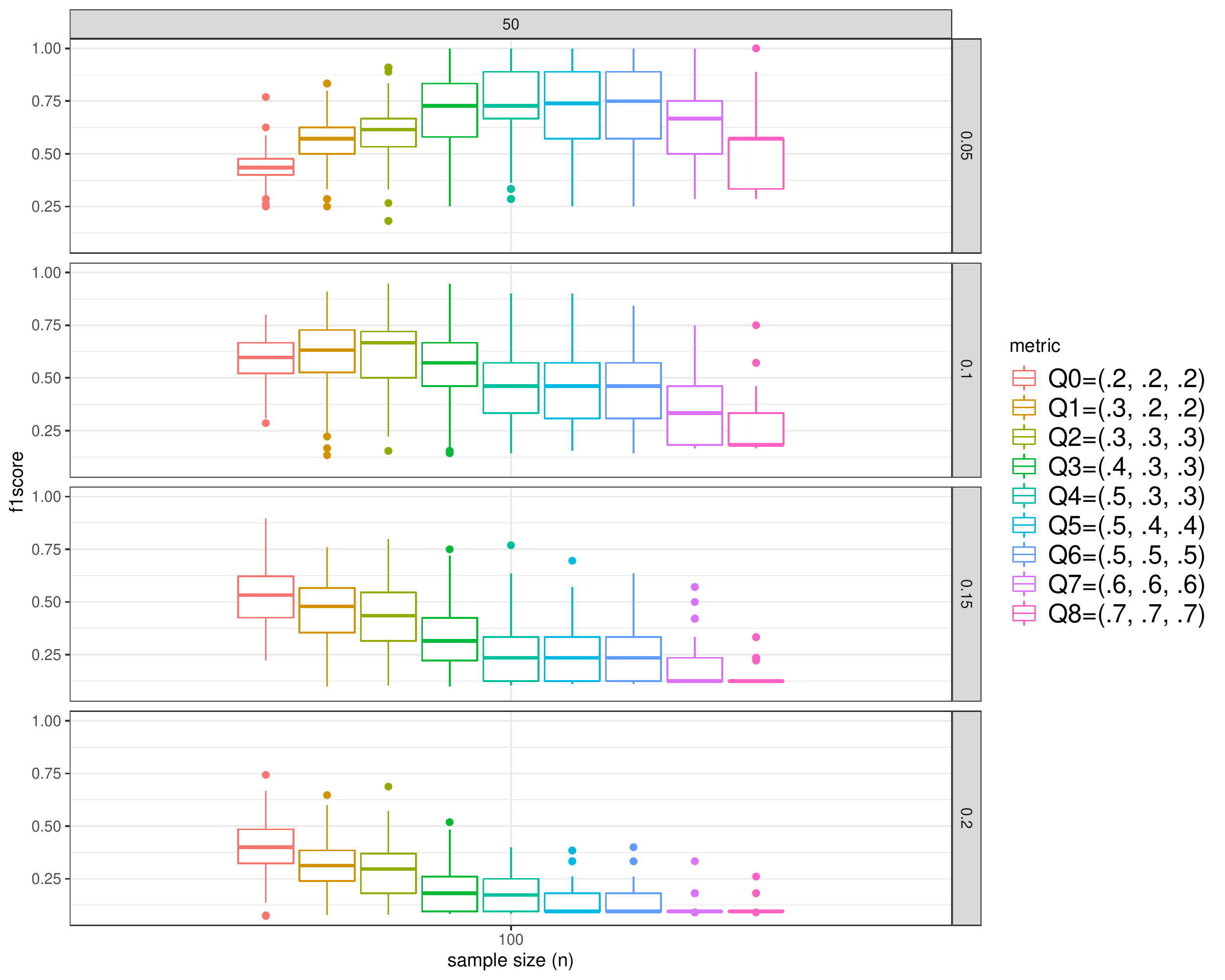}
\caption{F1 scores of FastMUOD with projections (FST-PRJ1) using different threshold values $Q = (\tau_S, \tau_A, \tau_M)$ on Model 4. The horizontal facets indicate the different contamination rates considered ($0.05, 0.1, 0.15, 0.2$).\label{fig::q_exp_model4}}
\end{figure}

\begin{figure}[htbp!]
	\centering
\includegraphics[scale = .75]{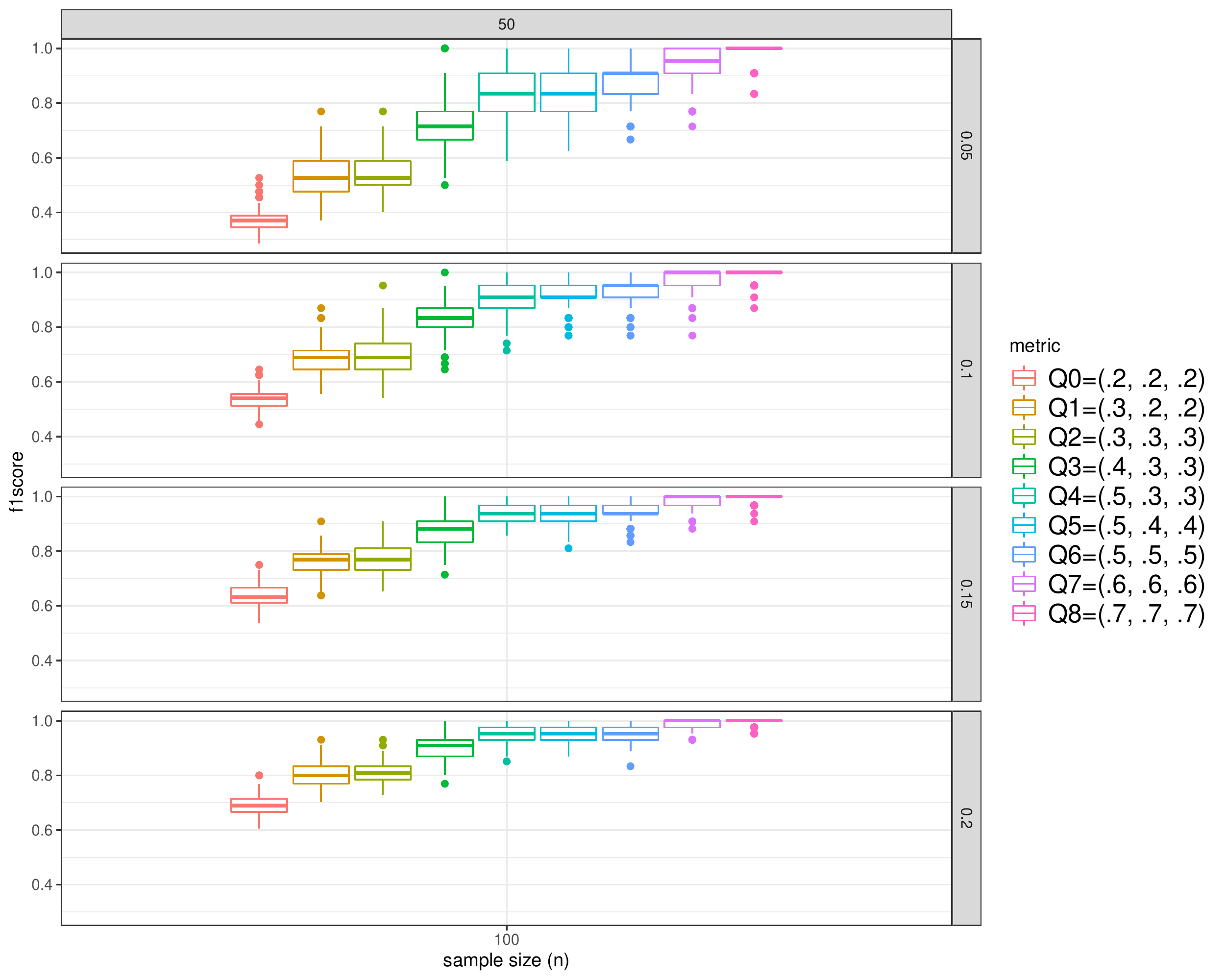}
\caption{F1 scores of FastMUOD with projections (FST-PRJ1) using different threshold values $Q = (\tau_S, \tau_A, \tau_M)$ on Model 5. The horizontal facets indicate the different contamination rates considered ($0.05, 0.1, 0.15, 0.2$).\label{fig::q_exp_model5}}
\end{figure}

\begin{figure}[htbp!]
	\centering
\includegraphics[scale = .75]{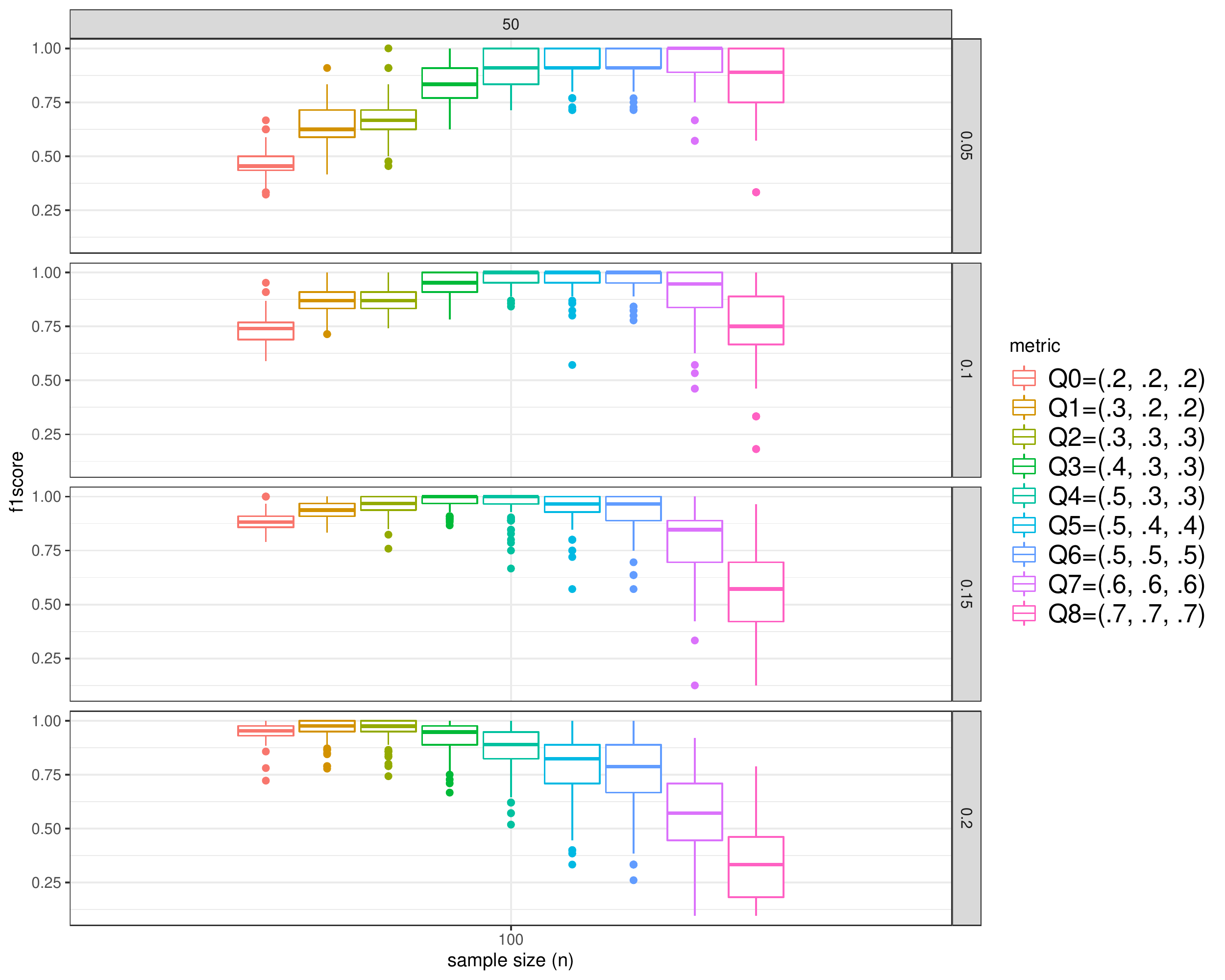}
\caption{F1 scores of FastMUOD with projections (FST-PRJ1) using different threshold values $Q = (\tau_S, \tau_A, \tau_M)$ on Model 6. The horizontal facets indicate the different contamination rates considered ($0.05, 0.1, 0.15, 0.2$).\label{fig::q_exp_model6}}
\end{figure}
\newpage
\newpage
\subsection{Character Data: Letter ``i"}
\begin{figure}[htbp!]
	\centering
\includegraphics[scale = .50]{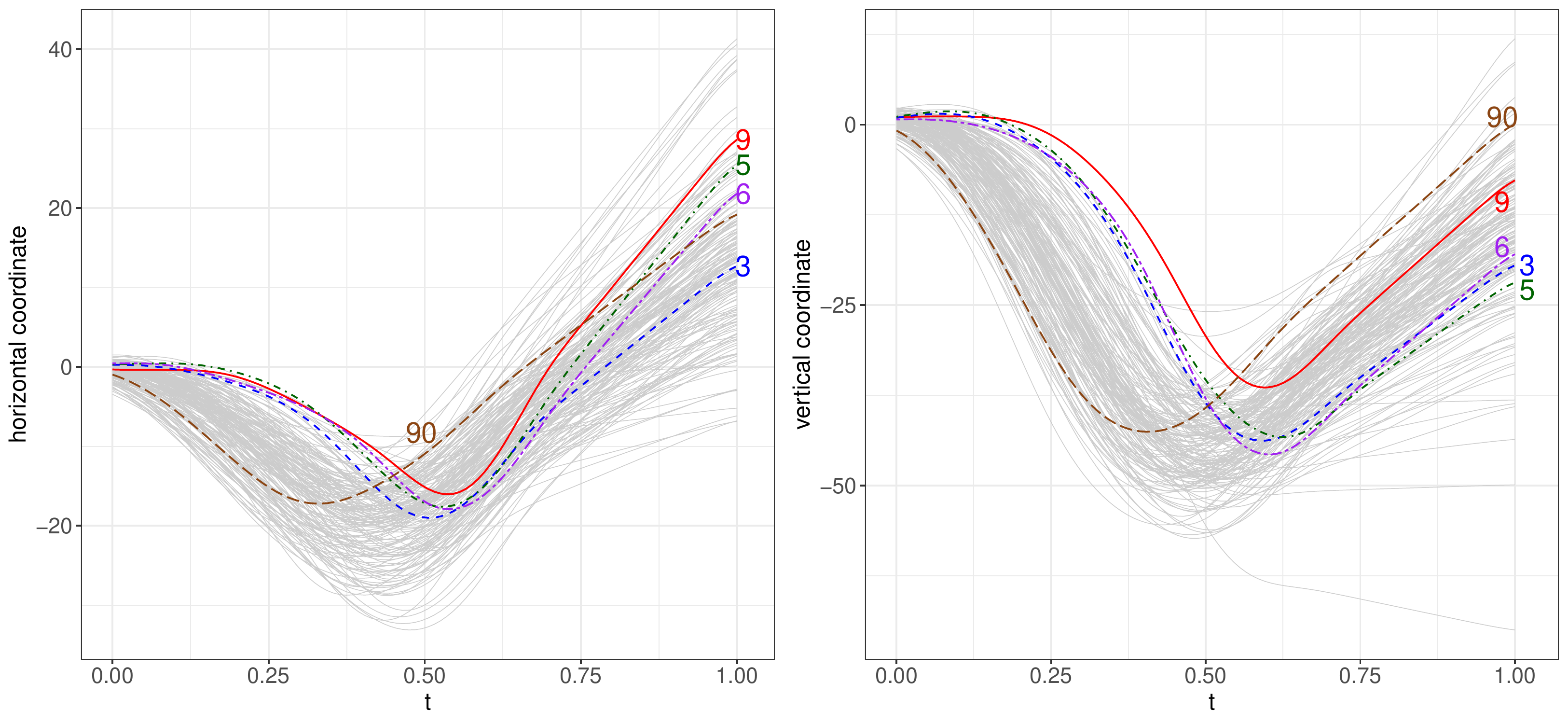}
\caption{Some shape outliers: curves 3, 5, 6, 9 and 90 with horizontal shift.\label{fig::charI_sha_shift}}
\end{figure}
\begin{figure}[htbp!]
	\centering
\includegraphics[scale = .50]{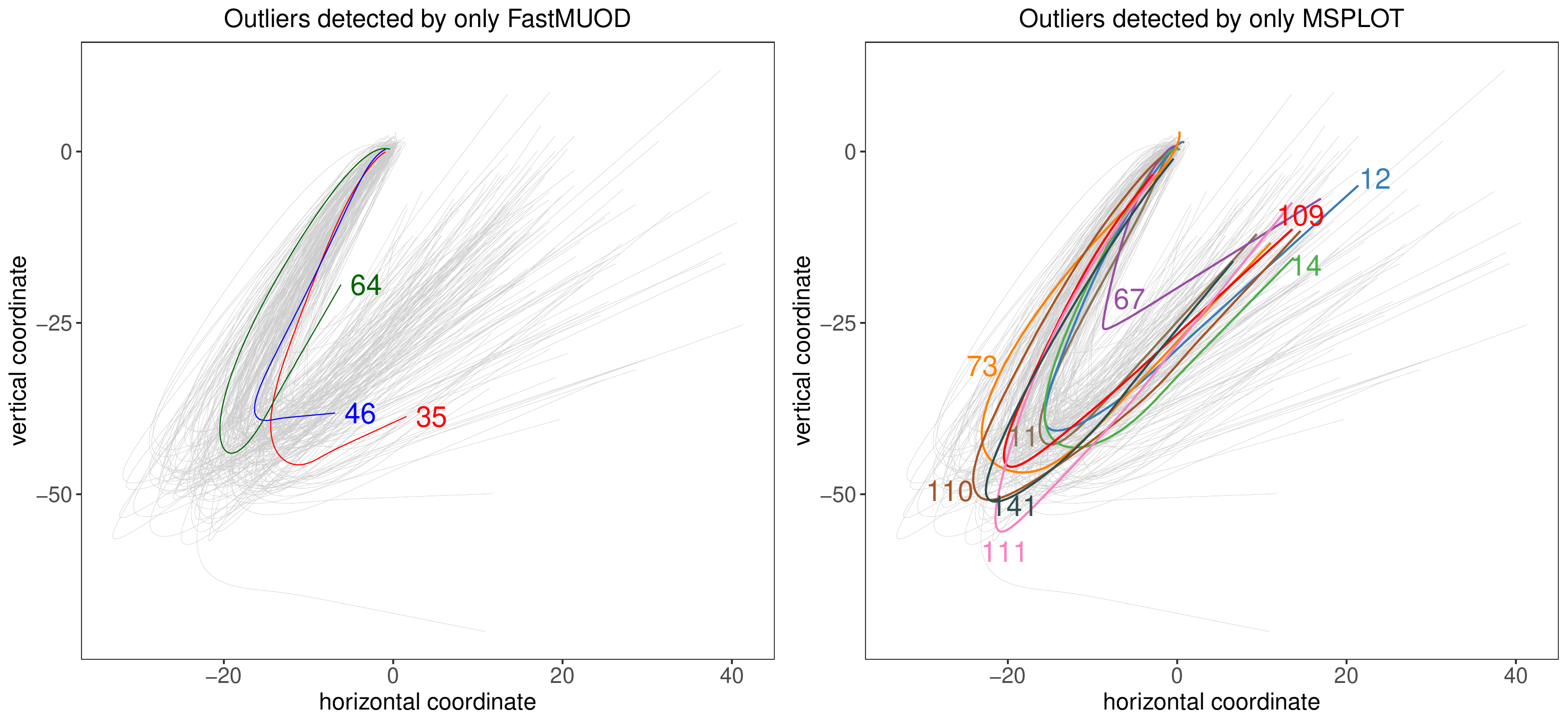}
\caption{Outliers detected by only FastMUOD and only MSPLOT.\label{fig::ex_outliers}}
\end{figure}
\begin{figure}[htbp!]
	\centering
\includegraphics[scale = .41]{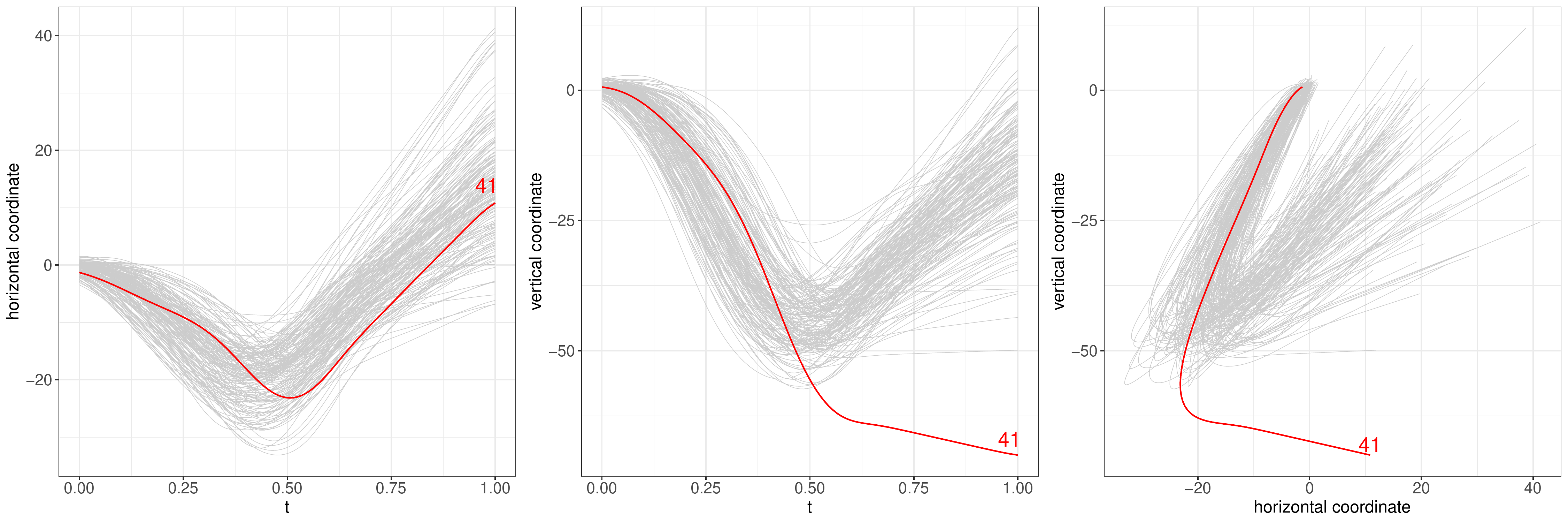}
\caption{Curve 41, the only outlier detected by FOM.\label{fig::curve_41_fom}}
\end{figure}

\newpage
\subsection{Character Data: Letter ``a"}
\begin{figure}[htbp!]
	\centering
\includegraphics[scale = .50]{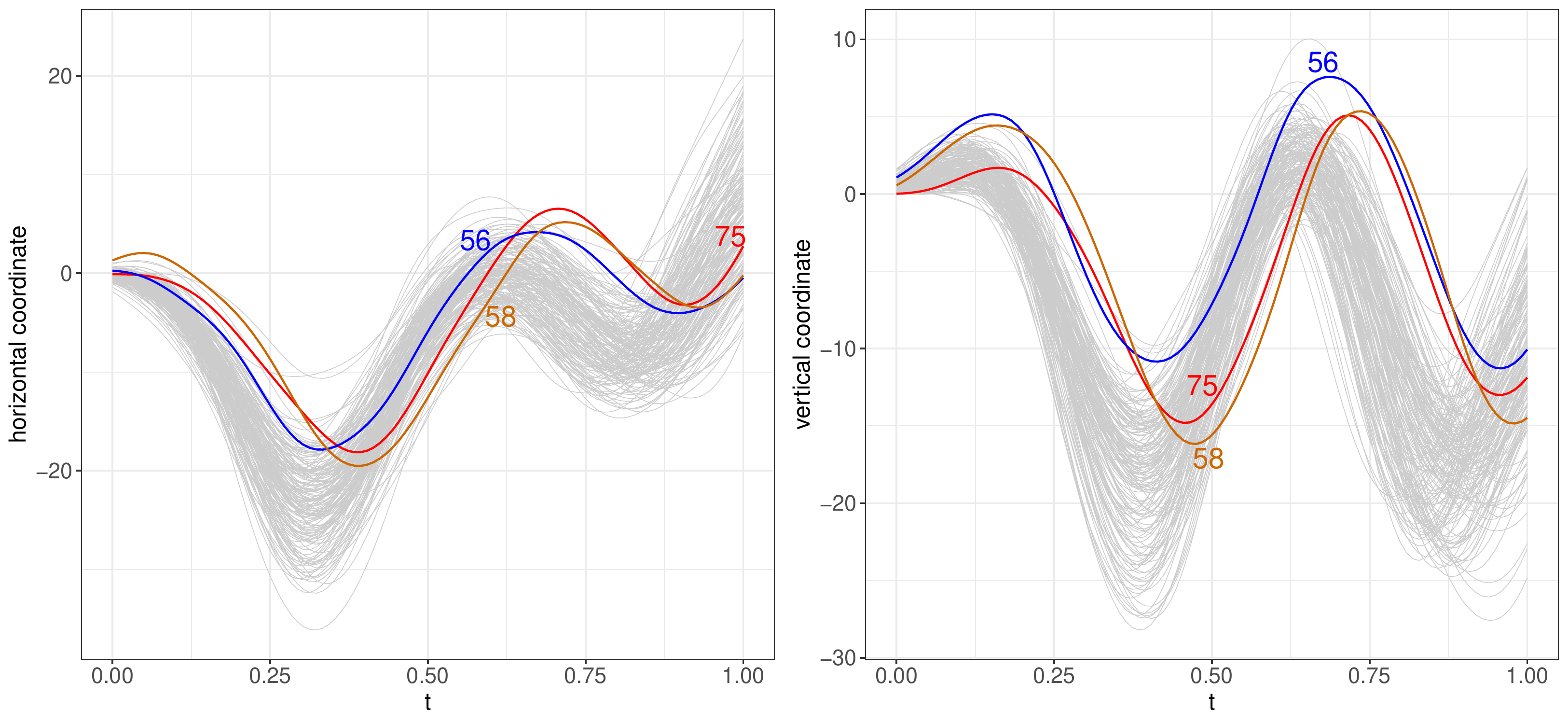}
\caption{Some shape outliers detected by FastMUOD  with a shift to the right in peaks resulting in ``short follow-throughs". See Figure \ref{fig::mag_amp_charA} of the Main Text.\label{fig::charA_sha_grp1}}
\end{figure}
\begin{figure}[htbp!]
	\centering
\includegraphics[scale = .50]{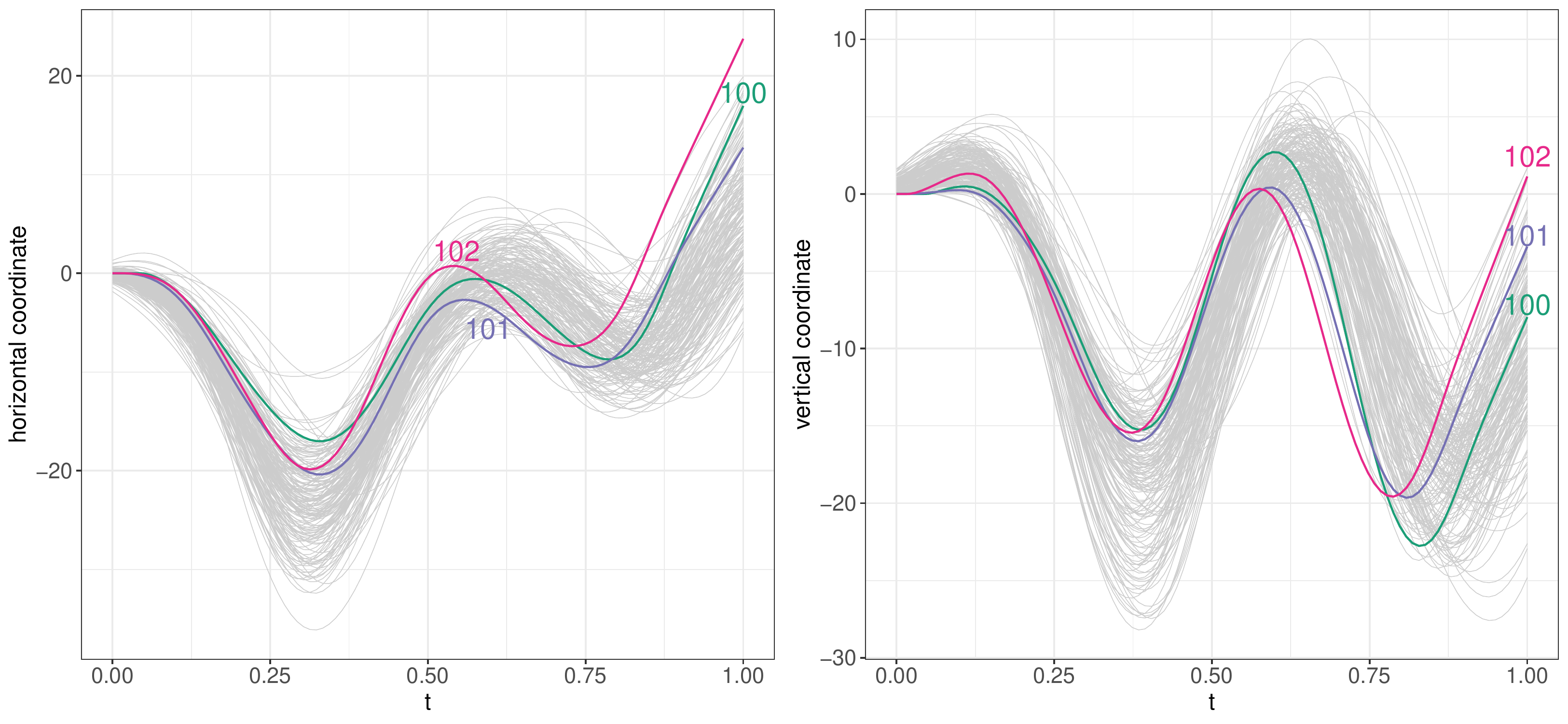}
\caption{Some shape outliers detected by FastMUOD with a shift to the left in peaks resulting in ``long follow-throughs". See Figure \ref{fig::mag_amp_charA} of the Main Text. \label{fig::charA_sha_grp2}}
\end{figure}

\begin{figure}[htbp!]
	\centering
\includegraphics[scale = .50]{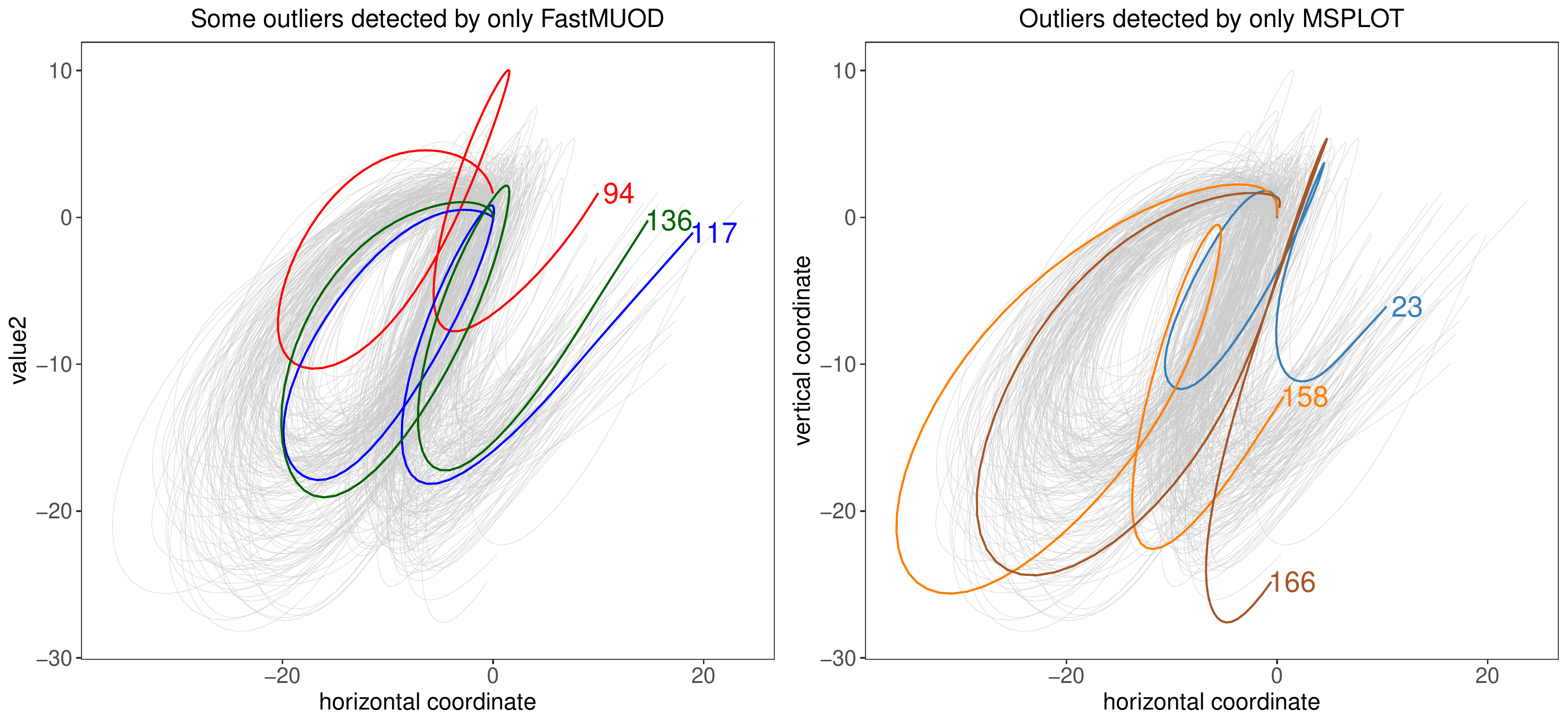}
\caption{Outliers detected by only FastMUOD and only MSPLOT.\label{fig::ex_outliers_carA}}
\end{figure}

\end{document}